\documentclass[twocolumn,preprintnumbers,amsmath,amssymb]{revtex4}
\usepackage{color}
\usepackage{bm, graphicx, amsmath}
\usepackage{hyperref}

\usepackage{algorithm}
\usepackage[noend]{algpseudocode}

\usepackage{fancyhdr}
\begin{document}
	
	\fancyhead[R]{\ifnum\value{page}<2\relax\else\thepage\fi}
	
	\title{Gaussian Amplitude Amplification for Quantum Pathfinding}
	\author{Daniel Koch$^{1,2}$$^{*}$, Massimiliano Cutugno$^{2}$, Samuel Karlson$^{3}$, Saahil Patel$^{2}$, Laura Wessing$^{2}$, Paul M. Alsing$^{2}$}
	\affiliation{$^{1}$The Griffiss Institute, Rome, NY }
	\affiliation{$^{2}$Air Force Research Lab, Information Directorate, Rome, NY }
	\affiliation{$^{3}$Air Force Academy, Colorado Springs, Co }
	\affiliation{$^{*}$Corresponding Author: daniel.koch.13@us.af.mil  }
	
	\begin{abstract}
		
		We study an oracle operation, along with its circuit design, which combined with the Grover diffusion operator boosts the probability of finding minimum or maximum solutions on a weighted directed graph.  We focus on a geometry of sequentially connected bipartite graphs, which naturally gives rise to solution spaces describable by gaussian distributions.  We then demonstrate how an oracle which encodes these distributions can be used to solve for the optimal path via amplitude amplification.  And finally, we explore the degree to which this algorithm is capable of solving cases which are generated using randomized weights, as well as a theoretical application for solving the Traveling Salesman problem.
		
	\end{abstract}

	\maketitle
	
	\thispagestyle{fancy}
	
	%%%%%%%%%%%%%%%%%%%%%%%%%%%%%%%%%%%%%%%%%%%%%%%%
	%%%%%%%%%%%%%%%%%%%%%%%%%%%%%%%%%%%%%%%%%%%%%%%%
	\section{Introduction}%                                     Introduction
	%%%%%%%%%%%%%%%%%%%%%%%%%%%%%%%%%%%%%%%%%%%%%%%%
	%%%%%%%%%%%%%%%%%%%%%%%%%%%%%%%%%%%%%%%%%%%%%%%%
	
	The use of quantum computers for tackling difficult problems is an exciting promise, but not one without its own set of challenges.  Qubits allows for incredible parallelism in computations via superposition states, but reliably pulling out a single answer via measurements is often difficult.  In 1996, Grover demonstrated one of the first mechanisms overcoming this weakness \cite{grover}, later shown to be optimal \cite{boyer,bennett}, and has since been refined into a broader technique in quantum algorithms known as `amplitude amplification' \cite{farhi,brassard1,brassard2,childs,ambainis,singleton}. In this study we seek to extend the capabilities of amplitude amplification as a means of pathfinding on a directed graph with weighted edges.
	
	The success of Grover's algorithm can be boiled down to two primary components: the oracle operation $U_{\textrm{G}}$ and diffusion operation $U_{\textrm{s}}$.  While $U_{\textrm{s}}$ is typically considered a straightforward mathematical operation --- achieving a reflection about the average amplitude --- critics of Grover's algorithm often point to $U_{\textrm{G}}$ as problematic \cite{lloyd,viamontes,reg,seidel}.  Neilsen and Chuang elegantly describe the dilemma of implementing $U_{\textrm{G}}$ as differentiating between an operation which $\textit{knows}$ the desired marked state, versus a true blackbox oracle which can $\textit{recognize}$ the answer \cite{nielsen}.  Only an oracle of the latter case can truly be considered a speedup for quantum, otherwise the solution to the unstructured search problem is already encoded into $U_{\textrm{G}}$, defeating the purpose of using a quantum computer in the first place.  We note this specific issue with Grover's algorithm because it is exactly the problem we aim to address in this study, specifically for the gate-based model of quantum computing.  In this study, we demonstrate an alternative to the standard Grover oracle, which we refer to as a `cost oracle' $U_{\textrm{P}}$, capable of solving weighted directed graph problems.
	
	Beyond the specific geometry used to motivate $U_{\textrm{P}}$ and build its corresponding quantum circuit, much of this study is aimed at formulating a deeper understanding of amplitude amplification.  The idea of using an oracle which applies phases different from the standard $U_{\textrm{G}}$ was first investigated by Long and Hoyer \cite{long1,long2,hoyer} and later others \cite{younes,li1,guo}, who showed the degree to which a phase other than $\pi$ on the marked state(s) could still be used for probability boosting.  Here, we study a $U_{\textrm{G}}$ replacement which affects $\textit{all}$ states with unique phases, not just a single marked state.  Consequently, the effect of $U_{\textrm{P}}$ is analogous to a cost function, whereby $U_{\textrm{P}}$ acting on any state results in a phase proportional to that state's representative weighted path.  The advantage for quantum is to utilize superposition, evaluating all costs simultaneously, and ultimately boosting the probability of measuring the solution to the optimization problem.  Using $U_{\textrm{P}}$ results in an amplitude amplification process which is more complex than standard Grover's, but still achieves high probabilities under ideal conditions.  And most importantly, we demonstrate the degree to which probability boosting is possible under randomized conditions which one would expect from realistic optimization problems \cite{song,pomeransky,janmark}.

	After demonstrating results for the success of $U_{\textrm{P}}$, the final topic of this study is a theoretical application of cost oracles for solving the Traveling Salesman problem (TSP) \cite{gutin}, or all-to-all connected directed graphs.  Notable strategies thus far for a quantum solution to the TSP are based on phase estimation \cite{srinivasan}, backtracking \cite{moylett}, and adiabatic quantum computing \cite{martonak,warren,warren2,chen}.  Here we approach the problem from an amplitude amplification perspective, continuing an idea which goes back over a decade \cite{bang}.   However, in order to realize the appropriate quantum states for this application of $U_{\textrm{P}}$, we must look beyond binary superposition states provided by qubits, in favor of a mixed qudit quantum computer which more naturally suits the problem.  Although still in their technological infancy compared to qubits, realization of qudit technologies \cite{kues,low,yurtalan,lu}, qudit-based universal computation \cite{niu,luo}, their fundamental quantum circuits \cite{li2,lanyon,gokhale,khan,muth,daboul}, and algorithm applications \cite{blok,hu} have all seen significant advancements over the last decade, making now an exciting time to consider their use for future algorithms.

	%%%%%%%%%%%%%%%%%%%%%%%%%%%%%%%%%%%%%%%%
	\subsection{Layout}%                                                                                                                                                                         Layout
	%%%%%%%%%%%%%%%%%%%%%%%%%%%%%%%%%%%%%%%%
	
	Section II. begins with an alternative oracle to Grover's $U_{\textrm{G}}$, which we use to introduce fundamental features of amplitude amplification and oracle operations.  The progression of this study then revolves around a specific directed graph problem, where the underlying characteristics of each graph's solution space are describable by the Central Limit Theorem \cite{laplace} and the Law of Large Numbers \cite{bernoulli}, resulting in solution space distributions which resemble a gaussian function \cite{gauss}.  Section III. covers specifics of this weighted directed graph problem, a graphical representation of all possible paths, and a proposed classical solving speed based on arguments of information access.  Sections IV. and V. show how each graph can be represented as a pathfinding problem, translated into quantum states, and ultimately solved using a modified Grover's algorithm.  In section VI. we present results from simulated perfect gaussian distributions, providing insight into fundamental properties of optimization problems which are viable for amplitude amplification.  In section VII. we explore the viability of using a cost oracle to solve optimization problems involving randomness.  Section VIII. explores a theoretical application of $U_{\textrm{P}}$ for solving the Traveling Salesman problem, and finally section IX. concludes with a summary of our findings and discussions of future research.

	%%%%%%%%%%%%%%%%%%%%%%%%%%%%%%%%%%%%%%%%%%%%%%%%
	%%%%%%%%%%%%%%%%%%%%%%%%%%%%%%%%%%%%%%%%%%%%%%%%
	\section{Gate-Based Grover's}%                                                                                                                                               
	%%%%%%%%%%%%%%%%%%%%%%%%%%%%%%%%%%%%%%%%%%%%%%%%
	%%%%%%%%%%%%%%%%%%%%%%%%%%%%%%%%%%%%%%%%%%%%%%%%
	
	Shown below in equation \ref{E1} is $U_{\textrm{s}}$, known as the diffusion operator, which is the driving force behind amplitude amplification.  The power of this operation lies in its ability to reflect every state in the quantum system about the average amplitude $without$ computing the average itself. 
	
	\begin{eqnarray}            
		U_{\textrm{s}}  = 2 | s \rangle \langle s | - \mathbb{I}
		\label{E1}
	\end{eqnarray} 
	
	In order to make use of this powerful geometric operation, we must pair it with an oracle operator in order to solve interesting problems.  For clarity, in order for an operator to qualify as an oracle, we require that the probability of measuring each state in the system must be the same before and after applying the oracle.  This requirement excludes any and all operations which cause interference, leaving only one type of viable operator: phase gates.  Thus, it is the aim of this study to investigate viable oracle operations which encode the information of problems into phases, and solve them using amplitude amplification.

	%%%%%%%%%%%%%%%%%%%%%%%%%%%%%%%%%%%%%%%%
	\subsection{Optimal Amplitude Amplification}%                                                        Optimal Amplitude Amplification
	%%%%%%%%%%%%%%%%%%%%%%%%%%%%%%%%%%%%%%%%
	
	It is important to understand why the standard Grover oracle $U_{\textrm{G}}$, given in equation \ref{E2} and used in algorithm \ref{A1}, is optimal in the manner in which it boosts the probability of the marked state to nearly $1$ \cite{boyer,bennett}.  Geometrically, this is because the entire amplitude amplification process takes place along the real axis in amplitude space (i.e. at no point does any state acquire an imaginary amplitude component).  Consequently, the marked state, origin, mean amplitude point, and non-marked states are all linearly aligned, which ensures that the marked state receives the maximal reflection about the average (mean point) at each step.  This property holds true for not only the real axis, but any linear axis that runs through the origin, so long as the marked and non-marked states differ in phase by $\pi$ as a result of the oracle operation.
	
	\begin{eqnarray}            
		U_{\textrm{G}} | \Psi \rangle = 
		\begin{cases}
			$\textrm{marked}$ &, e^{i \pi} | \Psi_i \rangle\\
			$\textrm{non-marked}$ &,  | \Psi_i \rangle
		\end{cases}  
		\label{E2}
	\end{eqnarray} 
	
	\begin{algorithm}
		\caption{Grover's Search Algorithm}
		\begin{algorithmic}[1]
			\State Initialize  Qubits: $|\Psi\rangle = |0\rangle ^{\otimes N}$
			\State Prepare Equal Superposition: $H^{\otimes N} |\Psi \rangle = |s\rangle$
			\For{ $k \approx  \frac{\pi}{4} \sqrt{ 2^N }$  }  
			\State Apply $U_\textrm{G} |\Psi\rangle$ (Oracle)
			\State Apply $U_\textrm{s} |\Psi\rangle$ (Diffusion)
			\EndFor
			\State Measure
		\end{algorithmic}
		\label{A1}
	\end{algorithm}
	
	We note the optimality of $U_{\textrm{G}}$ because it is directly tied to the nature of the problem which it solves, namely an unstructured search \cite{grover}.  The power of amplitude amplification using $U_{\textrm{G}}$ goes hand-in-hand with the rigidness of the operator.  Thus, if we want to expand the capabilities of amplitude amplification on gate-based quantum computers to more interesting problems, we must explore more flexible oracle operators, and consequently expect probability boosting that is less than optimal.

	%%%%%%%%%%%%%%%%%%%%%%%%%%%%%%%%%%%%%%%%
	\subsection{Alternate Two-Marked Oracle}%                                             Alternate Two-Marked Oracle
	%%%%%%%%%%%%%%%%%%%%%%%%%%%%%%%%%%%%%%%%
	
	Here we present an example analogous to Grover's search algorithm with two marked states, but with an oracle operator of our own design.    The purpose of this exercise is to illustrate several key ideas that will be prominent throughout the remainder of this study.  Firstly, the general idea of a multi-phased oracle operation \cite{satoh,bench}, or `non-boolean' oracles \cite{shyamsundar}.  Secondly, to demonstrate that the success of amplitude amplification can be directly traced back to inherent mathematical properties of an oracle.  And finally, to introduce terminology and features of amplitude amplification on discrete systems which will apply to later oracles.  All of the following results were verified using  IBM's Qiskit simulator as well as our own python-based simulator.
	
	\begin{eqnarray}            
		U'_{G2} | \Psi \rangle = \begin{cases}
			|0\rangle^{\otimes N}  &, |0\rangle^{\otimes N}  \\
			|1\rangle^{\otimes N}  &, e^{i \pi} |1\rangle^{\otimes N}  \\
			| \Psi_i \rangle \in | \textrm{G}_{\theta} \rangle &,  e^{i \theta}| \Psi_i \rangle         \\
			| \Psi_i \rangle \in | \textrm{G}_{ \textrm{-} \theta} \rangle &,  e^{\textrm{-}i \theta}| \Psi_i \rangle                                                                     
		\end{cases}  
		\label{E3}
	\end{eqnarray} 
	
	where
	
	\begin{eqnarray}         
		| \textrm{G}_{\theta} \rangle &\equiv&  \textrm{all} \hspace{0.2cm} | \Psi_i \rangle = |0\rangle \otimes | \psi \rangle,  \hspace{0.3cm}  | \psi \rangle \neq |0\rangle^{\otimes N-1} \nonumber \\
		| \textrm{G}_{\textrm{-}\theta} \rangle &\equiv&  \textrm{all} \hspace{0.2cm} | \Psi_i \rangle = |1\rangle \otimes | \psi \rangle, \hspace{0.3cm}  | \psi \rangle \neq |1\rangle^{\otimes N-1} 
		\label{E4}
	\end{eqnarray} 
	
\begin{figure}[h]               
		\centering
		\includegraphics[scale=.35]{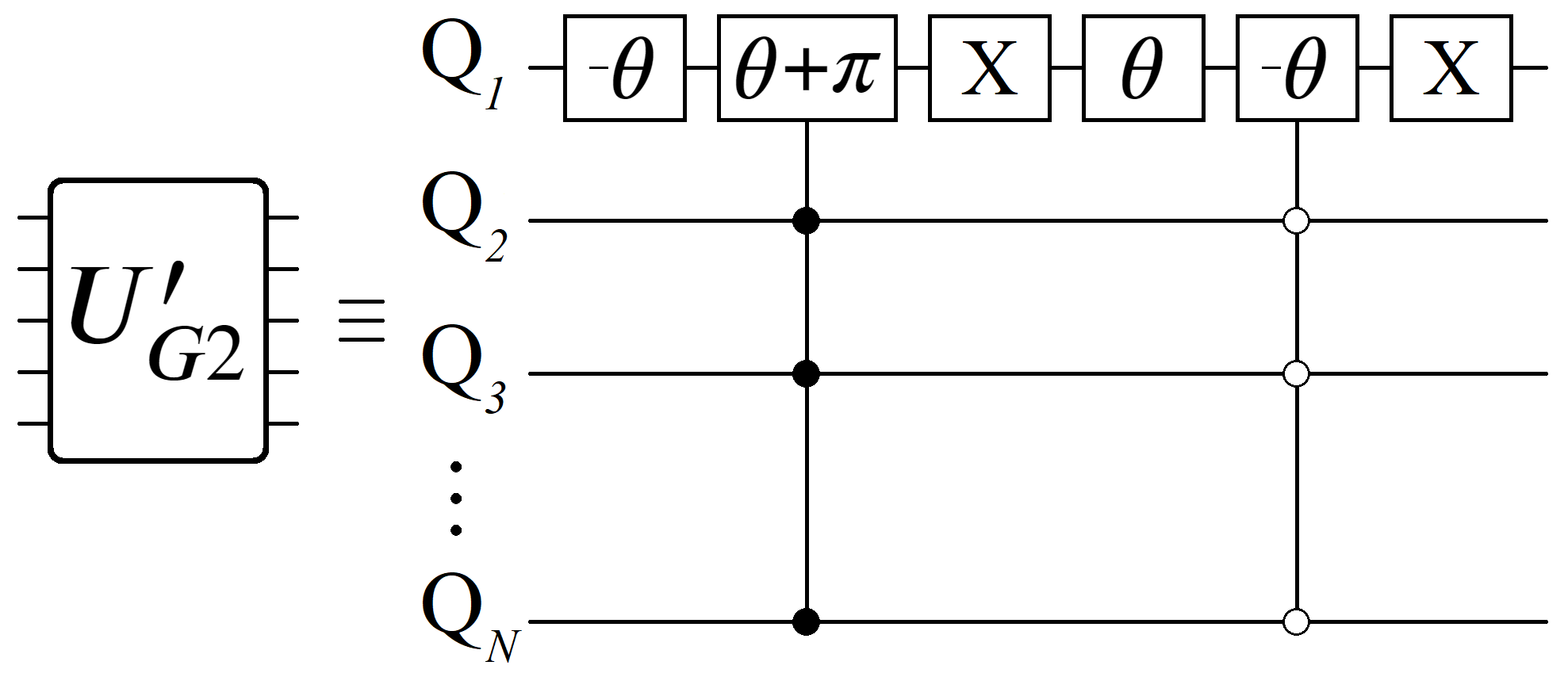}
		\caption{ Quantum circuit for implementing $U'_{G2}$.  Boxes with $\theta$ and $\pi$ are phase gates, both single and controlled. For the controlled operations, black dots indicate a $|1\rangle$ control state, and similarly white dots for $|0\rangle$.}
		\label{F1}
	\end{figure}

\begin{figure}[h]             
		\centering
		\includegraphics[scale=.35]{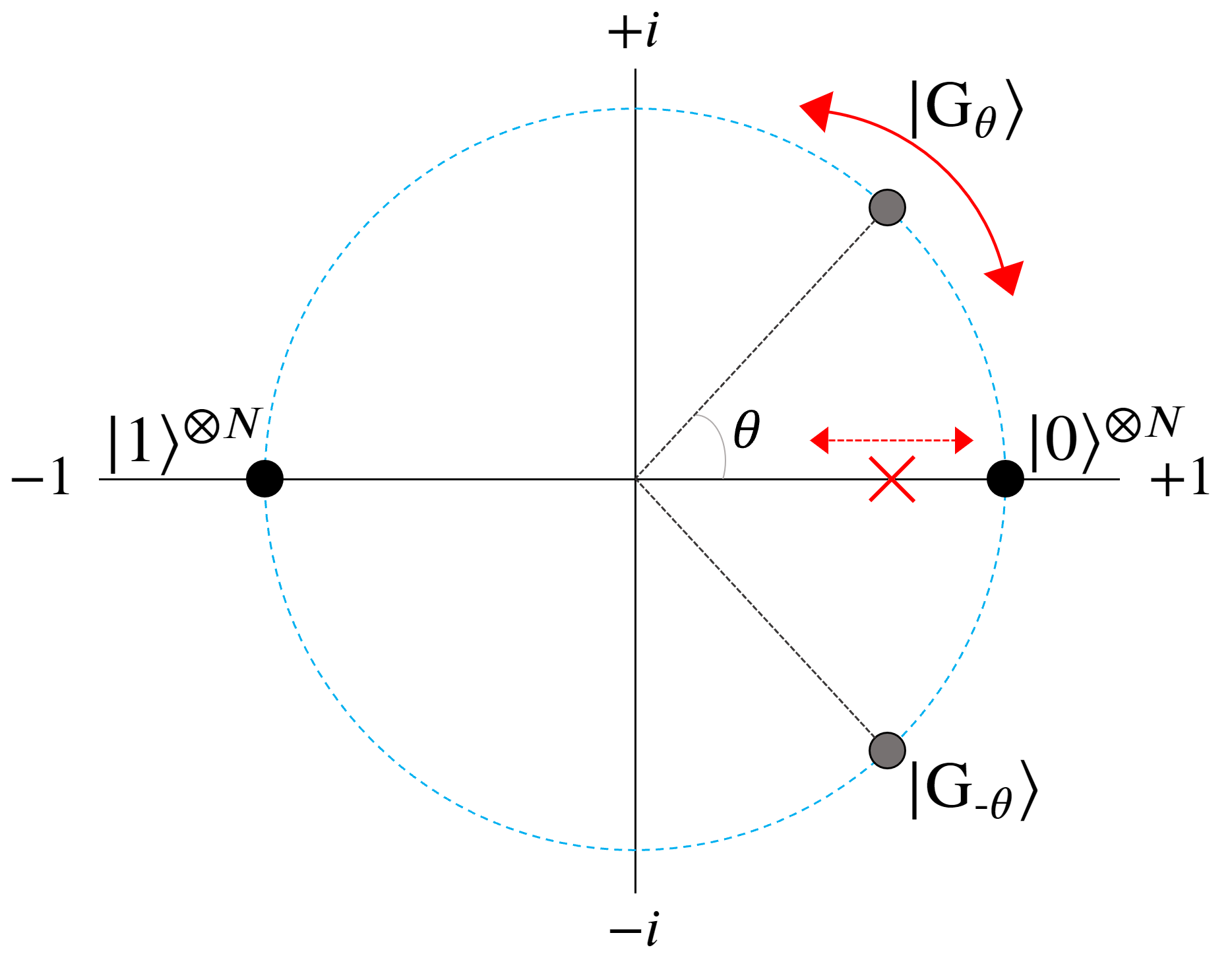}
		\caption{ An illustration of $U'_{G2} | s \rangle$.  A unit circle of radius 1/$\sqrt{2^N}$ is shown by the blue-dashed line, along with the point of average amplitude with a red `X'.  The parameter $\theta$ controls the phase acquired by the cluster of states $| \textrm{G}_{ \theta} \rangle$ and $| \textrm{G}_{ \textrm{-} \theta} \rangle$ , which in turn dictates the location of the mean point along the real axis.}
		\label{F2}
	\end{figure}

	We begin with the mathematical definition of our oracle function in equation \ref{E3} above, which we shall refer to as $U'_{G2}$, as well as its quantum circuit composition in figure \ref{F1}.  Contrary to equation \ref{E2}, we now have an oracle operation with four distinct outcomes depending on which state $| \Psi_i \rangle$ it is acting on.  Additionally, $U'_{G2}$ has a free parameter $\theta$, controlled by the experimenter, which dictates how the states $|0\rangle^{\otimes N}$ and $|1\rangle^{\otimes N}$ boost in probability.  Altogether, the effect of $U'_{G2}$ can be seen in figure \ref{F2}, which displays the position of each state in amplitude space (the complex plane) after the first application: $U'_{G2} | s \rangle$.
	
\begin{figure}[h]                 
		\centering
		\includegraphics[scale=.35]{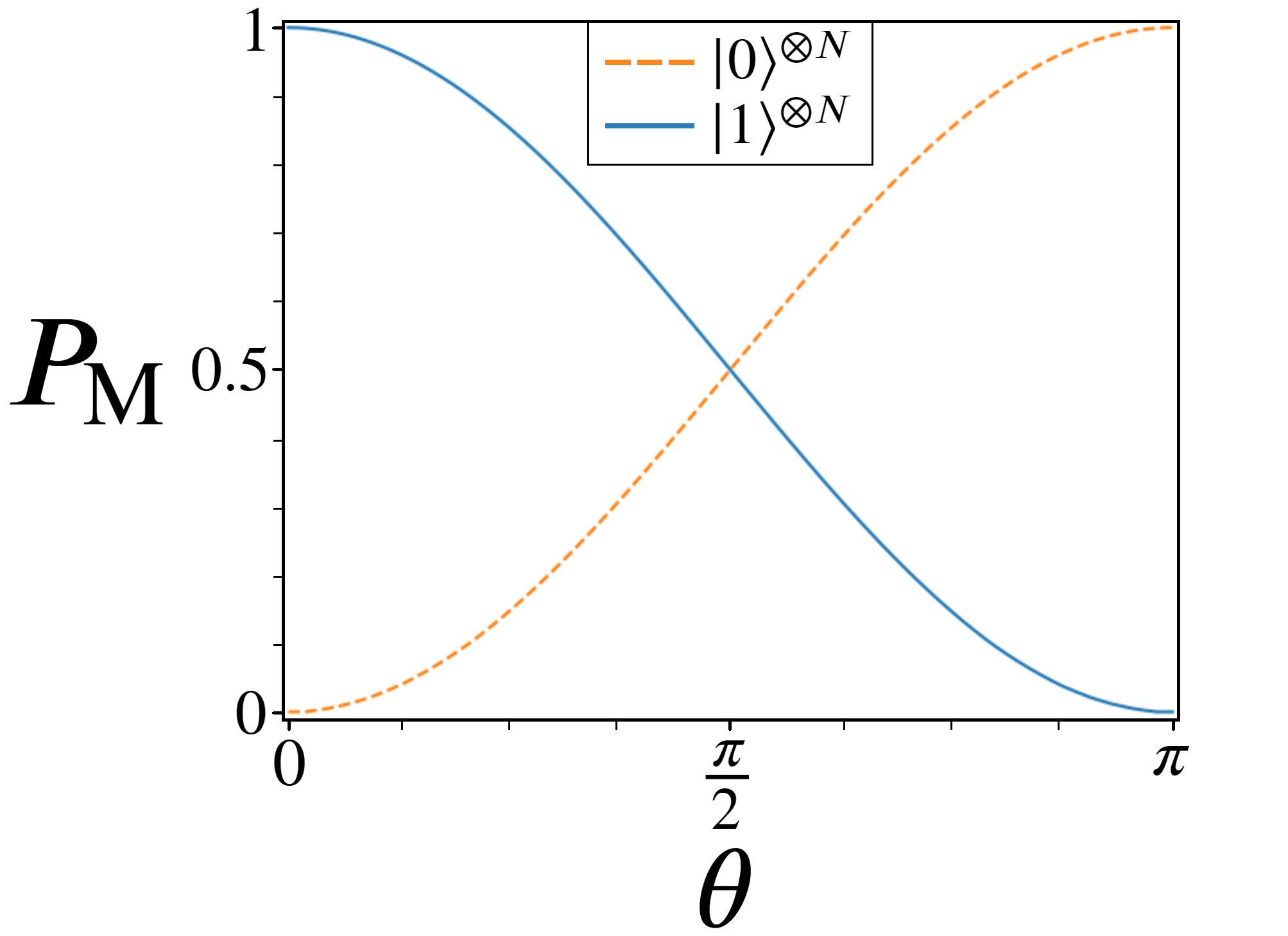}
		\caption{ A plot of $\theta$ vs. peak probability $P_{\textrm{M}}$ for the states $| 0 \rangle^{\otimes N}$ (orange-dashed) and $| 1 \rangle^{\otimes N}$ (blue-solid).  Approximate forms for the two plots are given in equations \ref{E5} and \ref{E6}.  }
		\label{F3}
	\end{figure}

	Before revealing how this alternate two-marked oracle performs at amplitude amplification, note the red `X' located along the real axis of figure \ref{F2}.  This `X' marks the mean point, or average amplitude, where every state in the system will be reflected about after the first diffusion operator $U_{\textrm{s}}$. Because $2^N - 2$ states are evenly distributed between $| \textrm{G}_{\theta} \rangle$ and $| \textrm{G}_{\textrm{-}\theta} \rangle$, this initial mean point can be made to lie anywhere along the real axis between $( -1/\sqrt{2^N}, 1/\sqrt{2^N} )$ as $\theta$ ranges from $0$ to $\pi$.  Shown in figure \ref{F3} below is the relation between $\theta$ and the resulting probability boosts for $| 0 \rangle^{\otimes N}$ and $| 1 \rangle^{\otimes N}$.
	
	We define the metric $P_{\textrm{M}}$, shown as the y-axis in figure \ref{F3}, to be the peak probability achievable through amplitude amplification as defined in algorithm \ref{A1} for a given state.  Here we track $P_{\textrm{M}}$ for the states $| 0 \rangle^{\otimes N}$ and $| 1 \rangle^{\otimes N}$ as a function of $\theta$, for the case of $N=20$.  Firstly, note the two extremes of $\theta$: $0$ and $\pi$, for which the resulting amplitude amplification processess are $exactly$ equal to standard Grover's for $| 1 \rangle^{\otimes N}$ and $| 0 \rangle^{\otimes N}$ respectively.  This is in agreement with the geometric picture of $U'_{G2}$ outlined in figure \ref{F2}, whereby all of the states of $| \textrm{G}_{\theta} \rangle$ and $| \textrm{G}_{\textrm{-}\theta} \rangle$ recieve phases of $0$ or $\pi$, isolating a single state to be $\pi$ phase different from the remaining $2^N - 1$ states.
	
	While $U'_{G2}$ is able to reproduce $U_{\textrm{G}}$ at the $\theta$ bounds, it is the intermediate $\theta$ values which are more revealing about the capabilities of amplitude amplification.  For sufficiently large $N$, the mean point produced from $U'_{G2}$ is dominated by the states making up $| \textrm{G}_{\theta} \rangle$ and $| \textrm{G}_{\textrm{-}\theta} \rangle$,  approximately equal to $\approx 1/\sqrt{2^N} \cdot \textrm{cos}(\theta)$ (the real axis).  We note this cos($\theta$) dependance because it also describes the two $P_{\textrm{M}}$ plots shown in figure \ref{F3}, given by equations \ref{E5} and \ref{E6} below.

	\begin{eqnarray}              
		P_{\textrm{M}}( | 1 \rangle^{\otimes N} ) &\approx& \frac{1}{2}( \textrm{cos}(\theta) + 1 )     \label{E5}     \\
		P_{\textrm{M}}( | 0 \rangle^{\otimes N} ) &\approx& \frac{1}{2}( \textrm{cos}(\theta - \pi) + 1 )
		\label{E6}
	\end{eqnarray} 
	
	The emphasis here is that we have a one-to-one correlation between a property of $U'_{G2}$, specifically $\theta$ , and the resulting peak probabilities $P_{\textrm{M}}$ achievable through amplitude amplification.  But more accurately, $\theta$ is just a parameter for controlling the mean amplitude point produced by $U'_{G2}$, which is the more fundamental indicator of successful amplitude amplification.  This is evidenced by the cos($\theta$) relation found in both $P_{\textrm{M}}$ plots here, as well as properties of oracle operators to come in this study, which can similarly be directly linked to the initial mean points they produce.

	%%%%%%%%%%%%%%%%%%%%%%%%%%%%%%%%%%%%%%%%%%%%%%%%
	%%%%%%%%%%%%%%%%%%%%%%%%%%%%%%%%%%%%%%%%%%%%%%%%
	\section{Pathfinding Geometry}%                                                                                                                                                  Pathfinding Geometry
	%%%%%%%%%%%%%%%%%%%%%%%%%%%%%%%%%%%%%%%%%%%%%%%%
	%%%%%%%%%%%%%%%%%%%%%%%%%%%%%%%%%%%%%%%%%%%%%%%%
	
	While the $U'_{G2}$ oracle is useful for gaining insight into non-boolean amplitude amplification processes, ultimately it does not correspond to a meaningful problem we would ideally look to a quantum computer to solve.  In particular, we want an oracle operation which boosts a quantum state unknown to the experimenter beforehand, yielding the answer to some unsolved problem.  To this end, we now introduce one such optimization problem which can be encoded as an oracle and ultimately solved through amplitude amplification.

	%%%%%%%%%%%%%%%%%%%%%%%%%%%%%%%%%%%%%%%%
	\subsection{Graph Structure}%                                                                                                                                Geometry Structure      
	%%%%%%%%%%%%%%%%%%%%%%%%%%%%%%%%%%%%%%%%
	
\begin{figure}[h]            
		\centering
		\includegraphics[scale=.35]{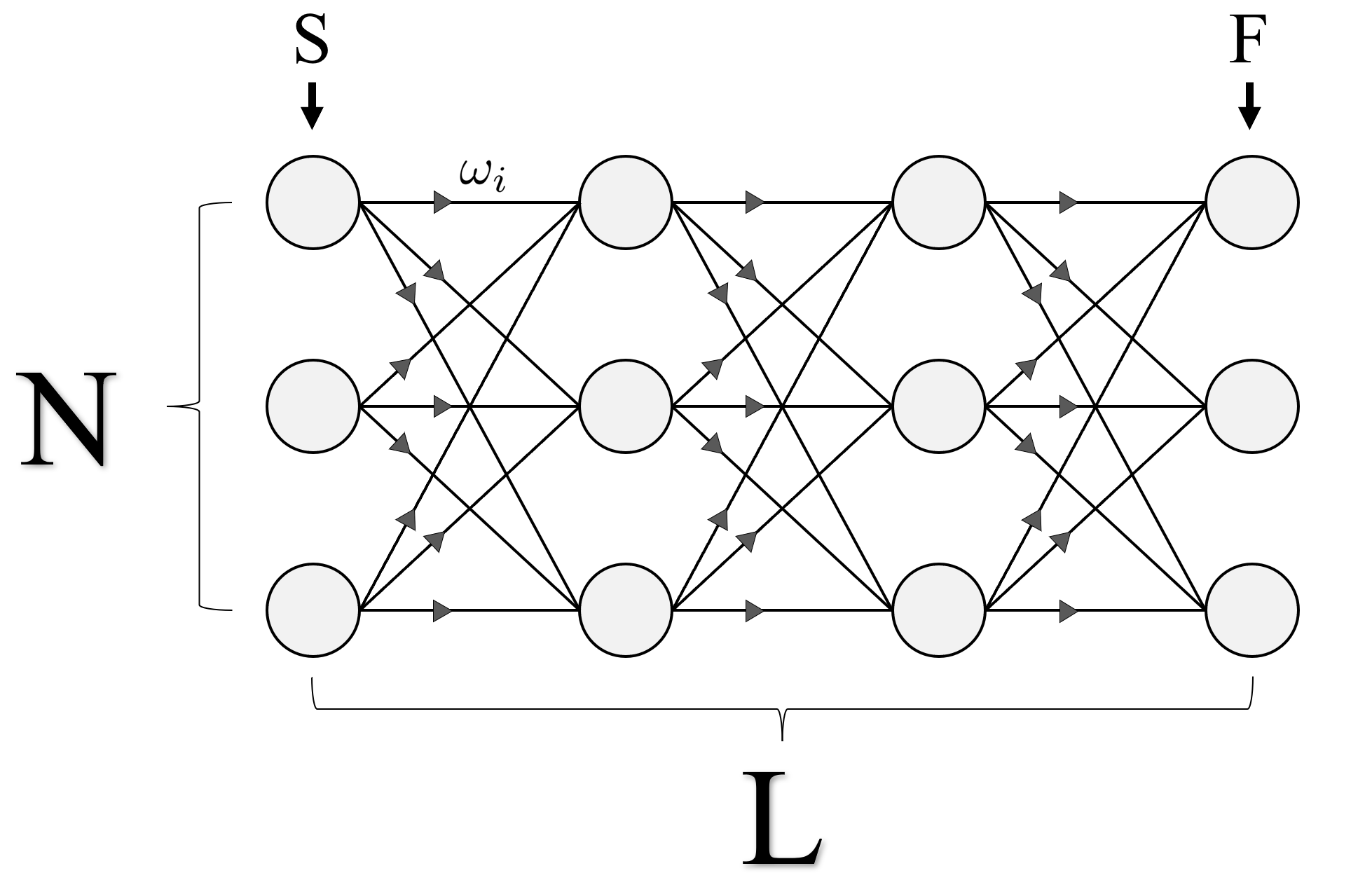}
		\caption{A geometry composed of sequentially connected bipartite directed graphs with weighted edges for which we are interested in finding the optimal path from layer S to layer F, touching exactly 1 node per layer. $N$ denotes the number of nodes per layer, while $L$ is the total number of layers.  With full connectivity between nearest neighboring layers, each geometry has a total of $N^{2} \cdot (L-1)$ edges, yielding $N^{L}$ possible paths from layer $S$ to $F$. }
		\label{F4}
	\end{figure}

	Shown in figure \ref{F4} is the general structure of the problem which will serve as the first primary focus for this study: a series of sequentially connected bipartite graphs with weighted edges, for which we are interested in finding the path of least or greatest resistance through the geometry.  More formally, we seek the solution to a weighted directed graph optimization problem.   Each geometry can be specified by two variables, $N$ and $L$, which represent the number of vertices per column and total number of columns respectively.  Throughout this study, we refer to vertices as `nodes', and each complete set of nodes in a vertical column as a `layer'.  For example, figure \ref{F4}'s geometry represents a 4-layer system ($L=4$), with 3 nodes per layer ($N=3$).
	
	Given the geometric structure shown above, we now assign a complete set of weights $\omega_{i}$, one for each of the total $N^{2} \cdot (L-1)$ edges throughout the geometry.  These weights are one directional, as we only consider solutions which span the full geometry from layer S to F in figure \ref{F4}.  In total, there are $N^L$ solutions to the directed graph, which we refer to as `paths'.  For clarity, a single path $\textrm{P}_j$ is defined as the collection of edges which span from the leftmost to rightmost layers (S to F), touching exactly one node in every layer (see figure \ref{F7} for an $N=2$ example). 
	
	\begin{eqnarray}             
		\omega_i &\in& [0,\textrm{R}], \hspace{0.5cm}  \omega_i \in \mathbb{Z} \label{E7}\\
		W_{j} &=& \sum_{i \hspace{.06cm}\in \hspace{.06cm} \textrm{P}_j} \omega_i   \label{E8}  \\
		\mathbb{P} &\equiv& \{\ \textrm{P}_1,  \textrm{P}_2, ..., \textrm{P}_{N^L} \}  \equiv \textrm{All Paths}    \label{E9} 
	\end{eqnarray} 
	
	For each path $\textrm{P}_j$, there is a cumulative weight $W_j$ that is obtained by summing the individual weighted edges that make up the path (Eqn. \ref{E8}).  The goal is to find the optimal solution path with a cumulative weight of either $W_{\textrm{min}}$ or $W_{\textrm{max}}$:

	\begin{eqnarray}        
		\mathbb{W}  &\equiv& \{\ W_1,  W_2, ..., W_{N^L} \}  \equiv \textrm{All Solutions}    \label{E10}  \\
		W_{\textrm{min}}  &=&  \textrm{min}(\mathbb{W}) \label{E11}  \\ 
		W_{\textrm{max}}  &=& \textrm{max}(\mathbb{W})   \label{E12}
	\end{eqnarray}

	For simplicity, we consider problems where each edge $\omega_i$ is an integer number between $0$ and some max R.  This will allow for a clearer picture when visualizing solution spaces $\mathbb{W}$ later on.  However, we note that all the results which follow are equally applicable to the continuous case $\omega_i \in \mathbb{R}$ (set of real numbers), which we discuss in section V.

	%%%%%%%%%%%%%%%%%%%%%%%%%%%%%%%%%%%%%%%%
	\subsection{Classical Solving Speed}%                                                                                                                                Classical Solving Speed  
	%%%%%%%%%%%%%%%%%%%%%%%%%%%%%%%%%%%%%%%%
	
\begin{figure}[h]              
		\centering
		\includegraphics[scale=.23]{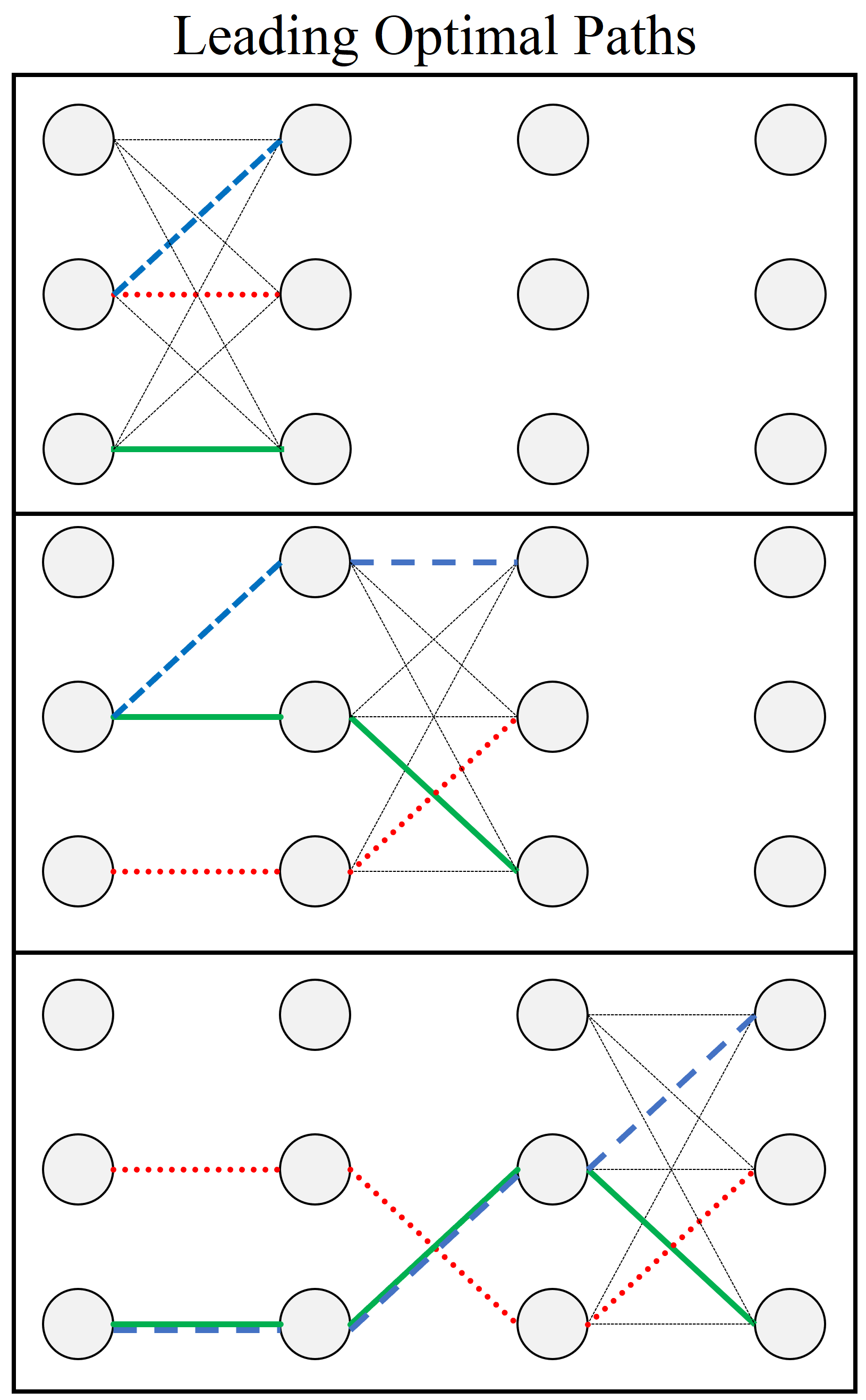}
		\caption{A layer by layer example of a classical approach to finding $W_{\textrm{min}}$ or $W_{\textrm{max}}$, for the case of $N=3$ and $L=4$. The blue-dashed, green-solid, and red-dotted lines each represent possible solutions for the optimal path ending on each of the three nodes per layer.}
		\label{F5}
	\end{figure}

	As outlined in equations \ref{E7}-\ref{E12}, we are interested in finding the path (collection of weighted edges) which corresponds to the smallest or largest $W_{i}$ value within the set $\mathbb{W}$.  However, the cumulative values $W_{i}$ are assumed to be initially unknown, and must be computed from a given directed graph like in figure \ref{F4}.  Importantly, this means that the base amount of information given to either a classical or quantum computer is the set of $\omega_i$ weights and their locations, for which either computer must then find an optimal solution.  For graphs defined according to figure \ref{F4}, yielding $N^2 \cdot (L - 1)$ total weights, we argue that the optimal classical solving speed is of this order.   Figure \ref{F5} below is an example of how a classical algorithm solves the pathfinding problem one layer at a time, checking each weighted edge exactly once.
	
	The steps illustrated in figure \ref{F5} can be summarized as the recursive process given in algorithm \ref{A2}.  The general strategy is to work through the graph one layer at a time, checking all $N^2$ edges between layers, and continually updating a list (labeled OP in Alg. \ref{A2}) of possible optimal paths as one moves through the geometry.  Importantly, the classical algorithm only needs to check each weighted edge one time in order to determine the optimal path. At each layer of the algorithm, $N$ candidate paths are stored in memory (the blue, red, and green lines in figure \ref{F5}) and used to compute the next $N^2$ possible paths (grey solid lines), repeating this process up to the final layer. 
	
	\begin{algorithm}
		\caption{Classical Pathfinding}
		\begin{algorithmic}[1]
			\State OP = \{ $0,0,...,0$ \}  (length $N$)
			\For{  $L-1$  }  
			\For{  $N^2$ }
			\State Check each edge $\textrm{OP}_k$ + $w_{i}$
			\If{$\textrm{OP}_k$ + $w_{i}$ $\textrm{is optimal}$}
			\State Update OP$_k$
			\EndIf
			\EndFor
			\EndFor
			\State $W_{\textrm{min} / \textrm{max}}$ = $\textbf{min}$/$\textbf{max}$  OP 
		\end{algorithmic}
		\label{A2}
	\end{algorithm}
	
	The algorithm shown above has an O($N^2 \cdot (L - 1)$) query complexity, which we will later compare with quantum.  However, this speed is specifically for directed graphs defined according to figure \ref{F4} and equations \ref{E7}-\ref{E12}.  And while quantum will offer a speedup for certain $N$ and $L$ ranges, this particular speedup is not the primary interest of this study.  As we demonstrate next, these sequential bipartite graphs were chosen to illustrate a problem with an efficient quantum circuit construction for the oracle.  Different graph structures will have varying classical speeds for quantum to compete against, but not all graph structures are easily encoded into quantum states and solvable using amplitude amplification.

	%%%%%%%%%%%%%%%%%%%%%%%%%%%%%%%%%%%%%%%%%%%%%%%%
	%%%%%%%%%%%%%%%%%%%%%%%%%%%%%%%%%%%%%%%%%%%%%%%%
	\section{Quantum Cost Oracle}%                                     Quantum Cost Oracle
	%%%%%%%%%%%%%%%%%%%%%%%%%%%%%%%%%%%%%%%%%%%%%%%%
	%%%%%%%%%%%%%%%%%%%%%%%%%%%%%%%%%%%%%%%%%%%%%%%%
	
	Having now outlined the problem of interest, as well as a classical solving speed, in this section we present the quantum strategy for pathfinding.  We begin by outlining the manner in which all $N^L$ possible paths are uniquely assigned a quantum state, with the goal of encoding each total path weight $W_i$ via phases.  Then later in section IV., we show how these phases can be used for amplitude amplification in order to solve for $W_{\textrm{min}}$ or $W_{\textrm{max}}$.
	
	%%%%%%%%%%%%%%%%%%%%%%%%%%%%%%%%%%%%%%%%
	\subsection{Representing Paths in Quantum}%                                                                                                                                Encoding Path Information
	%%%%%%%%%%%%%%%%%%%%%%%%%%%%%%%%%%%%%%%%
	
\begin{figure}[h]            
		\centering
		\includegraphics[scale=.25]{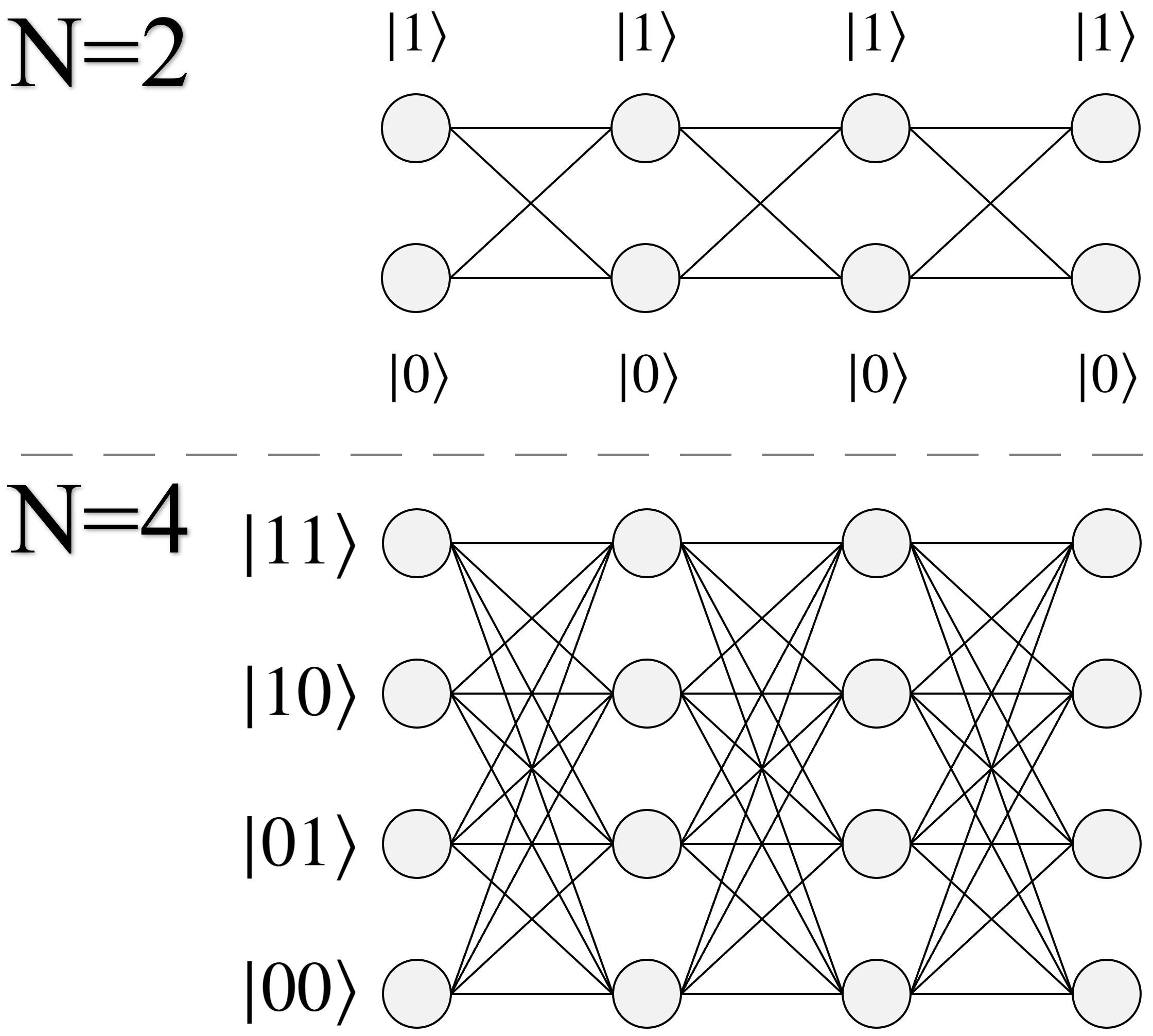}
		\caption{(top) An example geometry of size $N=2$, $L=4$.  For the case of $N=2$, a single qubit is sufficient for representing all possible node choices per layer via the states $|0\rangle$ and $|1\rangle$. (bottom) An example geometry of size $N=4$, $L=4$, requiring two qubits for representing the nodes in each layer.}
		\label{F6}
	\end{figure}

	For qubit-based quantum computing, the methodology put forth in this section is most naturally suited to problem sizes where $N=2^n$ (nodes per layer). This is because $N$ dictates how many quantum states are needed for encoding a layer, for which $2^n$ is achievable using qubits.  We begin by presenting two example cases in figure \ref{F6} of size $N=2$ and $N=4$, both $L=4$.  Accompanying each graph are the qubit states needed to represent each node per layer.
	
	Because we are interested in solving a quantum pathfinding problem, the manner in which the qubits' orthogonal basis states $|0\rangle$ and $|1\rangle$ are used needs to reflect this fact.  A final measurement at the end of the algorithm will yield a state $| \textrm{P}_i \rangle$, comprised of all $|0\rangle$'s and $|1\rangle$'s, from which the experimenter must then extrapolate its meaning as the path P$_i$.  We achieve this by encoding each individual qubit state (or group of qubits) as the location of a particular node in the geometry.  Using $\sqrt{N}$ qubits allows us to identify each of the $N$ nodes per layer (for problem sizes $N = 2^n$), for a total of $\sqrt{N} \cdot L$ qubits representing a complete graph.  For problems of size $N > 2$, multiple qubits are grouped together in order to represent all possible nodes per layer, such as in figure \ref{F6} (two qubits for representing four nodes).

	\begin{figure}[h]        
		\centering
		\includegraphics[scale=.25]{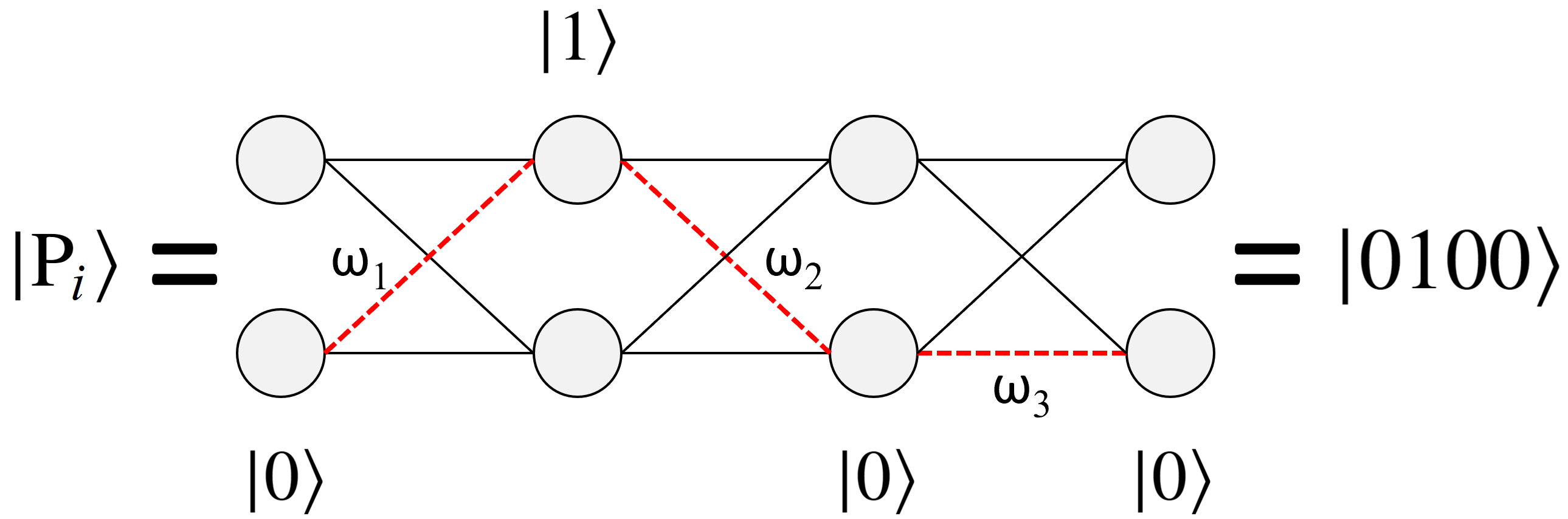}
		\caption{An example path (red-dashed) for a graph of size $N=2$, $L=4$.  The quantum state $|0100\rangle$ represents the path shown in red, using the single qubit states $|0\rangle$  and $|1\rangle$ for bottom and top row nodes respectively. }
		\label{F7}
	\end{figure}
	
	Figure \ref{F7} above shows an example path for $N=2$, and its corresponding state $|\textrm{P}_i \rangle$.  For this particular graph size there are a total of $16$ possible paths, which can be exactly encoded using the basis states $|0\rangle$ and $|1\rangle$ of four qubits.   Conversely for an $N=4$ geometry, two qubits are necessary for representing the four possible nodes per layer (states $|00\rangle$, $|01\rangle$, $|10\rangle$, and $|11\rangle$).  This yields a total of $8$ qubits for the complete graph ($N=4$, $L=4$), for a Hilbert space of size $2^8$, which is exactly equal to the total number of possible paths $4^4$.  With quantum states encoded in this manner, the goal of the algorithm is to measure $| \textrm{P}_{\textrm{min}} \rangle$ or $| \textrm{P}_{\textrm{max}} \rangle$, which will yield the answer $W_{\textrm{min}}$ or $W_{\textrm{max}}$ upon classically checking the path.

	%%%%%%%%%%%%%%%%%%%%%%%%%%%%%%%%%%%%%%%%
	\subsection{Cost Oracle $U_{\textrm{P}}$}%                                                                                                                                Phase Oracle Up
	%%%%%%%%%%%%%%%%%%%%%%%%%%%%%%%%%%%%%%%%
	
	The four qubit state shown in figure \ref{F7} corresponds to a single path, but a superposition state is capable of representing all $2^4$ solutions simultaneously (and more generally any $N^L$).  In order to use these states for finding the optimal path, we now need a mechanism for assigning each path state $|\textrm{P}_i\rangle$ its unique path weight $W_i$.  To achieve this, we implement an operator $U_{\textrm{P}}$, which we refer to as a `cost oracle', capable of applying the cumulative weights $W_i$ of each path through phases:

	\begin{eqnarray}            
		U_{\textrm{P}}  |0100\rangle &=& ( e^{i \omega_1} \cdot e^{i \omega_2} \cdot  e^{i \omega_3}) |0100\rangle   \nonumber \\
		&=&  e^{ i (\omega_1 + \omega_2 + \omega_3) } |0100\rangle  \nonumber \\
		&=&  e^{ i W_{\tiny{0100}} } |0100\rangle
		\label{E13}
	\end{eqnarray}

	In equation $\ref{E13}$ above, we've used the numerical weights $\omega_i$ from figure \ref{F7} as an example, where each edge is directly translated into a phase contribution.  In practice however, a scaling factor $p_{\textrm{s}}$ is necessary for meaningful results (which we discuss in sections IV. and V.).  The reason we refer to $U_{\textrm{P}}$ as a cost oracle is because the manner in which it affects quantum states is analogous to that of a cost function.  More specifically, applying $U_{\textrm{P}}$  to any state  $|\textrm{P}_i \rangle$ will cause the state to pick up a phase proportional to its cumulative weight $W_i$.  However, it is more accurate to call $U_{\textrm{P}}$ an oracle, because the exact manner in which phases are distributed throughout the quantum system is unknown to the experimenter.  That is to say, the experimenter is unaware of which $| \textrm{P}_i \rangle$ state is receiving the desired phase proportional to $W_{\textrm{min}}$ or $W_{\textrm{max}}$ until the conclusion of the algorithm.  The matrix representation of $U_{\textrm{P}}$ has the form of equation \ref{E14} below, where each phase $\phi_i$ is a scalar of the form $p_{\textrm{s}} \cdot W_i$. (the role of $p_{\textrm{s}}$ is discussed later).  The matrix for $U_{\textrm{P}}$ has dimensions $N^L$ x $N^L$, equal to the total number of possible solutions, with each path's phase along the main diagonal.  
	
	\begin{eqnarray}            
		U_{\textrm{P}} | \Psi \rangle 
		= \begin{bmatrix}   
			e^{ i \phi_1} & 0 & 0 & \cdot & \cdot & \hspace{.25cm}  \\
			0 &  e^{ i \phi_2} & 0 & & & \\
			0 & 0 & e^{ i \phi_3} & & & \\
			\cdot &  & & \cdot &  & \\
			\cdot &  & & & \cdot   &       \\
			& & & & &                                                                             
		\end{bmatrix}
		\begin{bmatrix}
			| \textrm{P}_1 \rangle \\
			| \textrm{P}_2\rangle \\
			| \textrm{P}_3\rangle  \\ 
			\cdot \\
			\cdot \\          
			\\                                                                           
		\end{bmatrix}
		\label{E14}
	\end{eqnarray}

	It is important to note that the matrix shown in equation \ref{E14} is $not$ necessary for the quantum circuit implementation of $U_{\textrm{P}}$.  In particular, computing all $N^L$ phases is already slower than the $O$($N^2 \cdot (L-1)$) approach laid out in section II.  Thus, as we demonstrate in the next subsection, a viable quantum approach needs to implement $U_{\textrm{P}}$ $without$ calculating any total path lengths $W_i$.
	
	%%%%%%%%%%%%%%%%%%%%%%%%%%%%%%%%%%%%%%%%
	\subsection{Quantum Circuit}%                                                                                                                                   Quantum Circuit
	%%%%%%%%%%%%%%%%%%%%%%%%%%%%%%%%%%%%%%%%
	
\begin{figure}[h]                    
		\centering
		\includegraphics[scale=.3]{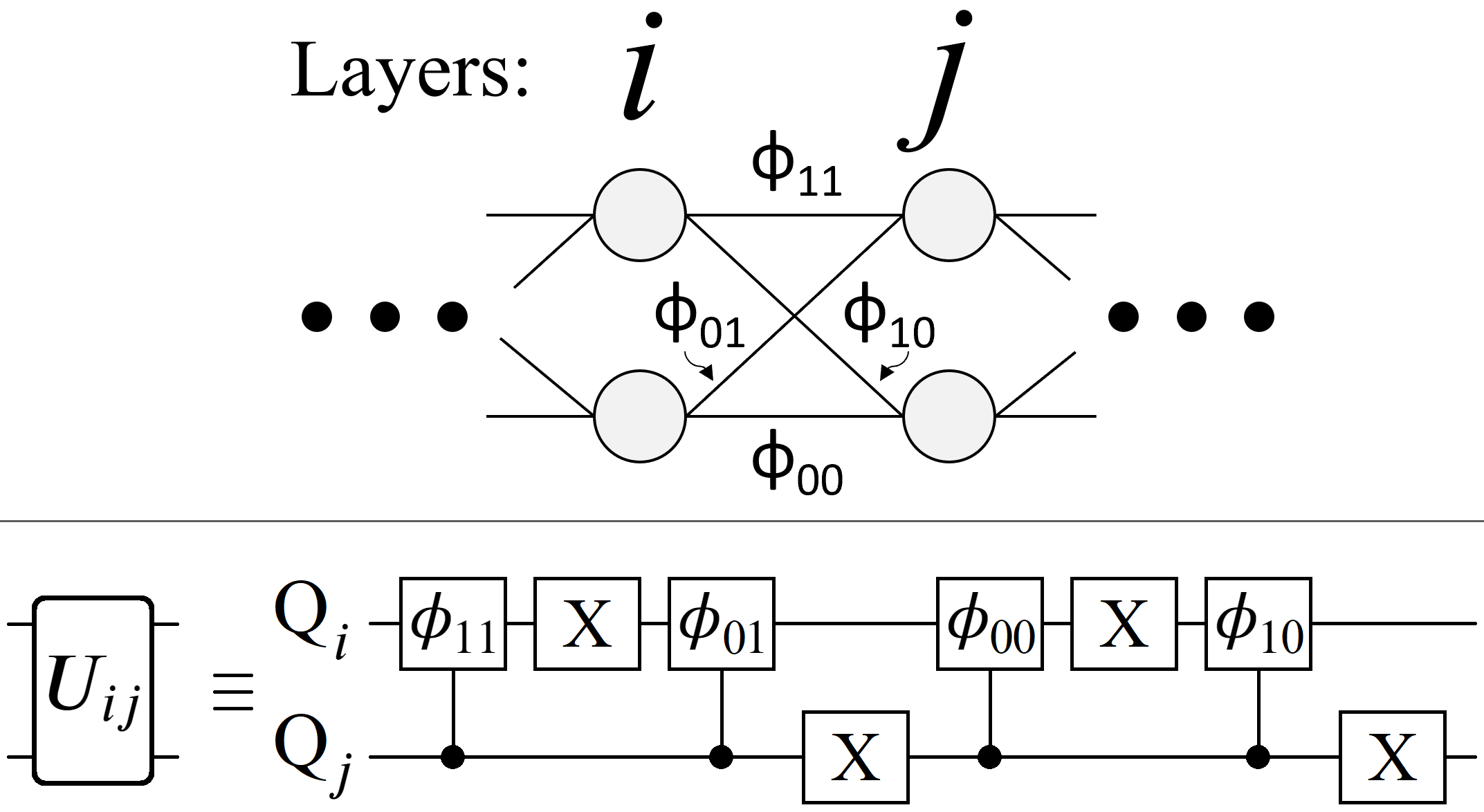}
		\caption{(top) Illustration of layers $i$ and $j$ for an $N=2$ graph, and the four weighted edges shared between them.  (bottom) Quantum circuit for achieving the $U_{ij}$ operation outlined in equation \ref{E15}.  }
		\label{F8}
	\end{figure}

	Having now seen the desired effect from $U_{\textrm{P}}$ (equation \ref{E14}), here we present a qubit-based quantum circuit design which efficiently achieves all $N^L$ unique phases, with no $\textit{a priori}$ classical computations of any $W_i$.  Here we will focus on the case $N=2$ for simplicity, leaving the general case for the next section. We begin by defining the operator $U_{ij}$ shown below in equation \ref{E15}, and its corresponding quantum circuit in figure \ref{F8}.  The operator $U_{ij}$ encodes all of the phases contained between layers $i$ and $j$, from which we can build up to the full $U_{\textrm{P}}$.
	
	\begin{eqnarray}               
		U_{ij} \equiv  	\begin{bmatrix}  e^{i \phi_{00}} & 0 & 0 & 0 \\    0 & e^{i \phi_{01}} & 0 & 0  \\  0 & 0 & e^{i \phi_{10}} & 0  \\ 0 & 0 & 0 & e^{i \phi_{11}}     \end{bmatrix}
		\label{E15}
	\end{eqnarray}	
	
	The circuit shown in figure \ref{F8} applies a unique phase to each of the 2-qubit basis states $| \textrm{Q}_i \textrm{Q}_j  \rangle$, one for each of the four edges connecting layers $i$ and $j$.  The complete information of all weighted edges connecting layers $i$ and $j$ is achieved with exactly one phase gate (controlled) per edge, which is a property that holds true for all geometry sizes.  Importantly, from a qubit connectivity view point, the qubits which make up layer $i$ only need to interact with the qubits making up layers $i \pm 1$.  This in turn can be used to significantly reduce circuit depth, demonstrated below in figure \ref{F9}.

	\begin{figure}[h]               
		\centering
		\includegraphics[scale=.4]{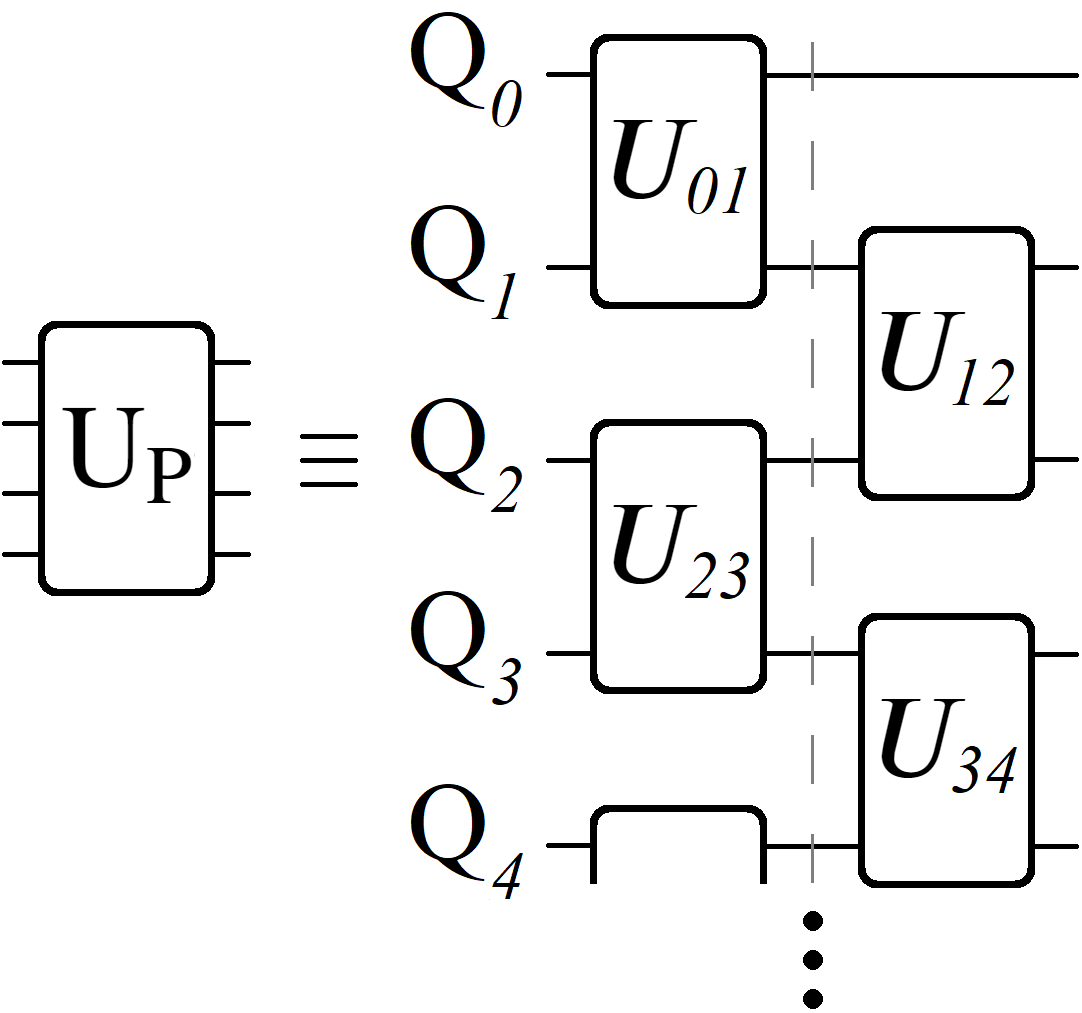}
		\caption{The complete circuit design for $U_{\textrm{P}}$, for the case of $N=2$.  Each $U_{ij}$ operation applies the four $\phi_i$ phases corresponding to the $\omega_i$ weights connecting layers $i$ and $j$.  Because of the way in which phases add exponentially, the order in which a total weight $W_i$ is applied to a state $| \textrm{P}_i \rangle$ can be done in two sets of parallel operations, shown by the dashed-grey line.}
		\label{F9}
	\end{figure}

	\begin{eqnarray}              
		U_{\textrm{P}} \equiv \prod_{i=1}^{L-1}  U_{i , i+1} 
		\label{E16}
	\end{eqnarray}	
	
	Let us now compare the desired effect of $U_{\textrm{P}}$ from equation \ref{E14}, with its layer by layer construction shown in figures \ref{F8} and \ref{F9}.  Each $U_{ij}$ operation applies phases proportional to the local weighted edges connecting layers $i$ and $j$, involving only the qubits representing those layers.  And due to the way in which phases add exponentially (equation \ref{E13}), the full path weight $W_i$ for each $| \textrm{P}_i \rangle$ state is achieved from the product of $U_{ij}$ operations, shown above in equation \ref{E16}.  Importantly, note that nowhere in $U_{\textrm{P}}$'s construction do we compute a single $W_i$ value.  As mentioned earlier, this is a necessary requirement of $U_{\textrm{P}}$ in order to truly consider it an oracle operation.  Here we have achieved exactly that by splitting $U_{\textrm{P}}$ up into localized $U_{ij}$ operations for each section of the graph.  For results on how an $N=2$ $U_{\textrm{P}}$ operation performs on IBM's superconducting qubits, please see appendix $\textbf{A}$.
	
	We would like to stress that the structure of figure \ref{F9} is general for all geometry sizes, which is one of the motivations for studying these sequential bipartite graphs.  The parameter $N$ dictates the number of quantum states per layer, which in turn determines the dimensionality of $U_{ij}$.  But for all graphs, the parameter $L$ has no impact on circuit depth, as the complete implementation of $U_{\textrm{P}}$ can always be achieved through two sets of parallel $U_{ij}$ operations, shown by the dashed-grey line in figure \ref{F9}.
	
	%%%%%%%%%%%%%%%%%%%%%%%%%%%%%%%%%%%%%%%%
	\subsection{Qudit Quantum Circuit}%                                                                                                                                   Quantum Circuit
	%%%%%%%%%%%%%%%%%%%%%%%%%%%%%%%%%%%%%%%%
	
	To compliment the results from the previous section for constructing $U_{\textrm{P}}$ on a qubit-based quantum computer, here we shall briefly mention how qudits can be used to greatly expand beyond simply $N = 2^n$ sized graphs, as well as further reduce circuit depth.  Since we will be interested in using qudits again in section VIII., let us now introduce the notation for a general $d$-level quantum bit:
	
	\begin{eqnarray}             
		| Q \rangle_d \equiv \sum_{i=0}^{d-1} \alpha_i | i \rangle_d
		\label{E17}
	\end{eqnarray}	
	
\begin{figure}[h]                 
		\centering
		\includegraphics[scale=.25]{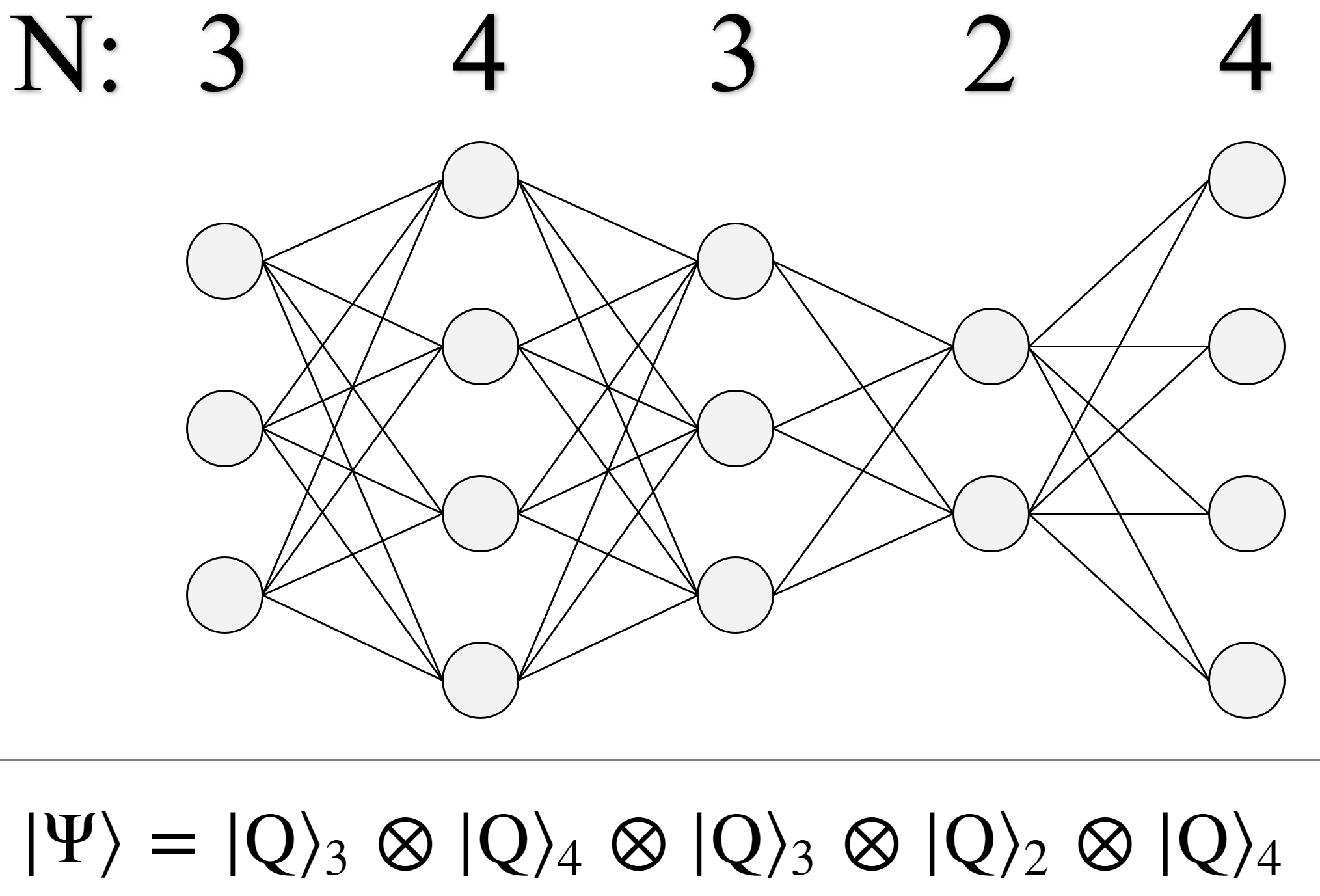}
		\caption{(top) A sequential bipartite graph of varying $N$ at each layer.  (bottom) A mixed qudit quantum state capable of representing all possible paths through the geometry.}
		\label{F10}
	\end{figure}

	As shown in equation \ref{E17}, the quantum state for any $d$-dimensional qudit can be expressed as a superposition of orthogonal basis states, spanning $|0\rangle$ through $|d-1\rangle$.  Experimentally, the realization of qudits has been steadily progressing over the past decade \cite{kues,low,yurtalan,lu}, which makes it an exciting time to start considering their applications for quantum algorithms.  Here, the use of qudits allows us to represent graphs beyond $N = 2^n$.  For example, a qutrit-based computer ($d=3$) can encode graphs of size $N = 3^n$.  Better still, a mixed qudit computer grants us the ability to encode graphs with a different $N$ at each layer, such as in figure \ref{F10}.
	
	Note that it is still possible to create a varying $N$ graph using qubits, so long as every layer has $N = 2^n$ nodes.  However, even for geometry sizes which are implementable using qubits, the use of qudits is still advantageous for several reasons.  Consider the two quantum circuits shown below in figure \ref{F11}, which both achieve a $U_{ij}$ operation connecting two $N=4$ layers, applying the same $16$ phases in total.
	
	\begin{figure}[h]                  
		\centering
		\includegraphics[scale=.18]{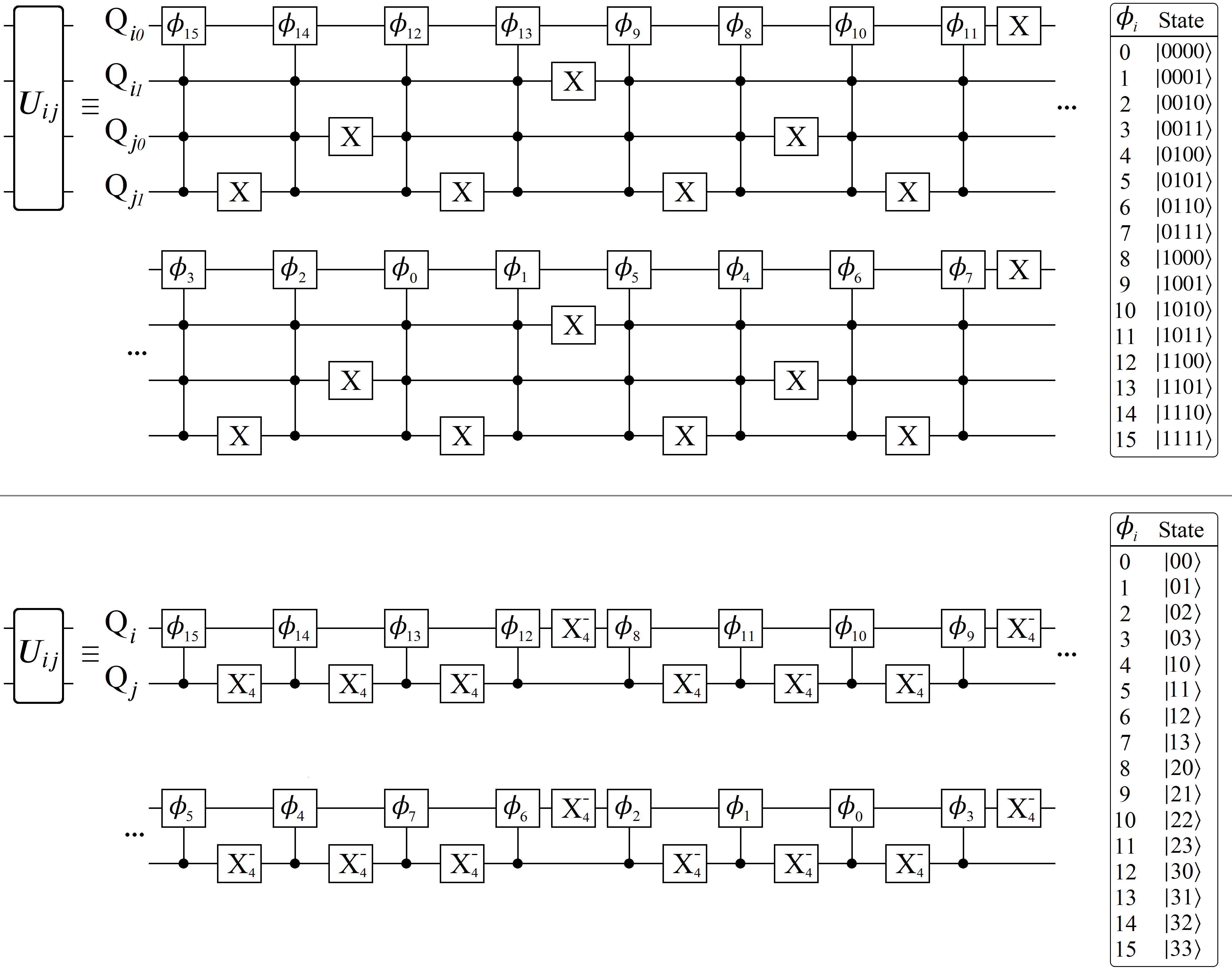}
		\caption{Quantum circuits for $U_{ij}$ connecting two layers of $N=4$ nodes. (top) A qubit-based quantum circuit (bottom) A $d=4$ qudit-based quantum circuit.  More information on qudit unitary operations and circuits can be found in the review study by Wang et. al \cite{wang}, such as the $X^{-}_d$ operator shown here.}
		\label{F11}
	\end{figure}
	
	The primary issue with using qubits is that there is a hidden resource cost when constructing higher order control operations.  In order to achieve an $N$-control phase gate, the true quantum circuit requires $N$ additional ancilla qubits to serve as intermediate excited states \cite{koch}.  This is because the qubit operations from which we build up higher order control-phase gates are P($\theta$) (single qubit phase), CX (control-X), and CCX (Toffoli).  The significant advantage that the $d=4$ qudit circuit has is the absence of Toffoli gates, as all $16$ control-phase operations only need to occur between the two qudits.  Thus, the qudit circuit is advantageous in both resource cost (two qudits vs. seven qubits) and circuit depth (reduction of four Toffoli gates per each of the $16$ phase operations).  Of course, the trade-off is that qudit technologies are still primitive compared to the more popular qubit, and as such would be expected to come with much higher error rates.  Nevertheless, we will return to the use of qudits again in section VIII., as the Hilbert space sizes they offer will be necessary for unlocking meaningful problems to solve.

	%%%%%%%%%%%%%%%%%%%%%%%%%%%%%%%%%%%%%%%%%%%%%%%%
	%%%%%%%%%%%%%%%%%%%%%%%%%%%%%%%%%%%%%%%%%%%%%%%%
	\section{Gaussian Amplitude Amplification}%                                  Gaussian Amplitude Amplification
	%%%%%%%%%%%%%%%%%%%%%%%%%%%%%%%%%%%%%%%%%%%%%%%%
	%%%%%%%%%%%%%%%%%%%%%%%%%%%%%%%%%%%%%%%%%%%%%%%%
	
	With the construction of $U_{\textrm{P}}$ outlined in section IV., here we discuss how this cost oracle operator can be used to solve for $W_{\textrm{min}}$ or $W_{\textrm{max}}$.  Because $U_{\textrm{P}}$ applies phases to every quantum state, substituting it for $U_{\textrm{G}}$ in Grover's algorithm has dramatic consequences on the way in which the amplitude amplification process plays out.

	%%%%%%%%%%%%%%%%%%%%%%%%%%%%%%%%%%%%%%%%
	\subsection{Solution Space Distributions}%                                                                                                                                  Naturally Gaussian
	%%%%%%%%%%%%%%%%%%%%%%%%%%%%%%%%%%%%%%%%
	
	The motivation for studying directed graphs according to figure \ref{F4} is only partially due to their circuit implementation (figures \ref{F8} - \ref{F11}).  Additionally, these sequential bipartite graphs possess a second important quality necessary for the success of the algorithm: their $\mathbb{W}$ distributions.  In equation \ref{E7} we restricted each edge weight $\omega_i$ to be an integer value, for a reason that we will now discuss.  By forcing each $\omega_i$ to be an integer within $[ 0, \textrm{R} ]$, we can create directed graphs which have a high likelihood of repeat $W_i$ values.  Consequently, two independent paths $| \textrm{P}_i \rangle$ and $| \textrm{P}_j \rangle$ will both yield the same cumulative weights $W_i = W_j$, from different contributing $\omega_i$'s.  As we let $N$ and $L$ increase, these repeat values lead to $\mathbb{W}$ distributions which become describable by a gaussian function, given in equation \ref{E18}, where the majority of $W_i$ values cluster around the expected mean $\mu \approx \frac{R}{2} (L - 1)$.  
	
	\begin{eqnarray}             
		G(x) = \alpha e^{ - \frac{(x - \mu)^2}{2 \sigma^2} } 
		\label{E18}
	\end{eqnarray}
	
	\begin{figure}[h]                   
		\centering
		\includegraphics[scale=.33]{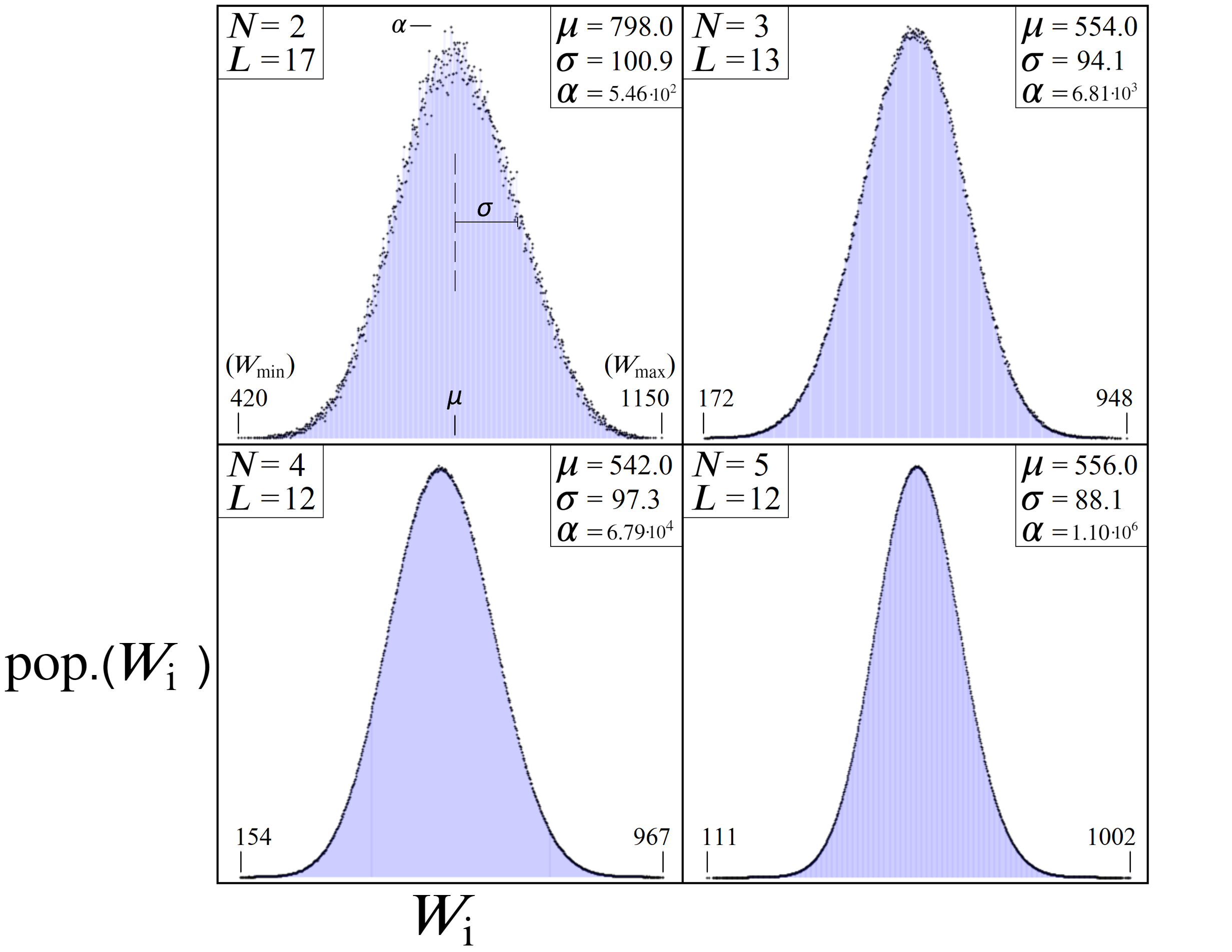}
		\caption{Histograms of $W_i$ for randomly generated graphs of various $N$ and $L$ sizes, with $R=100$. As $N$ and $L$ increase while keeping $R$ constant, the profile of these $\mathbb{W}$ distributions approach perfect gaussians, given by equation \ref{E18}.}
		\label{F12}
	\end{figure}
	
	Figure \ref{F12} illustrates a few example problem sizes for various $N$ and $L$, and their resulting $\mathbb{W}$ histogram distributions.  These distributions represent the range of expected outcomes from picking a path through the directed graph at random, and seeing what $W_i$ value one gets.  The odds of picking the optimal path are $1$ in $N^L$, while the most probable $W_i$ corresponds to the peak of the gaussian.  Importantly, the tail ends of the distribution represent our desired solutions $W_{\textrm{min}} $ and $W_{\textrm{max}} $ (top left plot in figure \ref{F12}), which are always maximally distanced from the cluster of states around the mean.  Also note that letting $\omega_i$ be continuous within $[ 0,\textrm{R} ]$ still produces the same gaussian effect, but discrete bin sizes are necessary for viewing the resulting $\mathbb{W}$ histogram distributions, hence our choice to let $\omega_i$ be integers only.
	
	\begin{eqnarray}             
		Y_i &=& \textrm{pop.}(W_i) \hspace{1cm} Y'_i = G( W_i) \nonumber \\
		\textrm{R}_{\textrm{corr}} &=& \sqrt{ \frac{ \sum_i (Y'_i - Y_i)^2}{ N^L } }
		\label{E19}
	\end{eqnarray}	
	
	\begin{figure}[h]                 
		\centering
		\includegraphics[scale=.4]{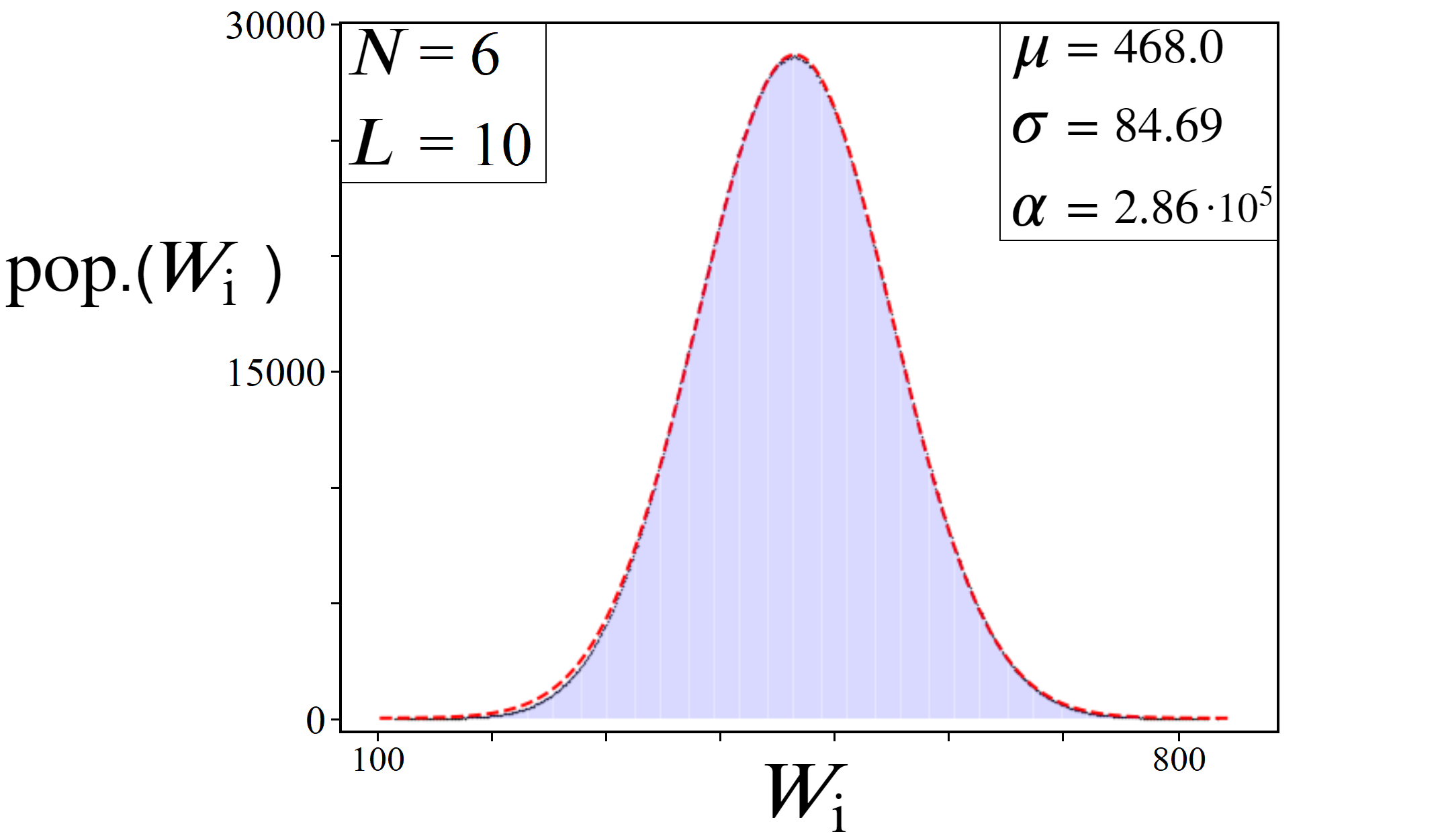}
		\caption{ (black circles / blue lines) A histogram of $\mathbb{W}$ for a randomly generated graph with parameters: $N=6$, $L=10$, $R=100$.  (red dash) A best-fit gaussian plot of the form given in equation \ref{E18}, minimizing equation \ref{E19} (R$_{\textrm{corr}}$ $\approx$ $3.981$), with gaussian parameter values reported in the top-right. }
		\label{F13}
	\end{figure}
	
	Shown above in figure \ref{F13} is an example distribution and accompanying gaussian best-fit.  This particular distribution was derived from a graph of size $N=6$, $L=10$, in anticipation of results later to come (figures \ref{F16}, \ref{F22}, and \ref{F24}).  With an R$_{\textrm{corr}}$ value of approximately $3.98$, given by equation \ref{E19}, it is clear that the gaussian approximation for this example is not perfect.  Even for a problem such as this one, composed of over $60$ million possible solutions, the resulting $\mathbb{W}$ distribution still has non-negligible deviations from a perfect gaussian, which will be a primary focus of section VII.  Nevertheless, these approximate gaussian profiles are sufficient for the success of the algorithm.

	%------------------------------------------------------------------------------------------------------------------- Here to generate on next page
	\begin{figure*}[t]               
		\centering
		\includegraphics[scale=.24]{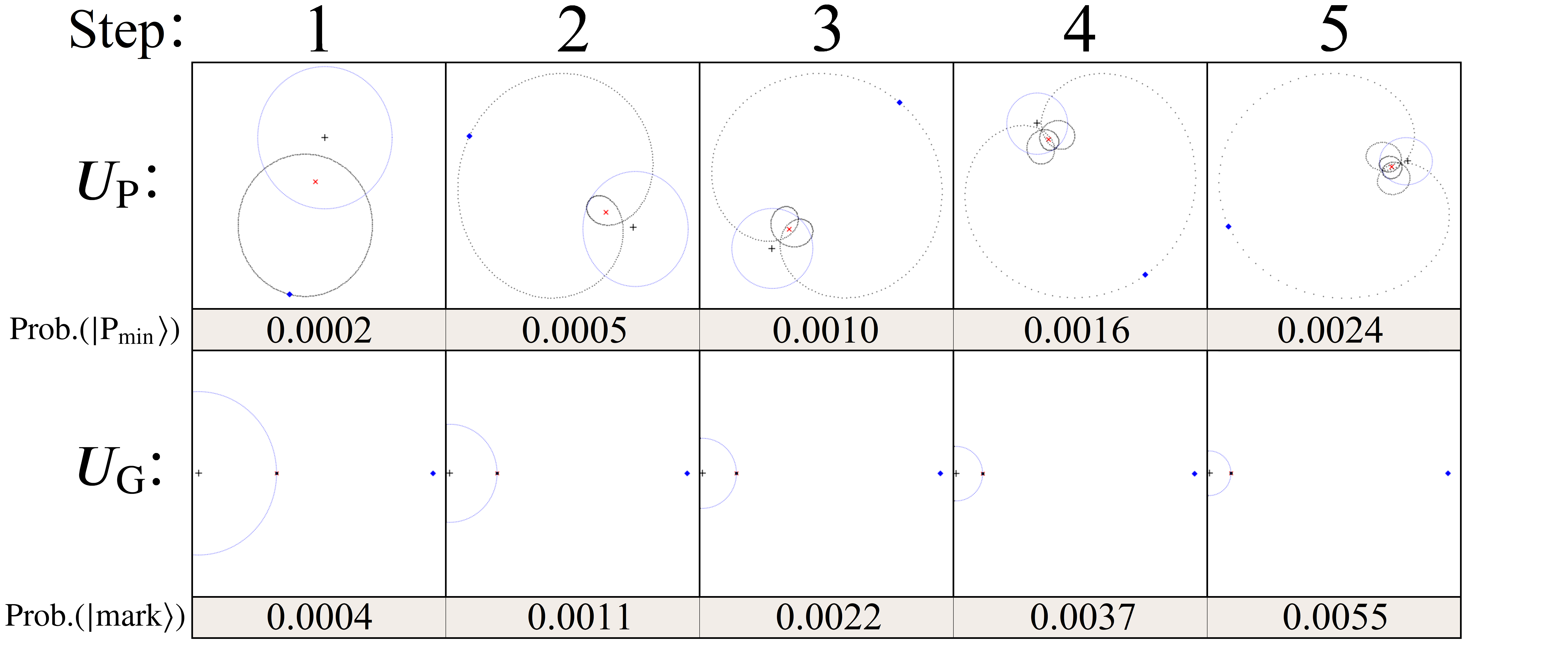}
		\caption{ Examples of amplitude amplification, comparing the use of $U_{\textrm{P}}$  vs. $U_{\textrm{G}}$ for five iterations, both with the same number of total states $N' = 24,000$.  In both plots, the origin (0,0) (black `+'), the mean point (red `x'), the desired boosted state (blue diamond), and all other points (black circles) are shown.  For scale, the radius of the equal superposition state $| \textrm{s} \rangle$ (blue circle) is also shown ($1/ \sqrt{N'}$), as well as the probability of measuring the blue diamond state (which can be used to infer distance to the origin).     }
		\label{F14}
	\end{figure*}
	%-------------------------------------------------------------------------------------------------------------------
	
	%%%%%%%%%%%%%%%%%%%%%%%%%%%%%%%%%%%%%%%%
	\subsection{Mapping to $2\pi$}%                                                                                                                                  Mapping to 2 pi
	%%%%%%%%%%%%%%%%%%%%%%%%%%%%%%%%%%%%%%%%
	
	When using the cost oracle as defined in equation \ref{E14}, one must be mindful that $U_{\textrm{P}}$ does not only mark the states corresponding to $W_{\textrm{min}}$ and $W_{\textrm{max}}$, but $\textit{all}$ states uniquely.  This is quite different from the standard Grover oracle $U_{\textrm{G}}$, which $\textit{only}$ marks the state(s) of interest.  For this reason, the use of $U_{\textrm{P}}$ for amplitude amplification can be viewed as less flexible than $U_{\textrm{G}}$.  While $U_{\textrm{G}}$ can in principle be used to boost any of the $N^L$ quantum states in $| \Psi \rangle$, $U_{\textrm{P}}$ on the other hand is better suited for boosting a much smaller percentage of states.  However, the states which $U_{\textrm{P}}$ $\textit{is}$  effective at boosting are $|  \textrm{P}_{\textrm{min}}  \rangle$ and $|  \textrm{P}_{\textrm{max}}  \rangle$, perfect for solving a directed graph problem. 
	
	In viewing the $\mathbb{W}$ histograms in figure \ref{F12}, let us now consider the effect of applying $U_{\textrm{P}}$ from equation \ref{E14} on an equal superposition state $| \textrm{s} \rangle \equiv H^{\otimes n} | 0 \rangle^{\otimes n}$.   Each point along the x-axis corresponds to a particular path length $W_j$, while the y-axis represents the total number of quantum states which will receive a phase proportional to that weight: $e^{i \phi_j} | \textrm{P}_j \rangle$.  Thus, the net result of $U_{\textrm{P}}$ will apply all $N^L$ phases in a gaussian-like manner, with the majority of states near the mean receiving similar total phases (from different contributing $\omega_i$'s).  And in order to capitalize on this distribution of phases, we will introduce a phase scaling constant $p_{\textrm{s}}$ into the oracle operation, which affects all states equally:

	\begin{eqnarray}               
		U_{\textrm{P}}( p_{\textrm{s}} ) |\Psi \rangle = \sum_{j}^{N^L}   e^{ i (p_\textrm{s} \cdot W_{\tiny{j}}) } | \textrm{P}_j \rangle 
		\label{E20}
	\end{eqnarray}

\begin{figure}[h]                  
		\centering
		\includegraphics[scale=.28]{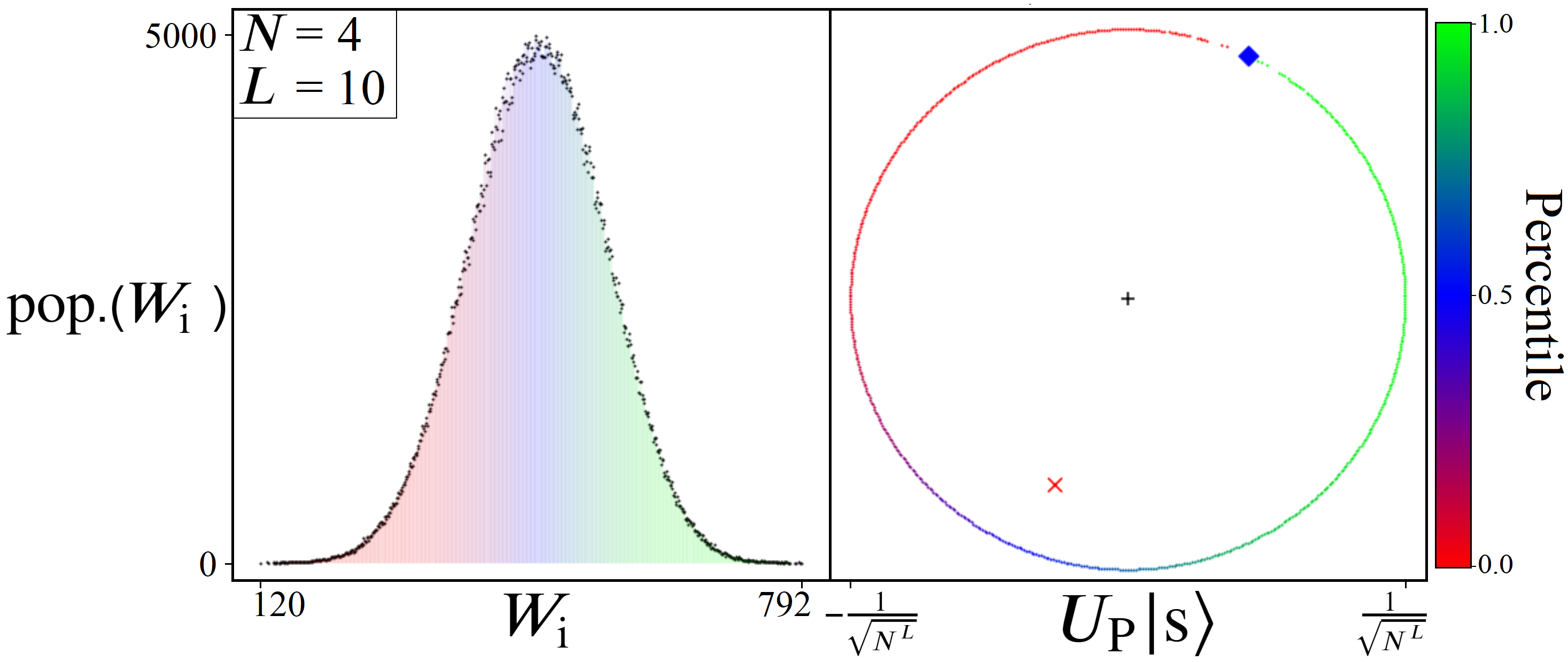}
		\caption{ (left)  An example histogram of all $W_i$ paths for the case of $N=4$, $L=10$, $R=100$.  (right) The same distribution mapped to a complete $2 \pi$ cycle of phases via the cost oracle $U_{\textrm{P}}$ acting on the equal superposition state $| \textrm{s} \rangle$. Additionally, the resulting mean (red `X') and $| \textrm{P}_{\textrm{min}} \rangle$ / $| \textrm{P}_{\textrm{max}} \rangle$ states (blue diamond) are shown.  An accompanying color scale is provided on the far right, illustrating the percentile distribution of states for both plots. }
		\label{F15}
	\end{figure}

	The scaling constant $p_{\textrm{s}}$ in equation \ref{E20} is a value which must be multiplied into every cumulative $W_j$ phase throughout the oracle.  This can be achieved by setting each individual phase in $U_{ij}$ to $ p_{\textrm{s}} \cdot \omega_i$, such that the cumulative operation of $U_{\textrm{P}}$ is equal to equation \ref{E20}.  The phase $p_{\textrm{s}}$ can be thought of as simply the translation of any problem's $\mathbb{W}$, for any scale of numbers used, into a regime of phases which can be used for boosting.  More specifically, a range of phases $[x, x+2 \pi]$ for which the state $| \textrm{P}_{\textrm{min}} \rangle$ or $| \textrm{P}_{\textrm{max}} \rangle$ is optimally distanced from the majority of states in amplitude space (complex plane).  See figure \ref{F15} for an illustrated example, and note the location of the red `x' corresponding to $| \Psi \rangle$'s collective mean after $U_{\textrm{P}}$.
	
	Without $p_{\textrm{s}}$, the numerical $W_i$ values from a given directed graph have no guarantee of producing any meaningful amplitude amplification.  However, when scaled properly with an optimal $p_{\textrm{s}}$ (which is discussed in sections VI. and VII. ), $U_{\textrm{P}}$ can be made to distribute phases like shown in figure \ref{F15}, where the phases picked up by $| \textrm{P}_{\textrm{min}} \rangle$ and $| \textrm{P}_{\textrm{max}}  \rangle$ form a range of $[x, x+2 \pi]$.  This in turn ensures that the majority of states will cluster near $x + \pi$, pulling the amplitude mean (red `X') away from $| \textrm{P}_{\textrm{min}} \rangle$ and $| \textrm{P}_{\textrm{max}}  \rangle$ (blue diamond).

	%%%%%%%%%%%%%%%%%%%%%%%%%%%%%%%%%%%%%%%%
	\subsection{$U_{\textrm{G}}$ vs $U_{\textrm{P}}$ Diffusion}%                                                                                 U_G vs U_P Diffusion
	%%%%%%%%%%%%%%%%%%%%%%%%%%%%%%%%%%%%%%%%
	
	As with the standard Grover search algorithm \cite{grover}, the $U_{\textrm{P}}$ oracle operation in isolation is not enough to solve for $W_{\textrm{min}}$ or $W_{\textrm{max}}$.  A second mechanism for causing interference is necessary in order to boost the probability of measuring the desired state.  For this, we once again use the standard Grover diffusion operator $U_{\textrm{s}}$, given in equation \ref{E1}. With $U_{\textrm{P}}$ distributing phases to each state, and $U_{\textrm{s}}$ causing reflections about the average, we now have the sufficient tools for quantum pathfinding, shown in algorithm \ref{A3}.
	
	\begin{algorithm}
		\caption{Quantum Pathfinding}
		\begin{algorithmic}[1]
			\State Initialize  Qubits: $|\Psi\rangle = |0\rangle ^{\otimes N}$
			\State Prepare Equal Superposition: $H^{\otimes N} |\Psi \rangle = |s\rangle$
			\For{ $k \approx \frac{\pi}{4} \sqrt{ N^L }$  }  
			\State Apply $U_\textrm{P} (p_{\textrm{s}}) |\Psi\rangle$ (Phase Oracle)
			\State Apply $U_\textrm{s} |\Psi\rangle$ (Diffusion)
			\EndFor
			\State Measure
		\end{algorithmic}
		\label{A3}
	\end{algorithm}
	
	As noted previously, the algorithm outlined here is identical to that of Grover's search algorithm, with $U_{\textrm{G}}$ swapped out for $U_{\textrm{P}}$.  However, this replacement significantly changes the way in which the states of $| \Psi \rangle$ go through amplitude amplification, illustrated in figure \ref{F14}.  For a comparison, the amplitude space when using the standard $U_{\textrm{G}}$ is also shown.
	
	Step 1 of figure \ref{F14} shows the effect of using the diffusion operator $U_{\textrm{s}}$ immediately following the first application of $U_{\textrm{P}}$ (see figure \ref{F15}).  The location of the mean point (red `X') causes states near $| \textrm{P}_{\textrm{min}} \rangle$ and $| \textrm{P}_{\textrm{max}} \rangle$ (blue diamond) to reflect further than those around the mean of the gaussian.  However, when compared with the lower plots using $U_{\textrm{G}}$, this increase in probability is always smaller than that of standard Grover's.  Geometrically, this is a consequence of having states with phases spread out over a $2 \pi$ range, resulting in a mean amplitude point which is closer to the origin (similar to $U'_{G2}$ from section II). 	
	
	What follows after step 1 for the case of $U_{\textrm{P}}$ is a process with no simple mathematical description.  As illustrated in steps 2 - 5, repeat applications of $U_{\textrm{P}}$ and $U_{\textrm{s}}$ result in quantum superposition states which exhibit a `spiraling'  effect around the mean point, which itself is also moving around the complex plane.  Although quite clearly different from standard Grover's, two key elements remain the same: 1) the distance between the mean point and the origin gradually decreases with each step, while 2) the distance between $| \textrm{P}_{\textrm{min}} \rangle$ / $| \textrm{P}_{\textrm{max}} \rangle$ and the origin increases (i.e. incremental probability gains with each step).  Just like Grover's, both of these statements hold true for O($ \sqrt{N^L}$) iterations, after which the process begins to rebound.

	\begin{figure}[h]        
		\centering
		\includegraphics[scale=.4]{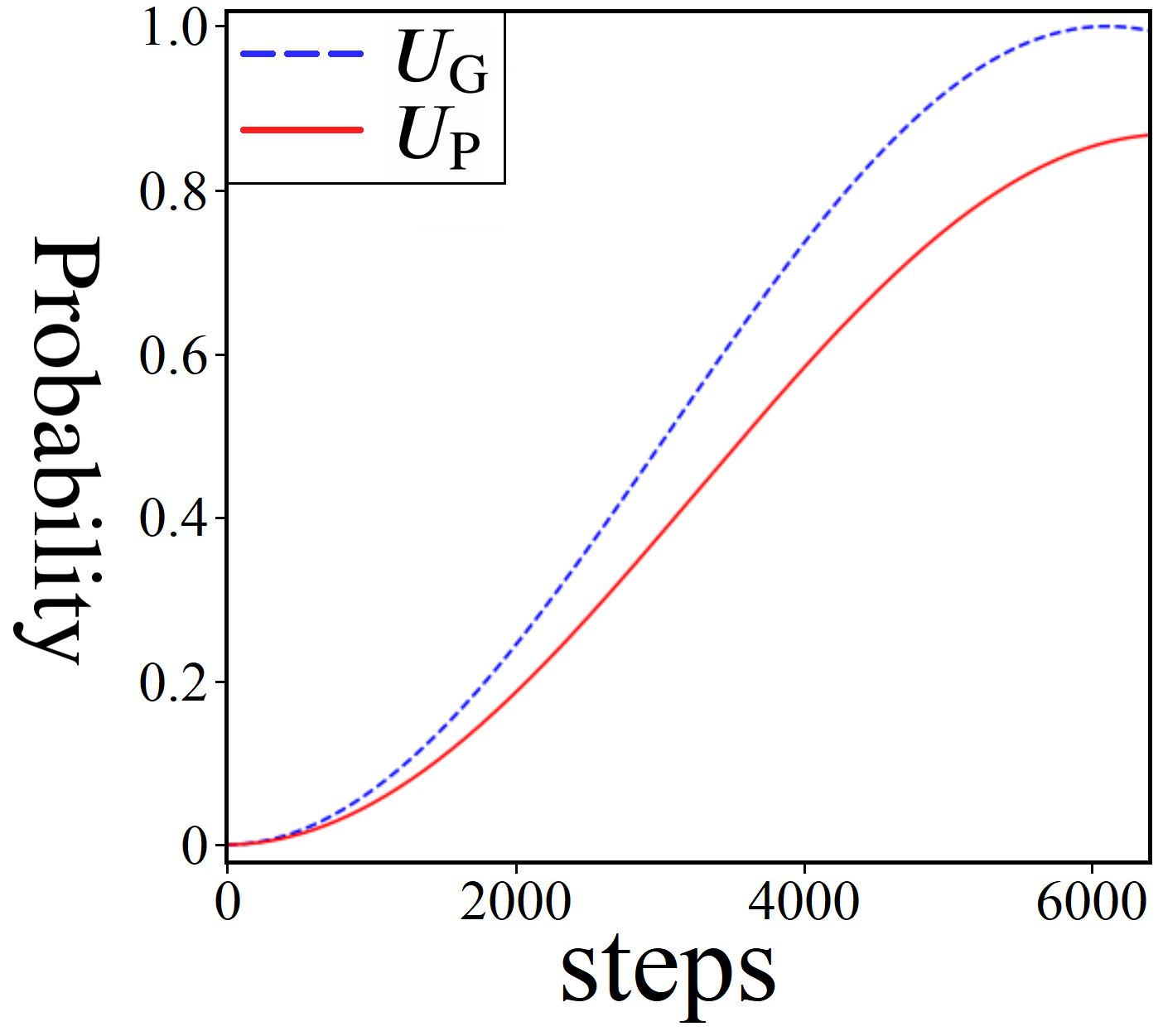}
		\caption{ A comparison of probability boosting using $U_{\textrm{G}}$ (blue-dashed) vs. $U_{\textrm{P}}$ (red-solid) as a function of steps (oracle + diffusion iterations), both acting on a quantum system of $6^{10}$ states.  For $U_{\textrm{G}}$ we track the probability of the marked state, while the $U_{\textrm{P}}$ case tracks the probability of measuring $| \textrm{P}_{\textrm{min}}\rangle$. }
		\label{F16}
	\end{figure}
	
	%------------------------------------------------------------------------------------------------------------------- Here to generate on next page
	\begin{figure*}[!t]               
		\centering
		\includegraphics[scale=.24]{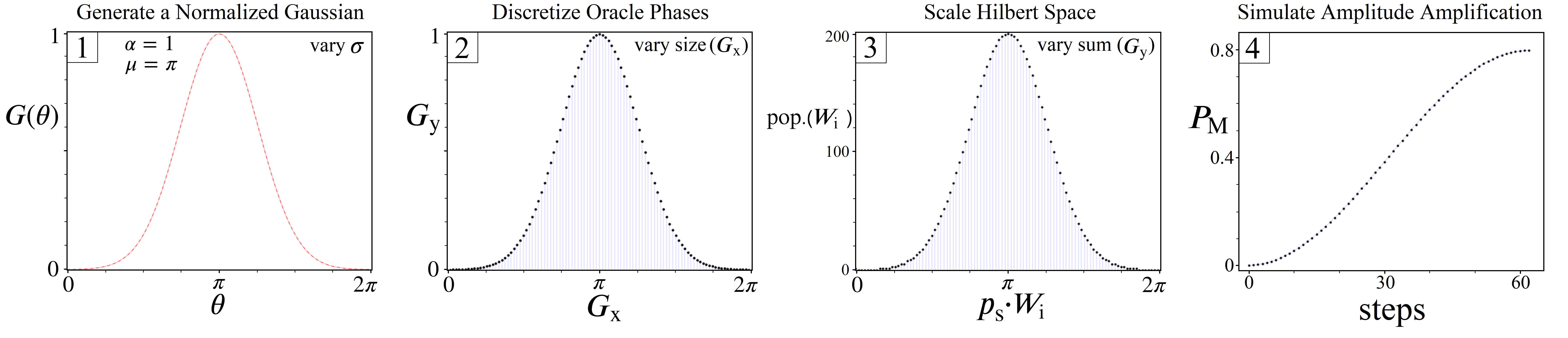}
		\caption{ (1-3) Illustrations of how our python-based simulator creates gaussian $\mathbb{W}$ distributions for testing.  In step 1, we pick a standard deviation $\sigma$ and create a continuous gaussian from $0$ to $2 \pi$, with $\alpha = 1$ and $\mu = \pi$.  In step2 we select how many unique $W_i$ phases we want to model, and use this number to discretize the continuous gaussian into two discrete arrays $G_{\textrm{x}}$ and  $G_{\textrm{y}}$.  In step3 we select a target Hilbert space size $N$ to model, and scale all of the values in $G_{\textrm{y}}$ up to integers, such that the sum($G_{\textrm{y}}$) is as close to $N$ as possible.  And finally in step 4 we similuate amplitude amplification using $G_{\textrm{x}}$ and  $G_{\textrm{y}}$, tracking the probability of $| \textrm{P}_{\textrm{min}} \rangle$.   }
		\label{F17}
	\end{figure*}
	%-------------------------------------------------------------------------------------------------------------------

	Shown above in figure \ref{F16} is a step-by-step comparison of probabilities for standard Grover's versus gaussian amplitude amplification (i.e. amplitude amplification using a $2 \pi$ gaussian distribution of phases), both for problem sizes of $6^{10}$ quantum states ($N=6$, $L=10$).  The blue-dashed line tracks the probability of measuring the marked state as it approaches $1$, while the red-solid line represents the probability of measuring $| \textrm{P}_{\textrm{min}}\rangle$.  Notably, the probability of $| \textrm{P}_{\textrm{min}}\rangle$ achieves a lower peak $P_{\textrm{M}}$, and at a later step count.  This is the trade-off for using $U_{\textrm{P}}$ versus $U_{\textrm{G}}$: a lower boost in probability, but a solution to an inherently different problem (unstructured search vs. weighted directed graph).  Importantly however, the combination of iterations and peak probability for $| \textrm{P}_{\textrm{min}}\rangle$ is still high enough for a potential quantum speedup under certain conditions, which we discuss in the next two sections.

	%%%%%%%%%%%%%%%%%%%%%%%%%%%%%%%%%%%%%%%%%%%%%%%%
	%%%%%%%%%%%%%%%%%%%%%%%%%%%%%%%%%%%%%%%%%%%%%%%%
	\section{Simulating Gaussian Amplitude Amplification}%                                               Quantifying Gaussian Amplitude Amplification
	%%%%%%%%%%%%%%%%%%%%%%%%%%%%%%%%%%%%%%%%%%%%%%%%
	%%%%%%%%%%%%%%%%%%%%%%%%%%%%%%%%%%%%%%%%%%%%%%%%
	
	Much like the analysis of $U'_{G2}$ from section II., here we present results which illustrate the capacity for successful amplitude amplification one can expect from a gaussian distribution of phases encoded by $U_{\textrm{P}}$.  To do this, we use a classical python-based simulator, capable of mimicking the amplitude amplification process outlined in Alg. \ref{A3}, allowing us to track quantum states and probabilities throughout.  Results from various simulations are provided in the coming subsections, as well as their significance for identifying properties of problems which are viable for amplitude amplification.
	
	%%%%%%%%%%%%%%%%%%%%%%%%%%%%%%%%%%%%%%%%
	\subsection{Modeling Quantum Systems}%                                                 Modeling Quantum Systems
	%%%%%%%%%%%%%%%%%%%%%%%%%%%%%%%%%%%%%%%%
	
	As illustrated in figure \ref{F15}, amplitude amplification is viable for solving optimization problems with naturally gaussian solution spaces $\mathbb{W}$, scaled down to a $2 \pi$ range of phases via $p_{\textrm{s}}$.  In the next section we address the challenges of finding $p_{\textrm{s}}$, while here we will focus solely on how the amplitude amplification process performs under ideal conditions. 
	
	\begin{eqnarray}             
		G(\theta) = \alpha e^{ - \frac{(\theta -  \pi)^2}{2 \sigma^2} },  \hspace{1cm} \theta \in [0,\pi]
		\label{E21}
	\end{eqnarray} 
	
	Let us now outline our methodology for creating and simulating discrete $U_{\textrm{P}}$'s derived from equation \ref{E21}, shown in figure \ref{F17}.  In step $1$, we begin with a normalized gaussian ($\alpha=1$) centered at $\pi$, with $\sigma$ (standard deviation) as the only free parameter.  Next we discretize the gaussian by using (x,y) points along the function ($\textrm{x}=\theta$, $\textrm{y}=G(\theta)$), taken in evenly spaced intervals of $\theta$ based on how many unique phases we want to model between $0$ and $2\pi$.  These x and y values are then stored in two vectors: $G_{\textrm{x}}$ and $G_{\textrm{y}}$.   At this stage, $G_{\textrm{x}}$ represents the various phases encoded by some $U_{\textrm{P}}$, but together with $G_{\textrm{y}}$ they do not represent a valid oracle yet.  This is because the values of $G_{\textrm{y}}$ need to model a histogram of states, which means: 1) every value in $G_{\textrm{y}}$ must be an integer, and 2) the sum of $G_{\textrm{y}}$ is the Hilbert Space size of the quantum system.  Analogous to the histograms shown throughout this study, $G_{\textrm{x}}$ represents the space of possible $W_i$ solutions, while $G_{\textrm{y}}$ represents how many states will receive a phase proportional to $W_i$.  Thus, a viable $U_{\textrm{P}}$ operator is finally achieved in step $3$ of figure \ref{F17}, after all the values of $G_{\textrm{y}}$ are multiplied by a constant factor and rounded to integers (preserving $\sigma$ from step $1$).

	For each simulation according to figure \ref{F17}, the full construction of $U_{\textrm{P}}$ is based upon three free parameters of our choosing: $\sigma$, size($G_{\textrm{x}}$), and the sum($G_{\textrm{y}}$), shown in steps $1$, $2$, and $3$ respectively.  The motivation for these three parameters is based on their direct ties to the quantities $N$, $L$, and $R$ from equations \ref{E7} - \ref{E12}.  For example, the combination of $N$ and $L$ determines the Hilbert space size of the quantum system needed to represent all possible paths, which we can control with the sum($G_{\textrm{y}}$). Simultaneously, $L$ and $R$ together dictate the maximum number of possible $W_{i}$ weights: [$0$,$R \cdot (L-1)$], which we can model with the size($G_{\textrm{x}}$).  And finally, $\sigma$ is impacted by all three parameters together, and as we show next, has the strongest correlation to whether or not amplitude amplification is viable.
	
	%%%%%%%%%%%%%%%%%%%%%%%%%%%%%%%%%%%%%%%%
	\subsection{Long Tail Model}%                                                        Long Tail Model
	%%%%%%%%%%%%%%%%%%%%%%%%%%%%%%%%%%%%%%%%
	
\begin{figure}[h]                
		\centering
		\includegraphics[scale=.2]{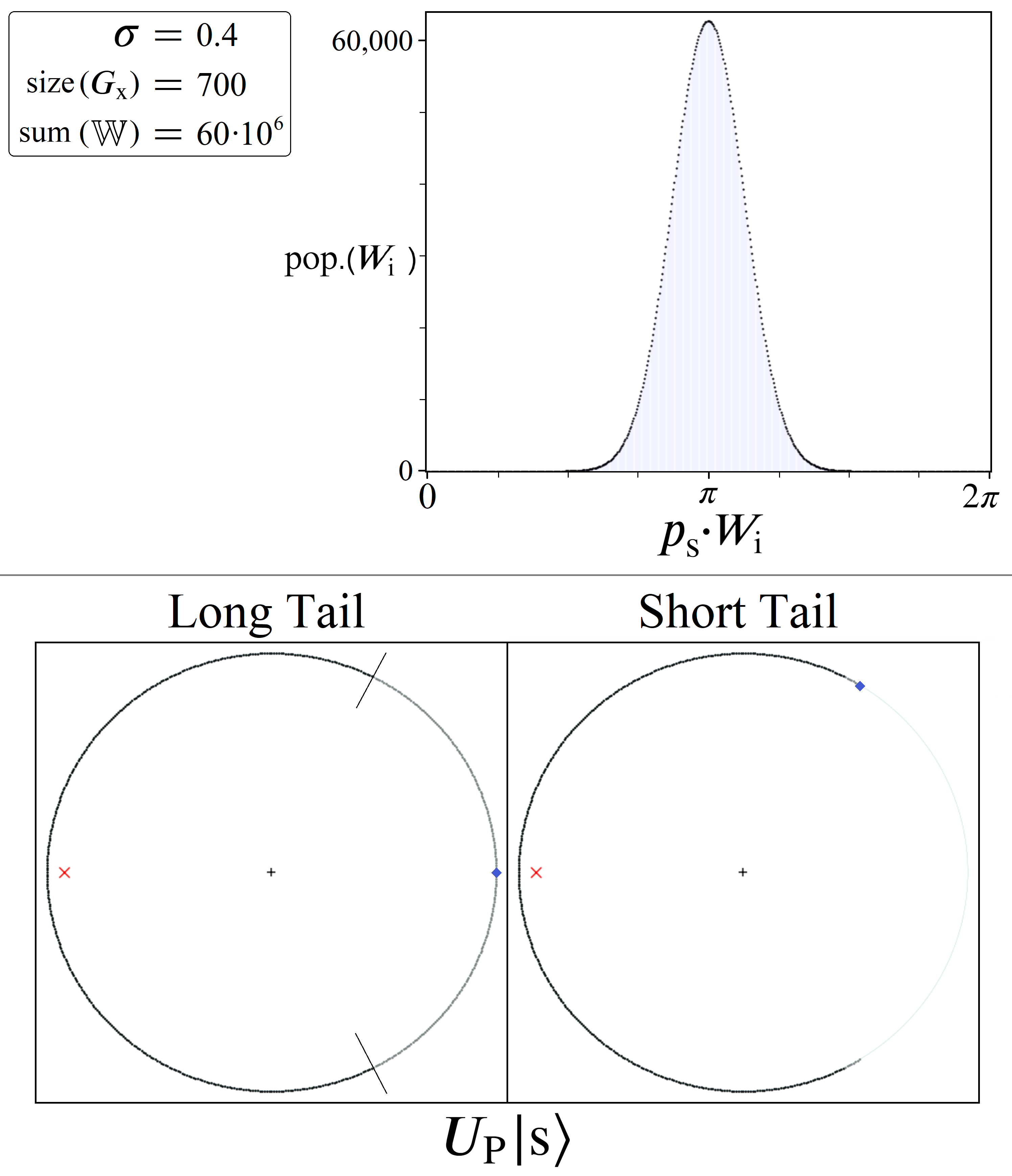}
		\caption{ (top) An example distribution created from our simulator, before rounding in stage 3, with properties of the distribution given on the left.  (bottom)  Two different $U_{\textrm{P}}$ interpretations of the distribution shown on top.  (left) The long tail model, whereby all values of $G_{\textrm{y}}$ are rounded up to the nearest integer.  Grey dashes indicate the region where pop($W_i$)$=1$. (right) The short tail model where all values are rounded down, causing pop($W_i$) values near the tails to be zero for small $\sigma$.}
		\label{F18}
	\end{figure}

	Using the methodology put forth in figure \ref{F17}, there is still one important choice that impacts the nature of the quantum system we are modeling, namely rounding.  In step $3$ of figure \ref{F17}, we must implement a rounding protocol in order to meet the requirement that all $G_{\textrm{y}}$ values be integers.  For phases near the central region of the gaussian, the choice in rounding is practically inconsequential for the amplitude amplification process, but not for the tails where $W_{\textrm{min}}$ and $W_{\textrm{max}}$ lie.  This can be seen in the two $U_{\textrm{P}} | s \rangle$ plots in figure \ref{F18}, where in one case all $G_{\textrm{y}}$ values are rounded up to the nearest integer (left), and one where all values are rounded down (right).
	
	In this subsection we shall focus on simulated distributions according to the left $U_{\textrm{P}} | s \rangle$ encoding in figure \ref{F18}, which we refer to as the `long tail' model.  Compared to the randomly generated distributions in figure \ref{F12}, this turns out to be an unrealistic model for problems where we expect $W_{\textrm{min}}$ to be larger than the theoretical minimum.  Nevertheless, this long tail model will serve to illustrate the most ideal case for gaussian amplitude amplification.  In particular, it allows us to simulate the theoretical limit of a gaussian distribution as $\sigma$ goes to zero, for which the resulting amplitude amplification process is most nearly a replication of standard Grover's.	
	
\begin{figure}[h]               
		\centering
		\includegraphics[scale=.32]{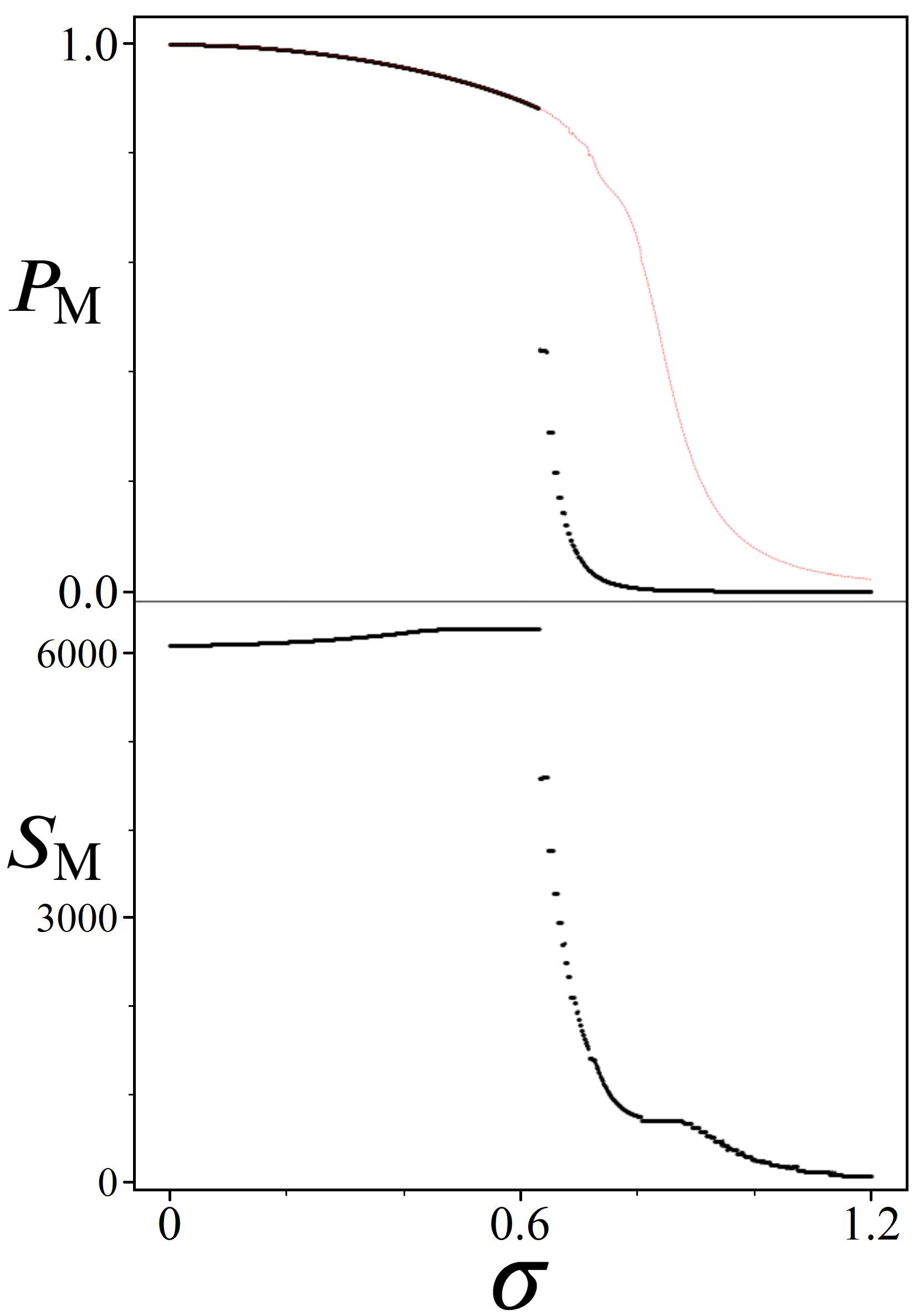}
		\caption{Results for simulated gaussian distributions of Hilbert space size $N = 60 \cdot 10^6$, following the long tail model, as a function of standard deviation $\sigma$. (top) Black data points indicate the highest achievable probabilities $P_{\textrm{M}}$ for $| \textrm{P}_{\textrm{min}} \rangle$, while the red-dashed line shows $P_{\textrm{M}} \cdot $pop($W_{\textrm{min}}$) for cases with multiple $W_{\textrm{min}}$ solutions. (bottom) The number of required iterations $S_{\textrm{M}}$ in order to reach $P_{\textrm{M}}$. }
		\label{F19}
	\end{figure}

	Shown in figure \ref{F19} are results from simulated amplitude amplifications for quantum systems of size $N \approx 60\cdot 10^6$ (sum($G_{\textrm{y}}$)).  Each $U_{\textrm{P}}$ oracle represents $700$ unique weights $W_i$ (size($G_{\textrm{x}}$)) scaled to a $2 \pi$ range, for $\sigma$ values ranging from [$0$,$1.2$].  The top plot shows the peak probabilities $P_{\textrm{M}}$ achievable for the  $| \textrm{P}_{\textrm{min}}\rangle$ state, while the bottom plot shows the corresponding number of needed $U_{\textrm{P}} U_{\textrm{s}}$ iterations $S_{\textrm{M}}$.  
	
	Beginning with $\sigma = 0$, we note how close the results from figure \ref{F19} are to that of standard Grover's: $P_{\textrm{M}}$ is $\sim 0.997$ vs. $\sim 1$, and $S_{\textrm{M}}$ is $6089$ vs. $6083$.  For this $\sigma$, we are modeling an oracle where $N-699$ states all receive $\pi$ phase, $| \textrm{P}_{\textrm{min}}\rangle$ receives a phase of $0$, and the remaining $698$ states all receive phases of varying $\pi/350$ multiples.  If instead these $698$ states were also set to receive phases of $\pi$, then $U_{\textrm{P}}$ would be exactly $U_{\textrm{G}}$.  But by having them evenly spread out over a full $2 \pi$ range, their impact on the amplitude amplification process can be seen in $P_{\textrm{M}}$ and $S_{\textrm{M}}$.
	
	While the special case of $\sigma = 0$ can be thought of as the theoretical limit where $U_{\textrm{P}}$ approaches $U_{\textrm{G}}$, the remaining results shown in figure \ref{F19} illustrate how gaussian amplitude amplification performs for $\sigma$ values which represent more realistic optimization problems.  As one might expect, the top plot shows a steadily decreasing trend in $P_{\textrm{M}}$ as $\sigma$ increases, accompanied by similar incremental increases in $S_{\textrm{M}}$.  These trends continue smoothly up to approximately $\sigma \approx 0.64$, which we shall refer to as $\sigma_{\textrm{cutoff}}$, at which point both plots change dramatically.  The critical difference between the quantum systems we are modeling above and below $\sigma_{\textrm{cutoff}}$ is that beyond this point the gaussian distributions of $U_{\textrm{P}}$ are so wide that they begin to populate multiple states with the value  $W_{\textrm{min}}$.  Consequently, if there are $M$ states all with the same $W_{\textrm{min}}$, then they will all share $1/M$$^{\textrm{th}}$ of the probability boosting from amplitude amplification.  For this reason, we've included the red-dashed line in the top plot of \ref{F19}, which multiplies each peak $P_{\textrm{M}}$ by the pop.($W_{\textrm{min}}$).  Thus, the red-dashed line is a more accurate representation of the relation between $P_{\textrm{M}}$ and $\sigma$ for this particular Hilbert space size, independent of how many $W_{\textrm{min}}$'s are present in the system.
	
	The value $\sigma_{\textrm{cutoff}}$ can be interpreted as the limit where a particular optimization problem is expected to have more than one optimal solution.  For sequential bipartite graphs, we can manipulate the odds of getting multiple $W_{\textrm{min}}$ paths by increasing $N$ while simultaneously decreasing $L$ and $R$.  Importantly, the presence of multiple $W_{\textrm{min}}$'s does not detract from a $U_{\textrm{P}}$'s aptitude for boosting states, as evidenced by the red-dashed line which represents the shared probability across all $| \textrm{P}_{\textrm{min}} \rangle$ states.   However, it does significantly impact the expected optimal number of iterations $S_{\textrm{M}}$, which can be seen in the bottom plot of figure \ref{F19}.  Having multiple states share the optimal phase is analogous to a result from 1998 \cite{brassard1}, where the step count for Grover's search algorithm is reduced from O($\frac{\pi}{4} \sqrt{N}$) to O($\frac{\pi}{4} \sqrt{N/M}$) for $M$ marked states.  Here the same effect can be observed in the $S_{\textrm{M}}$ plot, where each increase in the pop.($W_{\textrm{min}}$) results in a factional reduction to $S_{\textrm{M}}$.

	%%%%%%%%%%%%%%%%%%%%%%%%%%%%%%%%%%%%%%%%
	\subsection{Short Tail Model}%                                                       Short Tail Model
	%%%%%%%%%%%%%%%%%%%%%%%%%%%%%%%%%%%%%%%%
	
	One important trend from long tail model and figure \ref{F19}, which will continue throughout this study, is the inverse relation between the standard deviation $\sigma$ of a problem's solution space $\mathbb{W}$, and $U_{\textrm{P}}$'s ability to boost $| \textrm{P}_{\textrm{min}} \rangle$.  Thus, the ideal optimization problem for amplitude amplification is one with a naturally small $\sigma$, and $W_{\textrm{min}}$ as distanced from the mean as possible (i.e long tails).  More realistically however, these two conditions are contradictory to each other: the smaller $\sigma$ is for a given problem, the $closer$ we expect $W_{\textrm{min}}$ to be to the mean. 
	
	Returning now to the bottom right $U_{\textrm{s}} | s \rangle$ plot of figure \ref{F18}, here we present results from our simulator which model problems more akin to figure \ref{F12}.  We refer to these $\mathbb{W}$ distributions as the `short tail' model, by which we mean the expected number of solutions where pop.($W_i$)$=1$ is small, and the expected number of solutions where pop.($W_i$)$=0$ increases as $\sigma$ decreases.  Unlike the long tail model, this represents an optimization problem where $W_{\textrm{min}}$ is unknown (changing as a function of $\sigma$), making it more difficult to find an effective $p_{\textrm{s}}$ scaling factor, such as equation \ref{E24} below.

	\begin{eqnarray}            
		W_{\textrm{min}} \cdot p_{\textrm{s}}  &=&  x \label{E22}   \\
		W_{\textrm{max}} \cdot p_{\textrm{s}} &=&  x + 2 \pi \label{E23}  \\
		p_{\textrm{s}}  &=&  \frac{2 \pi}{ W_{\textrm{max}} - W_{\textrm{min}} }   \label{E24} 
	\end{eqnarray}
	
	Because we have full information of the quantum systems we are modeling, both $W_{\textrm{min}}$ and $W_{\textrm{max}}$ are known for every simulation so we are able to use equation \ref{E24} to find the optimal $p_{\textrm{s}}$ for each $U_{\textrm{P}}$.  In the long tail model no $p_{\textrm{s}}$ scaling was necessary, whereas here it is required in order to align $| \textrm{P}_{\textrm{min}} \rangle$ for optimal boosting.  Shown below in figure \ref{F20} is an illustration of this rescaling process, analogous to figure \ref{F15}.

	\begin{figure}[h]               
		\centering
		\includegraphics[scale=.2]{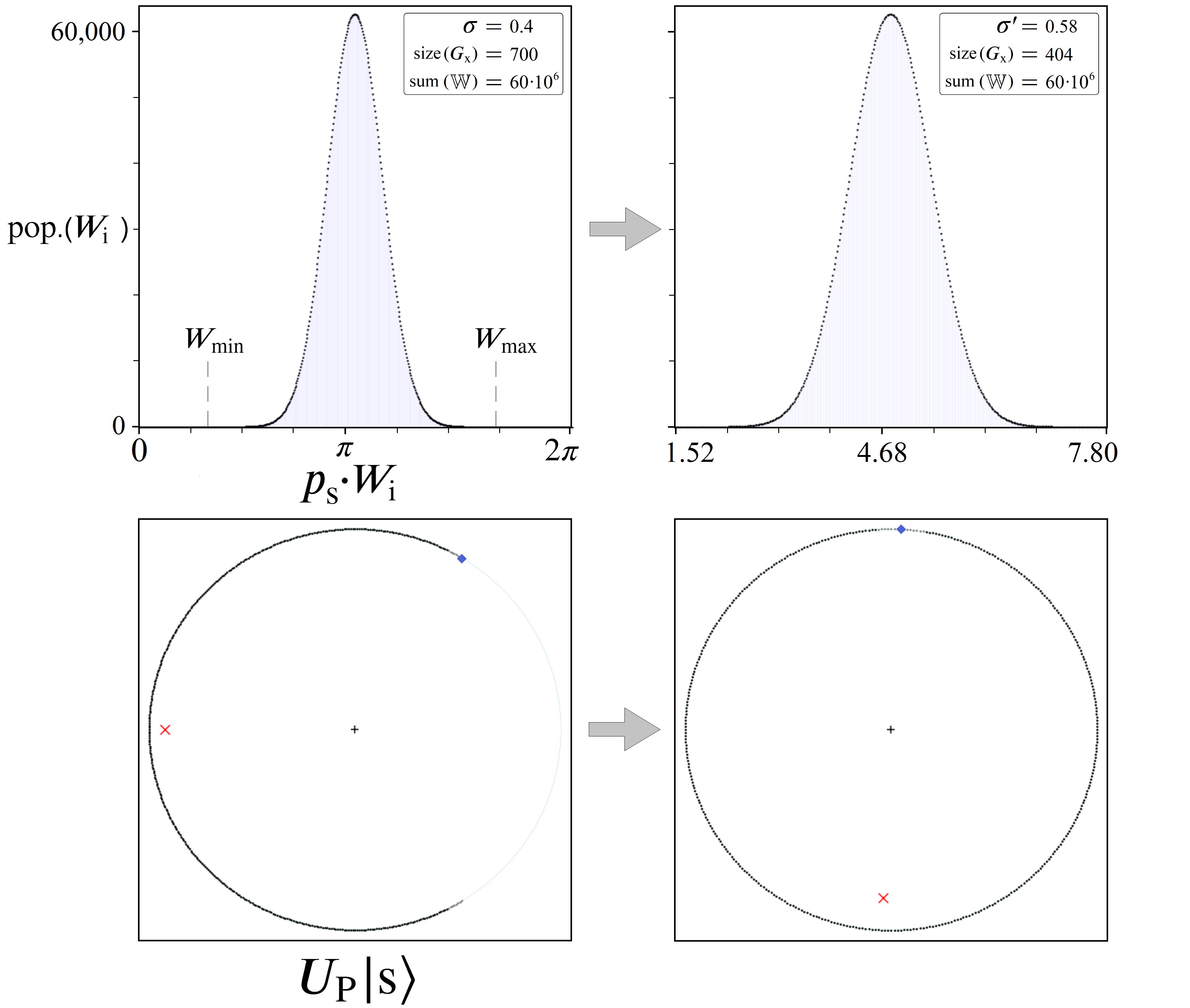}
		\caption{ (top left) An example distribution created from our simulator following the short tail model, causing $W_{\textrm{min}}$ and $W_{\textrm{max}}$ to be located away from $0$ and $2 \pi$. (top right) The same distribution scaled by $p_{\textrm{s}}$ to a full $2 \pi$ range.  (bottom) Below each histogram distribution is an amplitude space plot of $U_{\textrm{P}} | s \rangle$, tracking the location of $| \textrm{P}_{\textrm{min}} \rangle$ (blue diamond) and the mean point (red `X').}
		\label{F20}
	\end{figure}
	
	The process shown in figure \ref{F20} takes place in our simulations immediately following step $3$ of figure \ref{F17}, before simulating amplitude amplification for $P_{\textrm{M}}$ and $S_{\textrm{M}}$.  The consequence of this rescaling can be seen in the statistics of the top right distribution, resulting in new $\sigma '$ and size($G_{\textrm{x}}$) values from the original.  This rescaled size($G_{\textrm{x}}$) value comes from the number of $G_{\textrm{y}} \neq 0$ states (pop.($W_i$) $=0$) in the system, which have no impact on the amplitude amplification process.  Consequently, the boosting of $| \textrm{P}_{\textrm{min}} \rangle$ is driven by a new effective standard deviation $\sigma '$, which notably is always $ \sigma ' \geq  \sigma$.

	\begin{figure}[h]               
		\centering
		\includegraphics[scale=.22]{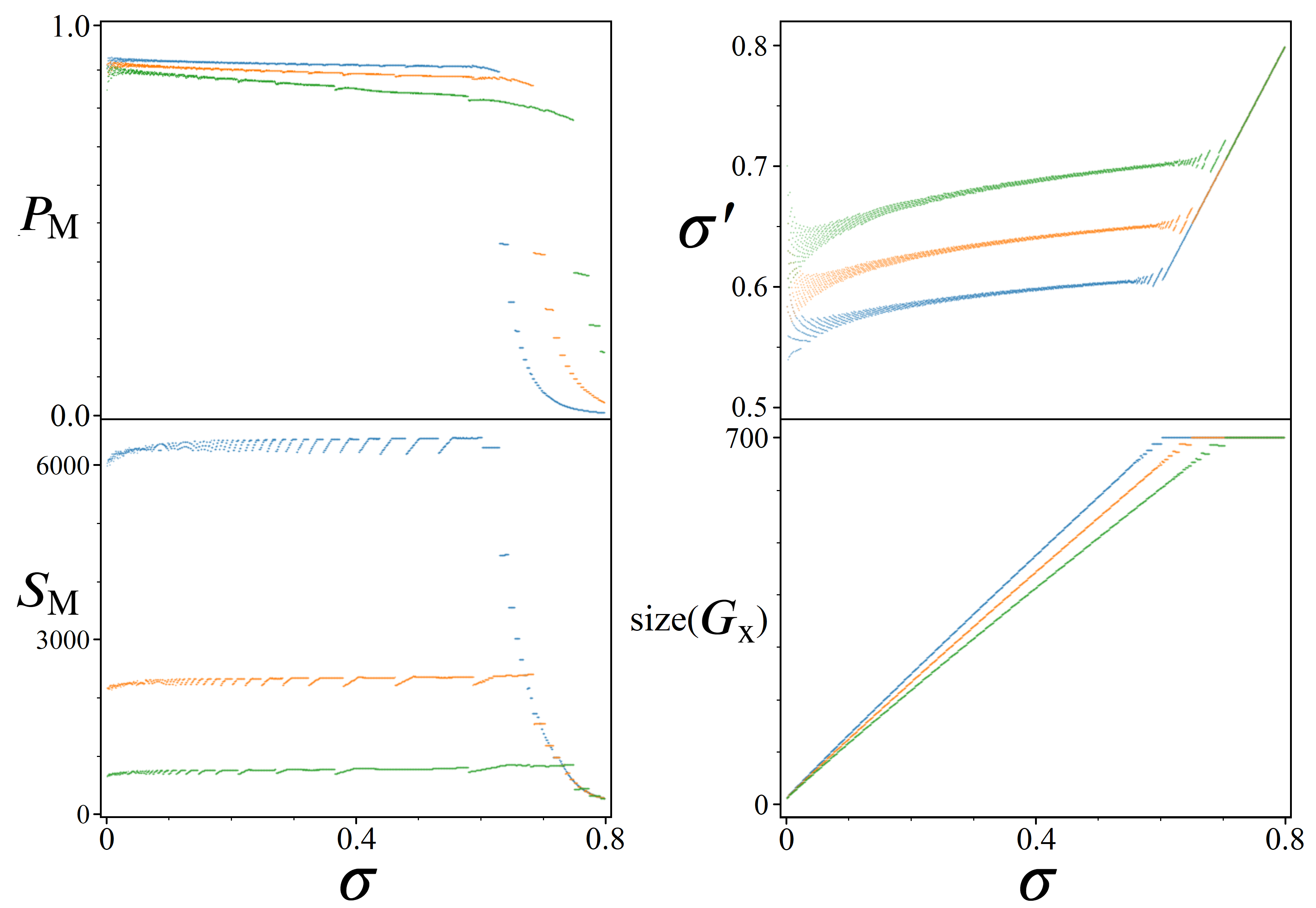}
		\caption{ Results for simulated gaussian distributions of various Hilbert space sizes (blue = $60 \cdot 10^6$, orange = $10 \cdot 10^6$, and green = $2 \cdot 10^6$), following the  short tail model, as a function of initial standard deviation $\sigma$. (left) $P_{\textrm{M}}$ and $S_{\textrm{M}}$ plots for boosting $| \textrm{P}_{\textrm{min}} \rangle$. (top right) The standard deviation $\sigma '$ after rescaling each distribution by the $p_{\textrm{s}}$ value which maximizes $P_{\textrm{M}}$ (see figure \ref{F20}). (bottom right) The total number of unique $W_i$ phases modeled by each distribution. }
		\label{F21}
	\end{figure}

	Shown in figure \ref{F21} are results of simulated amplitude amplification for the short tail model, for a range of initial $\sigma$ values [$0$,$0.8$] and initial size($G_{\textrm{x}}$)$=700$.  In all four plots there are three sets of data for various Hilbert space sizes: $N = 60 \cdot 10^{6}$ (blue), $N = 10 \cdot 10^{6}$ (orange), and $N = 2 \cdot 10^{6}$ (green).	In contrast to the long tail model results of figure \ref{F19}, figure \ref{F21} illustrates a different trend for $P_{\textrm{M}}$ vs. $\sigma$ up to $\sigma_{\textrm{cutoff}}$.  The highest $P_{\textrm{M}}$ achievable for $N = 60 \cdot 10^{6}$ at $\sigma = 0$ was previously $\sim 0.997$, now only $\sim 0.917$ following the short tail model.  However, if we look at the top right plot of $\sigma'$ vs. $\sigma$, we can see where this lower $P_{\textrm{M}}$ value comes from.  Over the range of initial $\sigma$ values [$0$ , $\sigma_{\textrm{cutoff}}$], the consequence of rescaling with $p_{\textrm{s}}$ are $\sigma '$ values between [$0.54$ , $0.59$].  Comparing these $\sigma '$ values with figure \ref{F19}, the long tail model predicts $P_{\textrm{M}}$ values around $0.89 \sim 0.92$, which is exactly what we find for $P_{\textrm{M}}$'s reported in figure \ref{F21}.
	
	To explain this new relation between  $\sigma '$ and $P_{\textrm{M}}$, we must note the two additional Hilbert sizes $N$ (orange and green data points) shown in figure \ref{F21}.  For any given initial $\sigma$, all three simulation sizes were derived from the same normalized gaussian in step $1$ of figure \ref{F17}.  Yet due to their differing $N$ values, each system size populates a different number of unique $W_{i}$ states, shown in the bottom right plot of size($G_{\textrm{x}}$) vs. $\sigma$.  For each $\sigma$, the largest Hilbert space $N = 60 \cdot 10^{6}$ always results in the biggest size($G_{\textrm{x}}$) after rounding, which consequently yields the largest distance between $W_{\textrm{min}}$ and $W_{\textrm{max}}$.  This distance dictates the necessary amount of rescaling by $p_{\textrm{s}}$ (equation \ref{E24}), resulting in different $\sigma '$ values, which in turn determine achievable $P_{\textrm{M}}$'s for $| \textrm{P}_{\textrm{min}} \rangle$.
	
	To summarize, the findings presented here for the long and short tail models demonstrate the range of success that gaussian amplitude amplification can produce.  For any optimization problem, we must consider not only the solution space $\mathbb{W}$'s natural $\sigma$, but how the distribution of $W_i$'s can be mapped to a $2 \pi$ range of phases for $U_{\textrm{P}}$.  This was the motivation for introducing $\sigma '$ via the short tail model, which demonstrated that problem size $N$ is just as important as $\sigma$.  Even for a problem that may possess a naturally small $\sigma$, if $N$ isn't sufficiently large enough to probabilistically produce $W_{\textrm{min}}$ / $W_{\textrm{max}}$ solutions away from the mean, then the problem may not be viable for a quantum solution.  Conversely, if we $are$ able to encode large optimization problems into $U_{\textrm{P}}$ oracles, then we can expect successes analogous to the long tail model with small $\sigma$.

	%%%%%%%%%%%%%%%%%%%%%%%%%%%%%%%%%%%%%%%%%%%%%%%%
	%%%%%%%%%%%%%%%%%%%%%%%%%%%%%%%%%%%%%%%%%%%%%%%%
	\section{Algorithmic Viability}%                                               Algorithmic Viability
	%%%%%%%%%%%%%%%%%%%%%%%%%%%%%%%%%%%%%%%%%%%%%%%%
	%%%%%%%%%%%%%%%%%%%%%%%%%%%%%%%%%%%%%%%%%%%%%%%%
	
	The hope of quantum computers isn't to solve artificially created ideal scenarios, but problems which arise naturally with inherent difficulties.  Following the simulated gaussian amplitude amplification results from the previous section, we now ask how reliable this boosting mechanism is for $\mathbb{W}$ distributions with imperfections that one would expect from realistic problems.  What follows in the coming subsections are observations and techniques for applying the quantum pathfinding algorithm \ref{A3} to randomly generated $\mathbb{W}$ distributions according to equations \ref{E7} - \ref{E12}.

	%%%%%%%%%%%%%%%%%%%%%%%%%%%%%%%%%%%%%%%%
	\subsection{Finding an optimal $p_{\textrm{s}}$}%                                                         Finding an Optimal ps
	%%%%%%%%%%%%%%%%%%%%%%%%%%%%%%%%%%%%%%%%
	
	In order to achieve a successful gaussian amplitude amplification on $| \textrm{P}_{\textrm{min}} \rangle$ / $| \textrm{P}_{\textrm{max}} \rangle$, for a $\mathbb{W}$ distribution with deviations from a perfect gaussian, the key lies in finding an optimal scaling parameter $p_{\textrm{s}}$.  In section V.B. we introduced $p_{\textrm{s}}$ as a necessary means for translating the full range of $\mathbb{W}$ down to $[x, x+2 \pi]$, and again in the short tail model for section VI.C.

\begin{figure}[h]             
		\centering
		\includegraphics[scale=.44]{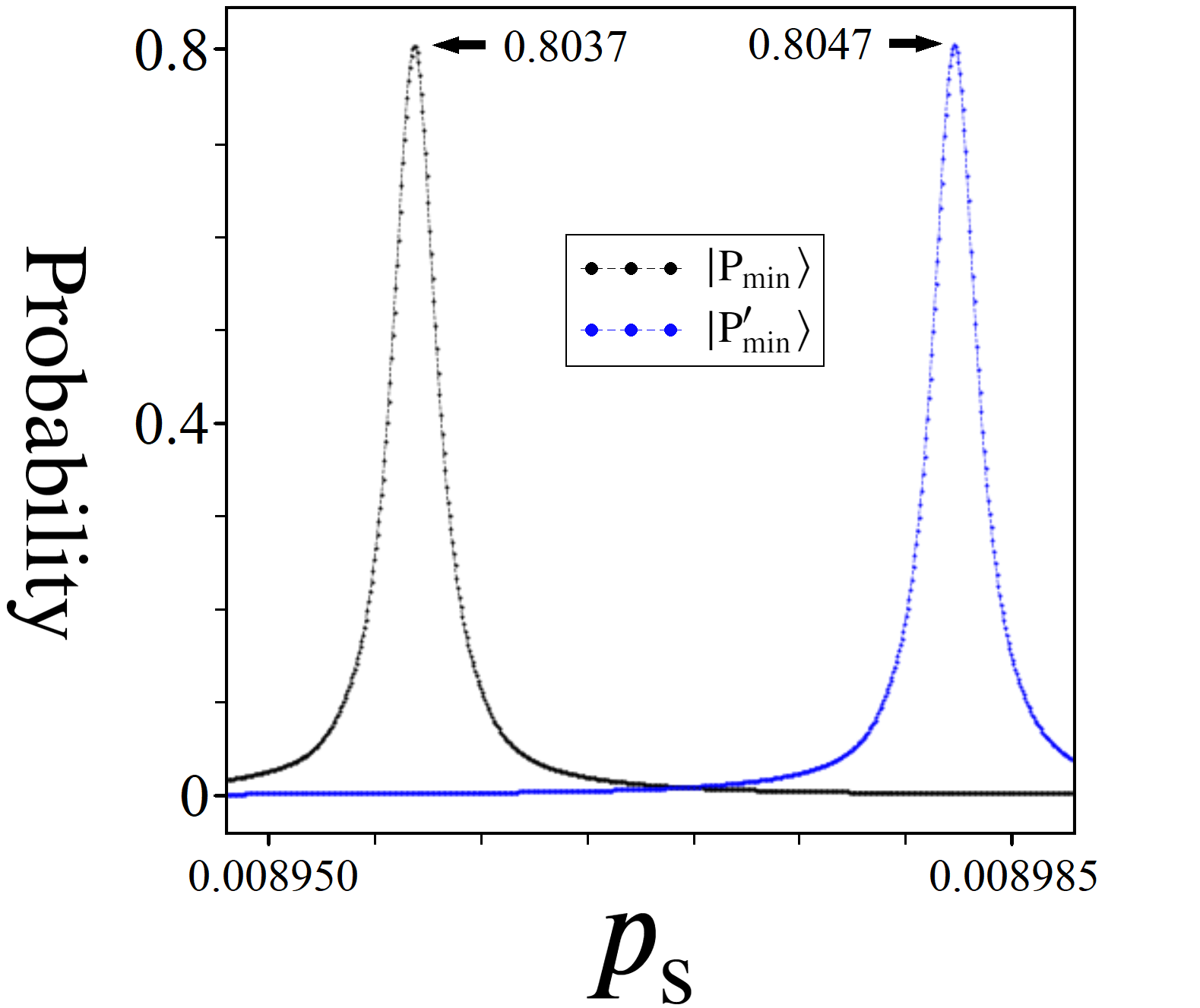}
		\caption{ A plot of $p_\textrm{s}$ vs. achievable probabilities via amplitude amplification, for the $\mathbb{W}$ distribution shown in figure \ref{F13}.   The state $| \textrm{P}_{\textrm{min}} \rangle$ represents the solution to the pathfinding problem $W_{\textrm{min}}$, while $| \textrm{P}_{\textrm{min}}' \rangle$ corresponds to the next smallest $W_{i}$.   }
		\label{F22}
	\end{figure}

	The approach outlined in equation \ref{E24} is a way of ensuring $| \textrm{P}_{\textrm{min}} \rangle$ and $| \textrm{P}_{\textrm{max}} \rangle$ form a complete $2\pi$ range, but it is not necessarily the optimal $p_{\textrm{s}}$ for amplitude amplification.  Firstly, it causes the states $| \textrm{P}_{\textrm{min}} \rangle$ and $| \textrm{P}_{\textrm{max}} \rangle$ to share the boosting effect equally, which is not ideal for problems where we are interested in finding only one or the other.  But more importantly, randomness in $\mathbb{W}$ means that the overall distribution of phases from $U_{\textrm{P}}$ is very likely to be not symmetric.  This means that the optimal $p_{\textrm{s}}$ for boosting $| \textrm{P}_{\textrm{min}} \rangle$ will differ from the optimal $p_{\textrm{s}}$ for $| \textrm{P}_{\textrm{max}} \rangle$.  These different $p_{\textrm{s}}$'s correspond to values which best align $| \textrm{P}_{\textrm{min}} \rangle$ / $| \textrm{P}_{\textrm{max}} \rangle$ with a $\pi$ phase difference from the mean.  Figure \ref{F22} illustrates an example of this, as well as the margin for error in finding the optimal $p_{\textrm{s}}$ value before accidentally boosting an unintended state.	
	
	Derived from the same directed graph used to produce figure \ref{F13}, the two plots shown in figure \ref{F22} were created by carefully simulating algorithm \ref{A3} over the range of $p_{\textrm{s}}$ values shown along the x-axis, for $| \textrm{P}_{\textrm{min}} \rangle$ as well as the second best solution state $| \textrm{P}_{\textrm{min}}' \rangle$.  It is clear by the two spikes in probability, and the space in between, that the role of $p_{\textrm{s}}$ for unlocking successful amplitude amplifications cannot be ignored. For this particular example, using a scaling factor of $p_{\textrm{s}} \approx 0.008957$ causes the state $| \textrm{P}_{\textrm{min}} \rangle$ to reach a peak probability of about 80.37$\%$, while using $p_{\textrm{s}} \approx 0.008982$ causes $| \textrm{P}_{\textrm{min}}' \rangle$ to boost to about 80.47$\%$.  Thus, a margin of error on the order of $\sim 3 \cdot 10^{-5}$ in $p_{\textrm{s}}$ is enough to change what state gets boosted.
	
	Additional notables from figure \ref{F22} are as follows: 1) Despite a single optimal $p_{\textrm{s}}$ for boosting $| \textrm{P}_{\textrm{min}} \rangle$, the plot shows a range of $p_{\textrm{s}}$ values around the optimal case for which the algorithm can still be successful.  2)  The range of $p_{\textrm{s}}$ values between the two peaks can be regarded as a `dead zone', where no state in the system receives a meaningful probability boost.  3)  Because states near $| \textrm{P}_{\textrm{min}} \rangle$ are also able to receive meaningful amplitude amplifications ($| \textrm{P}_{\textrm{min}}' \rangle$), this suggests that the algorithm may be viable for a heuristic technique. 4) From an experimental viewpoint, the scale of precision shown for $p_{\textrm{s}}$ must be achievable via phase gates, which means the size of implementable problems will be dictated by the technological limits of state-of-the-art quantum devices.

	%%%%%%%%%%%%%%%%%%%%%%%%%%%%%%%%%%%%%%%%
	\subsection{Single vs. Multiple $p_{\textrm{s}}$}%                                                     Single vs Multiple
	%%%%%%%%%%%%%%%%%%%%%%%%%%%%%%%%%%%%%%%%
	
	The two plots shown in figure \ref{F22} represent potential amplitude amplification peaks, where a single $p_{\textrm{s}}$ scaling factor is used for every iteration of $U_{\textrm{s}} U_{\textrm{P}}$.  However, in principle this is not necessarily the optimal strategy for boosting $| \textrm{P}_{\textrm{min}} \rangle$, as $p_{\textrm{s}}$ could theoretically be different with each iteration.  The choice of $p_{\textrm{s}}$ at each step is an extra degree of freedom available to the experimenter, which we explore here as a potential tool for overcoming randomness in $\mathbb{W}$. 
	
	In order to better quantify the advantage a step-varying $p_{\textrm{s}}$ approach has to offer, let us first define our metric for a successful amplitude amplification in equation \ref{E28} below.  We refer to this metric as `probability of success', labeled $P_{\textrm{succ}}$, which combines an amplitude amplification's peak probability and step count into a single number, quantifying the probability of a quantum speedup over classical.
	
	\begin{eqnarray}       
		C_{\textrm{steps}} &=& N^2 \cdot (L - 1) \label{E25} \\
		r &=& \lfloor   C_{\textrm{steps}} / Q_{\textrm{steps}}  \rfloor \label{E26} \\
		P_{\textrm{M}} &=& \textrm{Prob.}(  \hspace{.03cm}| \textrm{P}_{\textrm{min}} \rangle ) \label{E27} \\
		P_{\textrm{succ}} &=& 1 - (1-P_{\textrm{M}})^r
		\label{E28}
	\end{eqnarray}	
	
	To summarize the components making up equation \ref{E28}: $C_{\textrm{steps}}$ is the number of classical steps needed to find $W_{\textrm{min}}$ (equal to the total number of edges), $Q_{\textrm{steps}}$ is the number of $U_{\textrm{s}} U_{\textrm{P}}$ iterations needed in order to reach the peak probability $P_{\textrm{M}}$, and $r$ is the number of allowable amplitude amplification attempts to measure $| \textrm{P}_{\textrm{min}} \rangle $ before exceeding $C_{\textrm{steps}}$.  Altogether, $P_{\textrm{succ}}$ represents the probability that $| \textrm{P}_{\textrm{min}} \rangle $ will be successfully measured within $r$ attempts.  Using dice as a simple example, the probability of success that one will roll a 1-5 in four attempts is $P_{\textrm{succ}} = 1 - (1-\frac{5}{6})^4 \approx 99.92\%$.
	
	The quantity $P_{\textrm{succ}}$ is a simplified way of comparing quantum vs. classical speeds, more specifically query complexity, which ignores many of the extra complicating factors of a more rigorous speed comparison (classical CPU speeds, quantum gate times, quantum decoherence and error correction, etc.).  Here, we are simplifying one step in classical as the processing of information from a single weighted edge $\omega_i$ (steps 4-6 in Alg. \ref{A2}), versus one step in quantum as a single iteration of $U_{\textrm{s}} U_{\textrm{P}}$ (steps 4 \& 5 in Alg. \ref{A3}).  This is the typical manner in which Grover's search algorithm is considered a quadratic speedup, and is sufficient for our study's purpose.
	
\begin{figure}[h]             
		\centering
		\includegraphics[scale=.28]{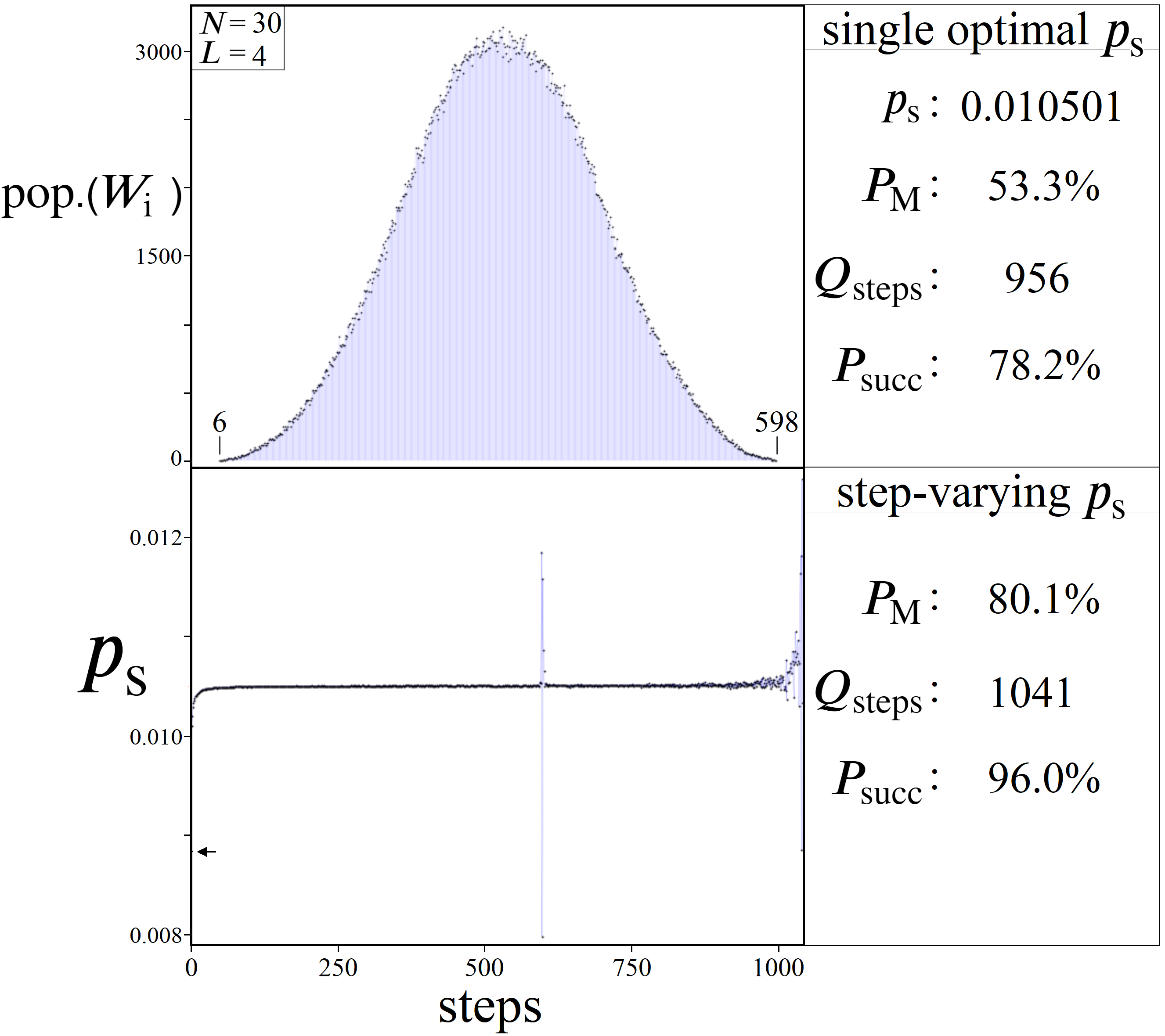}
		\caption{ (top) An example $\mathbb{W}$ histogram distribution for the case $N=30$, $L=4$, $R=200$. (bottom) A plot of all $p_{\textrm{s}}$ values used at each step in order to optimized the probability of measuring $| \textrm{P}_{\textrm{min}} \rangle $.  Note the small black arrow, marking the $p_{\textrm{s}}$ value at step 1.  To the right of each plot are accompanying details about the success of each amplitude amplification process for each approach.  }
		\label{F23}
	\end{figure}

	With $P_{\textrm{succ}}$ now defined, we return to the question of whether a step-varying approach to $p_{\textrm{s}}$ can improve gaussian amplitude amplification.  For details on how an optimal $p_{\textrm{s}}$ can be computed at each step of the algorithm, please see Appendix $\textbf{B}$ for our technique.  To summarize, we simulate a range of $p_{\textrm{s}}$ values at each step such that the distance in amplitude space between $| \textrm{P}_{\textrm{min}} \rangle $ and the mean point is maximized, resulting in the largest reflection about the average from $U_{\textrm{s}}$ per step.  Figure \ref{F23} shows an example for the case $N=30$, $L=4$, and resulting $P_{\textrm{M}}$ \& $P_{\textrm{succ}}$.

	%------------------------------------------------------------------------------------------------------------------- Here to generate on next page
	\begin{figure*}[!th]                
		\centering
		\includegraphics[scale=.5]{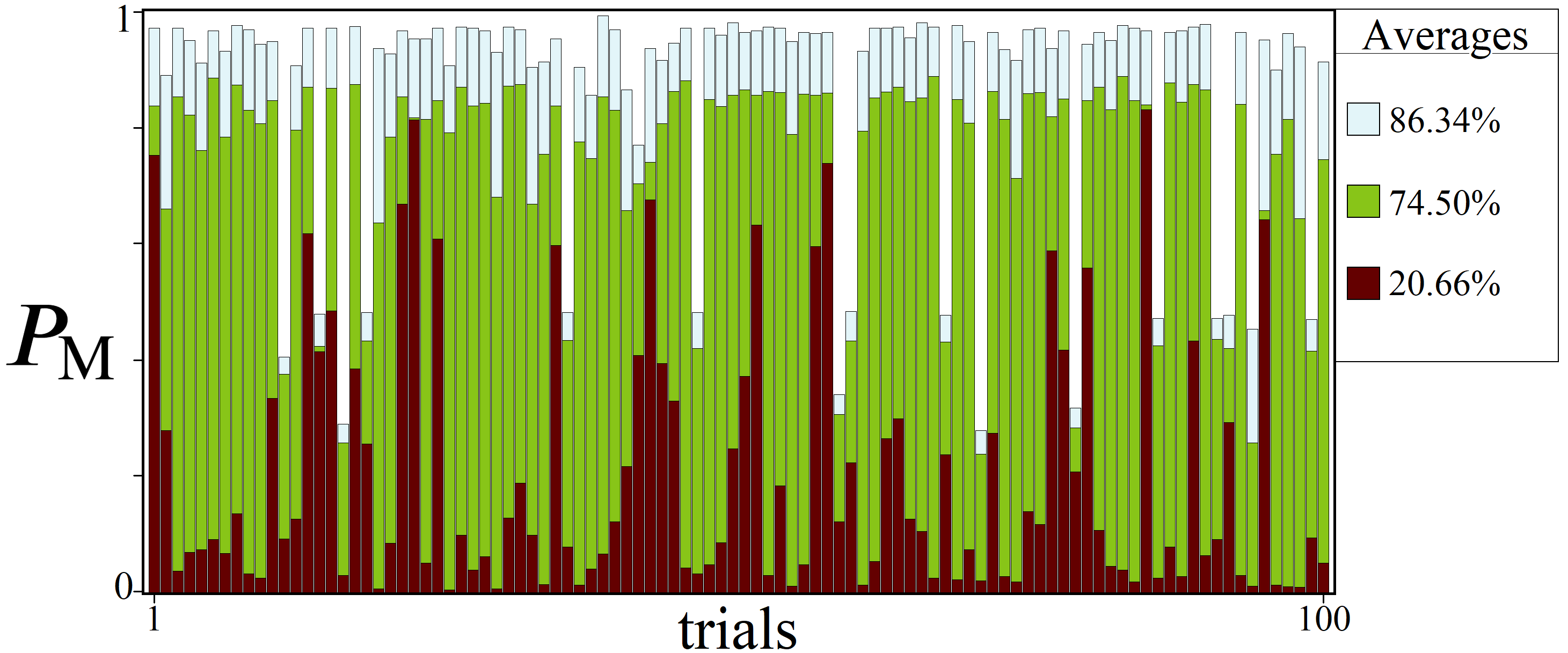}
		\caption{Results from testing on 100 randomly generated $\mathbb{W}$ distributions, for $N=6$, $L=10$, $R=100$.  For each trial, we report the highest $P_{\textrm{M}}$ probability found for the state $| \textrm{P}_{\textrm{min}} \rangle$ using (light blue) a step-varying $p_{\textrm{s}}$ approach, (green) a single optimal $p_{\textrm{s}}$ approach, and (dark red) an average $p_{\textrm{s}}$ approach.  Reported on the right side of the figure are the averages found for all three approaches.    }
		\label{F24}
	\end{figure*}
	
	%-------------------------------------------------------------------------------------------------------------------

	As evidenced by the accompanying numbers in figure \ref{F23}, a step-varying approach to $p_{\textrm{s}}$ is indeed advantageous for getting the maximal peak probability $P_{\textrm{M}}$ out of a given $\mathbb{W}$.  However, it is also clear that the exact sequence of $p_{\textrm{s}}$ values (bottom plot) is non-trivial, and likely unpredictable from an experimental perspective when dealing with randomized data.  Although the majority of $p_{\textrm{s}}$'s are near a single value, there are constant sharp fluctuations at every step, some small while others quite large.  These fluctuations can be understood as a signature of the $\mathbb{W}$ distribution, unique to every problem, actively counteracting the randomness of the graph's weighted edges at every step. 
	
	The result shown in figure \ref{F23} for improving $P_{\textrm{M}}$ was found to be very consistent. More specifically, $\textit{every}$ randomly generated graph that was studied, for all $N$ and $L$, could always be optimized to produce a higher $P_{\textrm{M}}$ using a step-varying $p_{\textrm{s}}$ approach versus only a single $p_{\textrm{s}}$.  However, in some cases it was found that the larger $P_{\textrm{M}}$ value did not directly translate to a better $P_{\textrm{succ}}$, as the resulting higher $Q_{\textrm{step}}$ count caused $P_{\textrm{succ}}$ to be lower (fewer attempts to measure $| \textrm{P}_{\textrm{min}} \rangle$).    In general, our tests found the step-varying $p_{\textrm{s}}$ approach to be most effective at improving $P_{\textrm{M}}$ and $P_{\textrm{succ}}$ for smaller problem sizes.  But these smaller cases oftentimes produced $p_{\textrm{s}}$ vs. step plots (bottom of figure \ref{F23}) which were highly chaotic and irregular from problem to problem, even for the same $N$ and $L$.  Conversely, as problem sizes increase, the difference between the single vs. step-varying approaches became more negligible, with much more regular and stable $p_{\textrm{s}}$ vs. step plots.

	%%%%%%%%%%%%%%%%%%%%%%%%%%%%%%%%%%%%%%%%
	\subsection{Statistical Viability}%                                                         Statistical Viability
	%%%%%%%%%%%%%%%%%%%%%%%%%%%%%%%%%%%%%%%%
	
	While the results from the previous subsection can be regarded as a more theoretical strategy for optimizing $P_{\textrm{M}}$, here we address the issue of finding $p_{\textrm{s}}$ from a more practical perspective.  In any realistic optimization problem, it is fair to assume that the experimenter has limited information about $\mathbb{W}$.  Consequently, using a strategy for finding a suitable $p_{\textrm{s}}$ such as equation \ref{E24} may be impossible, which then begs the question: how feasible is gaussian amplitude amplification when used blindly?  To help answer this question we conducted a statistical study, shown in figure \ref{F24}.  The general idea is to imagine a scenario in which the experimenter needs to solve the same sized directed graph problem numerous times, with randomized but similar values each time (for example, optimal driving routes throughout a city can change hourly due to traffic patterns).  Under these conditions, we are interested in whether a quantum strategy can use information from past directed graphs in order to solve future ones.
	
	The results shown in figure \ref{F24} illustrate the varying degrees of success one can expect using three different $p_{\textrm{s}}$ approaches.  The figure showcases $100$ randomly generated directed graphs of size $N=6$, $L=10$, $R=100$, and their resulting peak $P_{\textrm{M}}$ probabilities.  Optimal $P_{\textrm{M}}$  values for each graph were found through simulating amplitude amplification using 1) (light blue) a step-varying $p_{\textrm{s}}$ approach, 2) (green) a single optimal $p_{\textrm{s}}$, and 3) (dark red) an average $p_{\textrm{s}}$.  For the average $p_{\textrm{s}}$, this value was computed by averaging together the $100$ single optimal $p_{\textrm{s}}$ values: $ \sim 0.0083478$.
	
	Two notables from figure \ref{F24} are as follows: 1) Even for this appreciably large problem size (over $60$ million paths), about 15\% of the $\mathbb{W}$ distributions studied could not be optimized for $P_{\textrm{M}}$ values over $50$\%.  We found this to be of interest for a future study: what is it about these $\mathbb{W}$ distributions and their randomness that makes them inherently difficult to boost? (see section IX.A)  2) The large discrepancy between the single optimal and average $p_{\textrm{s}}$ plots can be seen quite clearly across the $100$ trials.  However, returning to the question posed at the top of the subsection, the average $P_{\textrm{M}}$ of these blind attempts is roughly $20\%$ (top right corner of figure \ref{F24}).  If a quantum computer could reliably be trusted to find $| \textrm{P}_{\textrm{min}} \rangle$ $20\%$ (or more) of the time using a single $p_\textrm{s}$, this could be a viable use case for quantum, used in conjunction with a classical computer for a hybrid approach.

	%%%%%%%%%%%%%%%%%%%%%%%%%%%%%%%%%%%%%%%%%%%%%%%%
	%%%%%%%%%%%%%%%%%%%%%%%%%%%%%%%%%%%%%%%%%%%%%%%%
	\section{The Traveling Salesman}%                                    The Traveling Salesman
	%%%%%%%%%%%%%%%%%%%%%%%%%%%%%%%%%%%%%%%%%%%%%%%%
	%%%%%%%%%%%%%%%%%%%%%%%%%%%%%%%%%%%%%%%%%%%%%%%%

	As the final topic of this study, here we present results for a theoretical application of gaussian amplitude amplification as a means to solve the Traveling Salesman problem \cite{gutin} (TSP).  Solving the TSP in this manner is an idea that goes back to 2012 \cite{bang}, which we build upon here using the new insights gained from this study, particularly sections VI. and VII.  Because the adaptation of $U_{\textrm{P}}$ discussed here relies on qudit technologies, which we will not explicitly cover, we encourage interested readers to see \cite{wang} for an overview of unitary operations and quantum circuits for qudits.

	%%%%%%%%%%%%%%%%%%%%%%%%%%%%%%%%%%%%%%%%%%%%%%%%
	\subsection{Weighted Graph Structure}%                                                                          Weighted Graph Structure
	%%%%%%%%%%%%%%%%%%%%%%%%%%%%%%%%%%%%%%%%%%%%%%%%

\begin{figure}[h]                  
		\centering
		\includegraphics[scale=.22]{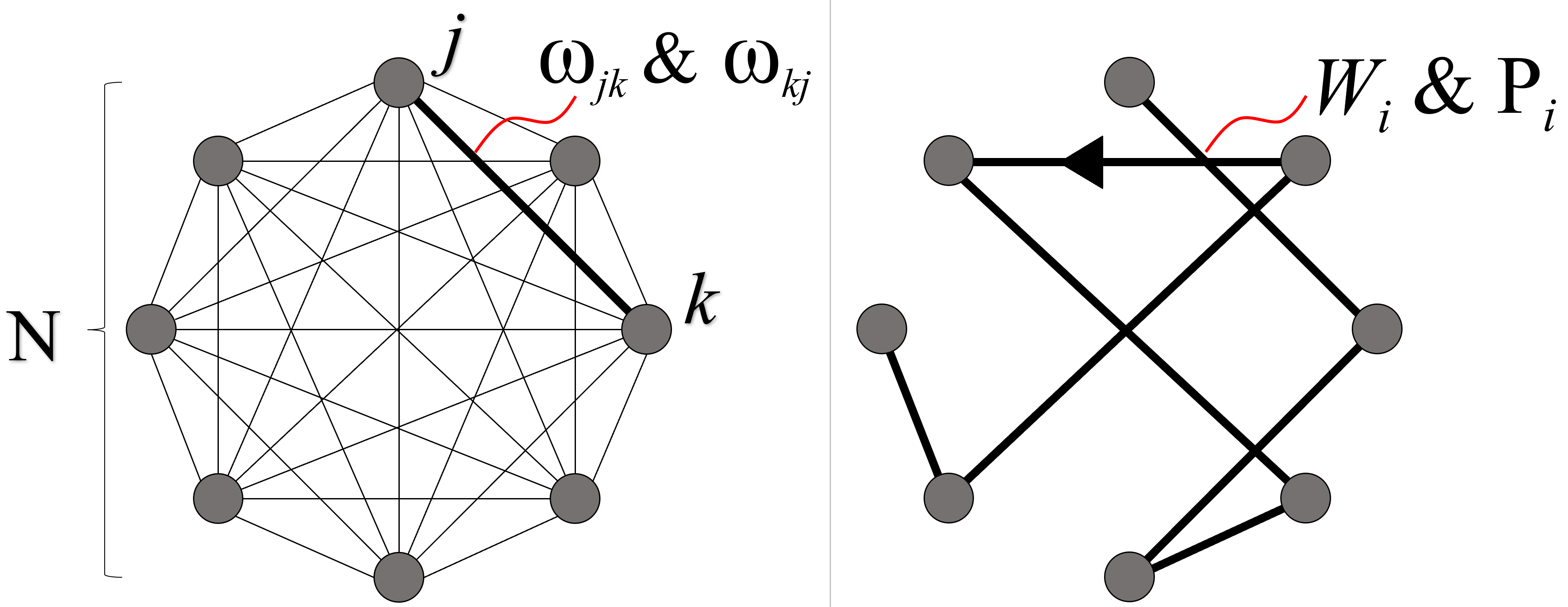}
		\caption{(left) Geometric structure for the Traveling Salesman Problem, for the case $N=8$.  Each edge contains a weighted value $w_{jk}$, where $j$ and $k$ are the two connected nodes. (right) An example path, touching each node exactly once.  Each path $\textrm{P}_i$ is defined by a unique ordering of all $N$ nodes ($N!$ in total), with $W_i$ corresponding to the sum of all weighted edges composing the path.}
		\label{F25}
	\end{figure}

	Let us begin by defining the exact formalism of the Traveling Salesman problem that we seek to solve using amplitude amplification.  Shown in figure \ref{F25} is an example TSP for the case of $N=8$, where $N$ corresponds to the total number of cities (nodes).  Just as with the sequential bipartite graphs from sections III. - VI., a TSP can be represented as a weighted directed (or undirected) graph.  Here we are interested in the most general case, an asymmetric TSP, where each edge has two unique weights $w_{ij}$ and $w_{ji}$, one for traveling in either direction across the edge.
	
	Once again, the solution we seek is $W_{\textrm{min}}$ or $W_{\textrm{max}}$, given in equations \ref{E29} - \ref{E30}.  For clarity, here we are defining a path P$_i$ as shown in figure \ref{F25}, traversing every node in the graph exactly once (and not returning to the starting node). In total this produces a solution space of $N!$ unique path permutations for a given TSP (for a symmetric TSP the number of permutations is the same, but the number of unique solutions is halved).  We will continue to denote the set of all possible paths as $\mathbb{P}$, and similarly the set of all possible solutions as $\mathbb{W}$.

	\begin{eqnarray}            
		\omega_{jk} &\in& [0,\textrm{R}]    \label{E29}\\
		W_{i} &=& \sum_{jk \hspace{.06cm}\in \hspace{.06cm} \textrm{P}_i} \omega_{jk}   \label{E30}  
	\end{eqnarray}

	%%%%%%%%%%%%%%%%%%%%%%%%%%%%%%%%%%%%%%%%%%%%%%%%
	\subsection{Encoding Mixed Qudit States}
	%%%%%%%%%%%%%%%%%%%%%%%%%%%%%%%%%%%%%%%%%%%%%%%%
	
	In order to realize a Hilbert space of size $N$! such that every possible path P$_i$ can be encoded as a quantum state $| \textrm{P}_i \rangle$, we require a mixed qudit quantum computer.  Given in equation \ref{E17} is the quantum state of a $d$-dimensional qudit, capable of creating superposition states spanning $|0\rangle_d$ through $|d-1\rangle_d$.  When using qudits of different dimensions together, their combined Hilbert space size is the product of each qudit's dimensionality, like shown in equation \ref{E31} below.  If one is restricted to a quantum computer composed of a single qudit size $d$, then only quantum systems of size $d^n$ are achievable.  Thus, a single $d$-qudit computer can never produce the needed $N!$ Hilbert space size (unless $d = N!$, which is impractical) for solving the TSP.
	
	\begin{eqnarray}           
		| \Psi \rangle_{24} = | Q \rangle_4 | Q' \rangle_3 | Q'' \rangle_2  = \sum_{i=0}^{3} \sum_{j=0}^{2} \sum_{k=0}^{1} \alpha_{ijk} | i \rangle_4 | j \rangle_3  | k \rangle_2 \hspace{0.2cm}
		\label{E31}
	\end{eqnarray}	
	
\begin{figure}[h]              
		\centering
		\includegraphics[scale=.24]{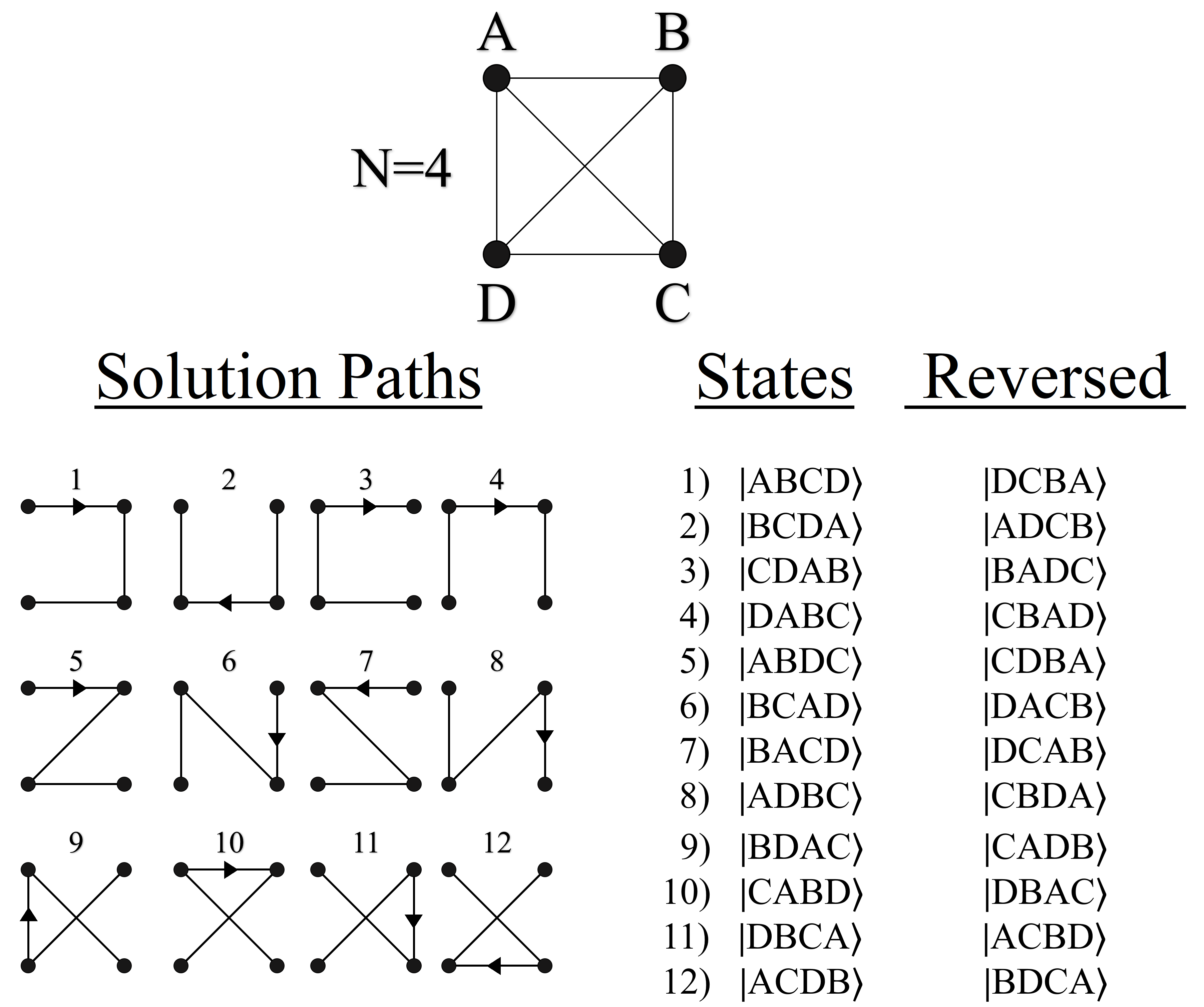}
		\caption{(left) Geometric illustrations for $12$ of the possible solution paths for an $N=4$ TSP weighted graph. (right)  Quantum state representations for the $12$ paths shown, plus $12$ additional states with opposite direction.}
		\label{F26}
	\end{figure}

	The quantum state shown above in equation \ref{E31} is the mixed qudit composition which can encode an $N=4$ TSP, capable of creating a superposition of $4! = 24$ states.  These $24$ states span every combination from the lowest energy state $ | 0 \rangle | 0 \rangle  | 0 \rangle$, up to the highest energy level for each qudit $ | 3 \rangle| 2 \rangle  | 1 \rangle$.  Each of these basis states will serve as a $| \textrm{P}_{i} \rangle$, receiving a phase proportional to its total path weight W$_{i}$ via the oracle $U_{\textrm{P}}$.  See figure \ref{F26} for an $N=4$ TSP example.
	
\begin{figure}[h]         
		\centering
		\includegraphics[scale=.2]{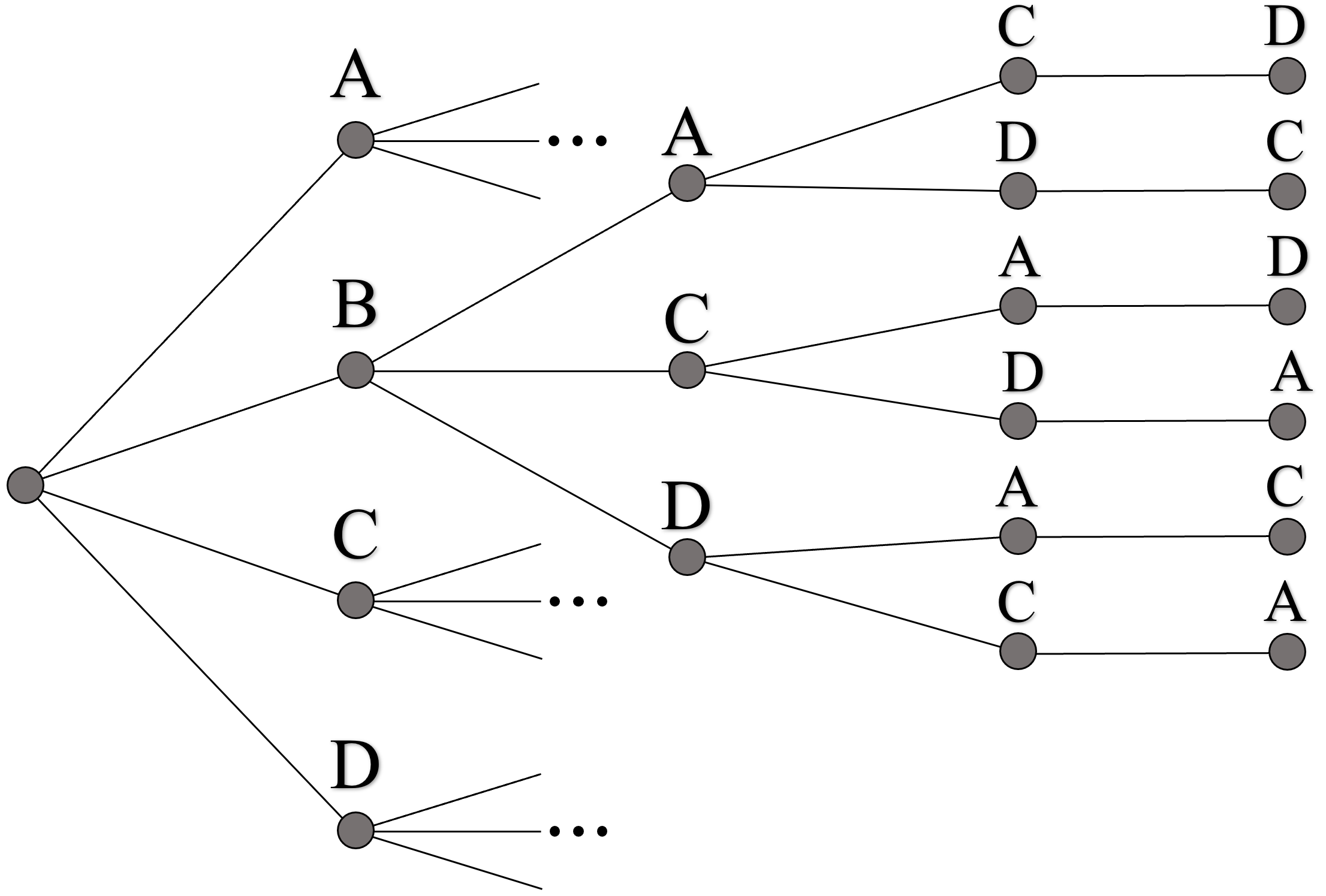}
		\caption{ Spanning tree representation of all possible paths for an $N=4$ Traveling Salesman problem.}
		\label{F27}
	\end{figure}

	The quantum states shown in figure \ref{F26} are meant to be symbolic, representing the information needed to specify each of the $24$ unique paths (order of nodes traversed).  For the realization of $U_{\textrm{P}}$ however, we must encode the information of these $24$ paths into the orthogonal basis states $ | i \rangle | j \rangle  | k \rangle$ via phases.  But unlike the convention used in figure \ref{F7}, where individual qubit states represent a single node in the graph, here we cannot use qudits in the same manner.  To understand why, it is helpful to visualize the problem from a different geometric perspective, shown in figure \ref{F27}.  
	
	The spanning tree representation shown in figure \ref{F27} is equivalent to the weighted directed graph in figure \ref{F25}, with the same solution $W_{\textrm{min}}$.  The motivation for looking at the problem in this manner is to highlight the decreasing number of possible choices with each successive layer.	Returning now to equation \ref{E31}, $| \Psi \rangle_{24}$'s mixed qudit composition was chosen to exactly mimic the dimensionality of choices at each layer in figure \ref{F27}.  For example, the largest qudit $| Q \rangle_4$ in the system has four available states, one to represent each of the four possible starting nodes.  Similarly, the next largest qudit $| Q' \rangle_3$ provides three possible states, one for each of the remaining untouched nodes, and so forth until the final qubit.  However, while the four states of $| Q \rangle_4$ can all be exactly assigned to one of the four starting nodes, the same cannot hold true for the states of $| Q' \rangle_3$ and $| Q'' \rangle_2$.
	
	%------------------------------------------------------------------------------------------------------------------- Here to generate on next page
	\begin{figure*}[!t]               
		\centering
		\includegraphics[scale=.19]{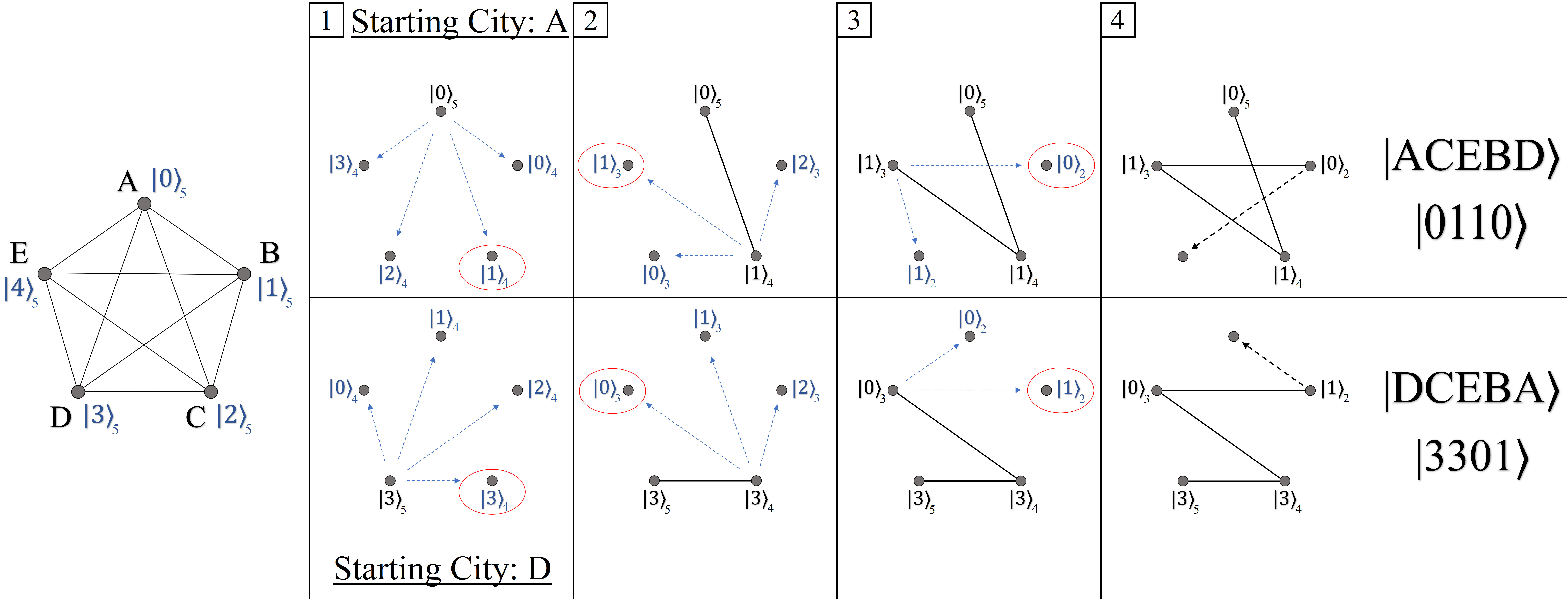}
		\caption{ (leftmost) Initial mapping of an $N=5$ TSP to the quantum states $|0\rangle$ - $|4\rangle$, and their accompanying city names.  (panels 1-4) Step by step outline of two different paths through the geometry, illustrating the `clockwise' nomenclature outlined in this section.  At each step, the path thus far is illustrated in solid black lines / states, while potential next nodes are shown in blue arrows / states.  }
		\label{F28}
	\end{figure*}
	%-------------------------------------------------------------------------------------------------------------------
	
	If we want to repeat the strategy for labeling $| \textrm{P}_i \rangle$ states like in figure \ref{F7}, then we require $N$ total $d=N$ qudits, such that each $| i \rangle_d$ basis state can be uniquely specified as a particular node in the graph.  However, this leads to a Hilbert space size of $N^N$, which is more than the number of total possible paths (for $N=4$, this is $256$ states for only $24$ paths).  These extra states are problematic because they represent invalid solutions to the TSP we want to solve, i.e. paths that traverse a single node more than once.  Thus, in order to solve an $N!$ sized problem, we must use a Hilbert space created from a mixed qudit approach like in equation \ref{E31}.
	
	Our solution to this $N!$ path/state encoding problem is outlined in figure \ref{F28}, for the case $N=5$.  The strategy for identifying each basis state of $| \Psi \rangle$ as a particular $| \textrm{P}_i \rangle$ follows from two rules: 1) initially label all nodes in the TSP graph with a unique $| i \rangle_d$ basis state for the $d=N$ largest qudit (leftmost graph).  2)  For subsequent $d < N$ qudits, each $| j \rangle_d$ basis state corresponds to one of the remaining untraversed nodes, ordered clockwise from the position of the $previous$ qudit state.  See figure \ref{F28} for two example paths, where possible qudit states at each step are shown in blue, and previous qudit states in black.
	
	The two rules specified above are enough to guarantee every $| \textrm{P}_i \rangle$ is unique, even though the meaning of individual qudit states are not.  while this encoding is sufficient, we note that other encodings are equally valid as well.  So long as $U_{\textrm{P}}$ is able to apply each phase $p_{\textrm{s}} \cdot W_{i}$ to the correct basis state $| \textrm{P}_{i} \rangle$, then the amplitude amplification results of the following subsection are applicable.

	%%%%%%%%%%%%%%%%%%%%%%%%%%%%%%%%%%%%%%%%
	\subsection{Simulated TSP Results}%                                                                                          Simulated TSP Results
	%%%%%%%%%%%%%%%%%%%%%%%%%%%%%%%%%%%%%%%%
	
	To conclude this discussion of the Traveling Salesman problem, here we present results which demonstrate how amplitude amplification performs as a function of $N$.  To do this, we analyzed each problem size from two approaches: 1) Analogous to figure \ref{F24}, find the optimal single $p_{\textrm{s}}$ for randomly generated graphs of each size, and record $P_{\textrm{M}}$ values. 2) Compare these results against our simulator from section VI.C. by gathering average statistics for $W_{\textrm{min}}$, $W_{\textrm{max}}$, and $\sigma '$, and use these along with $N!$ to predict expected $P_{\textrm{M}}$ values.  Results for method (1) are shown in figure \ref{F29} below.
	
	\begin{figure}[h]                    
		\centering
		\includegraphics[scale=.4]{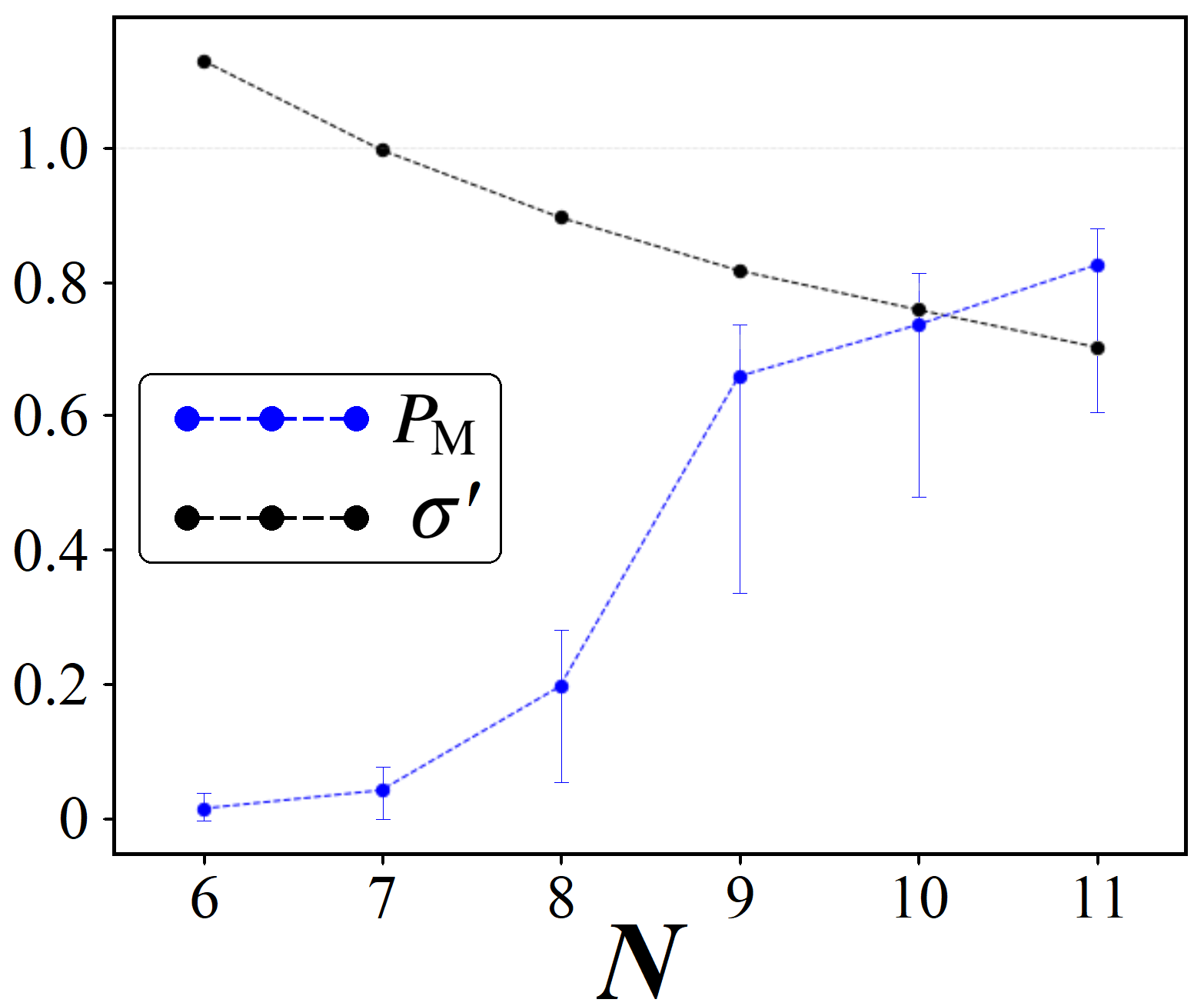}
		\caption{ Results from using a single optimal $p_{\textrm{s}}$ for randomly generated TSP weighted graphs as a function of problem size $N$, $R = 100$. (dots) Average values for $\sigma '$ (black) and $P_{\textrm{M}}$ (blue). (bars) Intervals indicating the top $90\%$ of all $P_{\textrm{M}}$ values found.}
		\label{F29}
	\end{figure}
	
	Starting with $\sigma '$, indicated by the black dots in figure \ref{F29}, we find a trend which is consistent with the sequential bipartite graphs from earlier in this study. As $N$ increases, the rescaled standard deviation $\sigma '$ of the solution space distribution $\mathbb{W}$ decreases, and consequently we find higher $P_{\textrm{M}}$ values (blue dots). Accompanying each average $P_{\textrm{M}}$ are intervals which represent the top 90$\%$ of all values found.  These bars are in agreement with figure \ref{F24}, whereby the average values may be high, but working with randomized data is always subject to occasional $\mathbb{W}$ distributions which are inherently difficult to boost $| \textrm{P}_{\textrm{min}} \rangle$. Even for $N=11$, which was the largest size studiable with our computing resources, we still found the effects of randomness to be strong enough to cause $P_{\textrm{M}}$ values to be under $40\%$.  
	
	Finally, using average $\mathbb{W}$ statistics in our simulator, we found predicted $P_{\textrm{M}}$ values which were in strong agreement with those shown in figure \ref{F29}.  For problem sizes $N=9,10,11$, the simulator predicted $P_{\textrm{M}}$ values which were all within $5\%$ of the averages found experimentally.  For smaller $N$ sizes, the resulting $\mathbb{W}$ distributions become less and less resemblant of gaussian profiles, making their comparison to our perfect gaussian simulator less meaningful. Overall, the two trends shown in figure \ref{F29} are positive for quantum, indicating that as $N$ increases so too does the viability of boosting $| \textrm{P}_{\textrm{min}} \rangle $.

	%%%%%%%%%%%%%%%%%%%%%%%%%%%%%%%%%%%%%%%%%%%%%%%%
	%%%%%%%%%%%%%%%%%%%%%%%%%%%%%%%%%%%%%%%%%%%%%%%%
	\section{Conclusion}%                                     Conclusion
	%%%%%%%%%%%%%%%%%%%%%%%%%%%%%%%%%%%%%%%%%%%%%%%%
	%%%%%%%%%%%%%%%%%%%%%%%%%%%%%%%%%%%%%%%%%%%%%%%%
	
	Amplitude amplification is a powerful tool for the future success of quantum computers, but it is not strictly limited to the unstructured search problem proposed by Grover over two decades ago \cite{grover}.  In this study, we've demonstrated the viability of amplitude amplification as a means for solving a completely different problem type, namely pathfinding through a weighted directed graph.  This was made possible by two key factors: 1) a cost oracle capable of encoding all possible solutions via phases, and 2) the gaussian-like manner in which the solution space naturally occurs.  It is because of these gaussian-like distributions that we are able to boost the desired solution state to high probabilities.  More specifically, we are able to utilize the central cluster of states around the mean of the gaussian to create an oracle $U_{\textrm{P}}$ which produces a mean point away from the desired solution state in amplitude space.  This in turn allows for reflections about the average at each step via $U_{\textrm{s}}$ to incrementally increase the probability of the desired solution state up to some maximum ($P_{\textrm{M}}$), which can be related back to the distribution encoded into $U_{\textrm{P}}$.  And finally, we've demonstrated that such oracles are implementable for the gate-based model of quantum computing, such that the answer to the optimization problem is not directly encoded into the quantum circuit for $U_{\textrm{P}}$.

	%%%%%%%%%%%%%%%%%%%%%%%%%%%%%%%%%%%%%%%%
	\subsection{Future Work}%                                                                                                                                   Future Work
	%%%%%%%%%%%%%%%%%%%%%%%%%%%%%%%%%%%%%%%%
	
	The algorithmic potential for gaussian amplitude amplification presented in this study is a promising first step, but there is still much to be learned.  We view the process illustrated in figure \ref{F14} as an open question for a more rigorous mathematical study.  Throughout this study we were able to simulate gaussian amplitude amplification classically because each Hilbert Space had a finite number of states.  However, studying a truncated continuous gaussian function as it undergoes $U_{\textrm{s}} U_{\textrm{P}}$ through many steps is more difficult, but could lead to improved success of the algorithm.  Additionally, studying the same process but with a skewed gaussian could yield highly valuable insight into more realistic problem cases, such as why certain $\mathbb{W}$ distributions in figure \ref{F24} performed better than others.
	
	Much of the discussion in section VII. was centered around the scaling constant $p_{\textrm{s}}$ and its role for unlocking successful amplitude amplifications.  This is arguably the biggest unknown for the future success of the algorithm.  We demonstrated that given an optimal $p_{\textrm{s}}$ the algorithm can solve for the desired solution, but it is still unclear under what circumstances an experimenter can reliably obtain $p_{\textrm{s}}$ since it changes from problem to problem.  We also showed the degree to which an average $p_{\textrm{s}}$ could be used, which we believe is a viable application for quantum under certain circumstances, requiring further research.  Alternatively, it is possible that an optimal $p_{\textrm{s}}$ could be found through a learning style algorithm, such as QAOA \cite{qaoa,qaoa2} or VQE \cite{vqe}, whereby the results of each attempted amplitude amplification are fed back to a classical optimizer.
	
	Finally, the Traveling Salesman oracle in section VIII. is a theoretical application, but with the highest upside for a quantum speedup (O($\sqrt{N!}$)), relying on future qudit technology for realization.  Critically, we neglected to provide an efficient quantum circuit for $U_{\textrm{P}}$ (an inefficient circuit is easy to construct, but too cumbersome to provide a quantum speedup), which is an open question we are still pursuing.  Beyond the TSP however, we plan to investigate more optimization problems which also naturally give rise to gaussian solution space distributions, making them candidates for amplitude amplification.
	
	%%%%%%%%%%%%%%%%%%%%%%%%%%%%%%%%%%%%%%%%
	\section*{Acknowledgments}
	%%%%%%%%%%%%%%%%%%%%%%%%%%%%%%%%%%%%%%%%
	
	We gratefully acknowledge support from the Griffiss Institute.  Any opinions, findings, conclusions or recommendations expressed in this material are those of the author(s) and do not necessarily reflect the views of AFRL.
	
	%%%%%%%%%%%%%%%%%%%%%%%%%%%%%%%%%%%%%%%%
	\section*{Data \& Code Availability}
	%%%%%%%%%%%%%%%%%%%%%%%%%%%%%%%%%%%%%%%%
	
	The data and code files that support the findings of this study are available from the corresponding author upon reasonable request.

	\clearpage
	
	%%%%%%%%%%%%%%%%%%%%%%%%%%%%%%%%%%              Bibliography      
	    %%%%%%%%%%%%%%%%%%%%%%%%%%%%%%%%%% 
	
	\clearpage

	%%%%%%%%%%%%%%%%%%%%%%%%%%%%%%%%%%     
	\appendix

	\section{$U_{\textrm{P}}$ Fidelity Results}
	
	Here we present experimental results which demonstrate the viability of implementing $U_{\textrm{P}}$ on IBM's state-of-the-art qubit architectures `Casablanca' and `Lagos'\cite{ibmq}.  Because $U_{\textrm{P}}$ only applies phases (which are undetectable through measurements), each experiment consists of an application of $U_{\textrm{P}}$ followed by $U_{\textrm{P}}^{\dagger}$, ensuring that each experiment has a definitive measurement result for calculating fidelity (the state of all $|0\rangle$'s).  Equation \ref{EA1} below shows the fidelity metric used.
	
	\begin{eqnarray}             
		|\Psi \rangle &=&  H^{\otimes L} U_{\textrm{P}}^{\dagger} U_{\textrm{P}}  H^{\otimes L} | 0^{\otimes L} \rangle  \\
		f &=& \langle 0^{\otimes L} | \Psi \rangle
		\label{EA1}
	\end{eqnarray}
	
\begin{figure}[h]                 
		\centering
		\includegraphics[scale=.4]{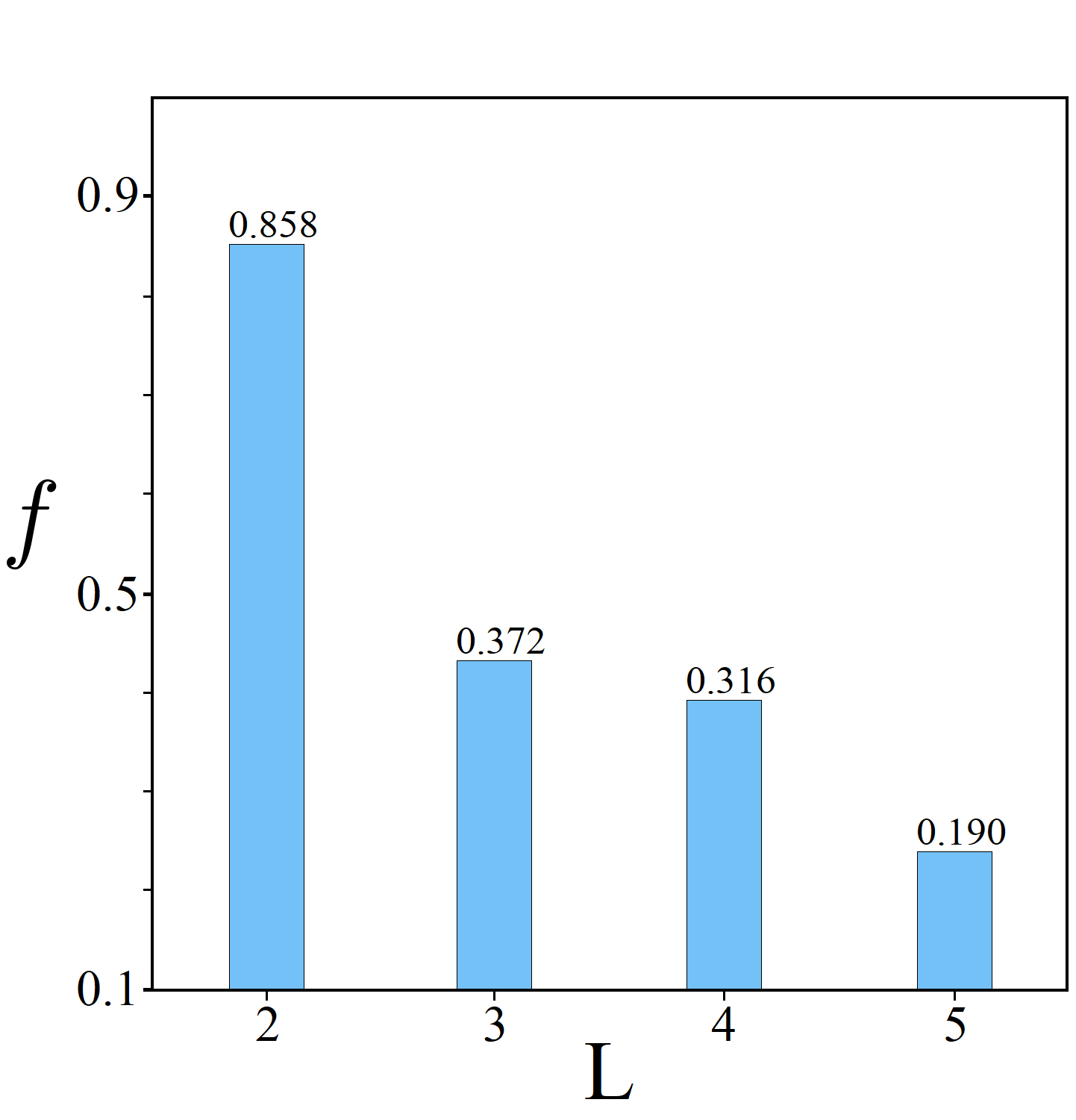}
		\caption{Fidelity results as defined in equation \ref{EA1}, for the case $N=2$,  $L \in {2,3,4,5}$, performed on IBM's superconducting qubits. }
		\label{F30}
	\end{figure}

	Because of the multiplicative nature of fidelities, the actual fidelity of a single $U_{\textrm{P}}$ application can be estimated as higher than the values shown in figure \ref{F30}.  Also note the dramatic decrease in fidelity between experiments $L=2$ and $L=3$.  This drop off can be explained by revisiting figure \ref{F9}, and noting the difference in circuit depth for $U_{\textrm{P}}$ when using $2$ versus $3$ qubits.  For the special case of $L=2$, we have $U_{\textrm{P}}$ $=$ $U_{ij}$ (equation \ref{E15}), while for all other cases $U_{\textrm{P}}$ requires two sets of $U_{ij}$ operations (figure \ref{F9}).  This difference in circuit depth explains the high fidelity for $L=2$ versus $L=3,4,5$.
	
	\section{Step-varying $p_{\textrm{s}}$}
	
	In order to compute the maximal $P_{\textrm{M}}$ values displayed in figures \ref{F23} and \ref{F24}, we used a classical simulation of the quantum state $| \Psi \rangle$ at each step of the amplitude amplification process in order to determine optimal $p_{\textrm{s}}$ values.  At each step of the algorithm we test a range of $p_{\textrm{s}}$ values when applying $U_{\textrm{P}}$, tracking the distance in amplitude space between the state $| \textrm{P}_{\textrm{min}} \rangle$ and collective mean, given in equation \ref{EB4}.  Once a maximal $D$ is found at each step, the corresponding $p_{\textrm{s}}$ value is stored, the diffusion operator $U_{\textrm{s}}$ is applied to $| \Psi \rangle$, and the resulting probability $P_{\textrm{M}}$ for $| \textrm{P}_{\textrm{min}} \rangle$ is recorded.  This process is repeated until the simulation finds a $P_{\textrm{M}}$ value which is smaller than the previous step, signaling the rebound point of the algorithm.
	
	\begin{eqnarray}          
		| \Psi \rangle &=& \sum_k^{N^L} \alpha_{k} | \textrm{P}_k \rangle  \label{EB1} \\
		\textrm{Dist}(\alpha,\beta) &\equiv& \sqrt{ \textrm{real}(\alpha - \beta)^2 + \textrm{imag}(\alpha - \beta)^2 } \label{EB2} \\
		\alpha_{mean} &=& \frac{1}{N^L} \sum_k^{N^L} \alpha_k \label{EB3}\\
		D &=& \textrm{Dist}( \alpha_{\textrm{mean}},\alpha_{\textrm{min}}  )  \label{EB4}
	\end{eqnarray}
	
	\begin{figure}[h]              
		\centering
		\includegraphics[scale=.27]{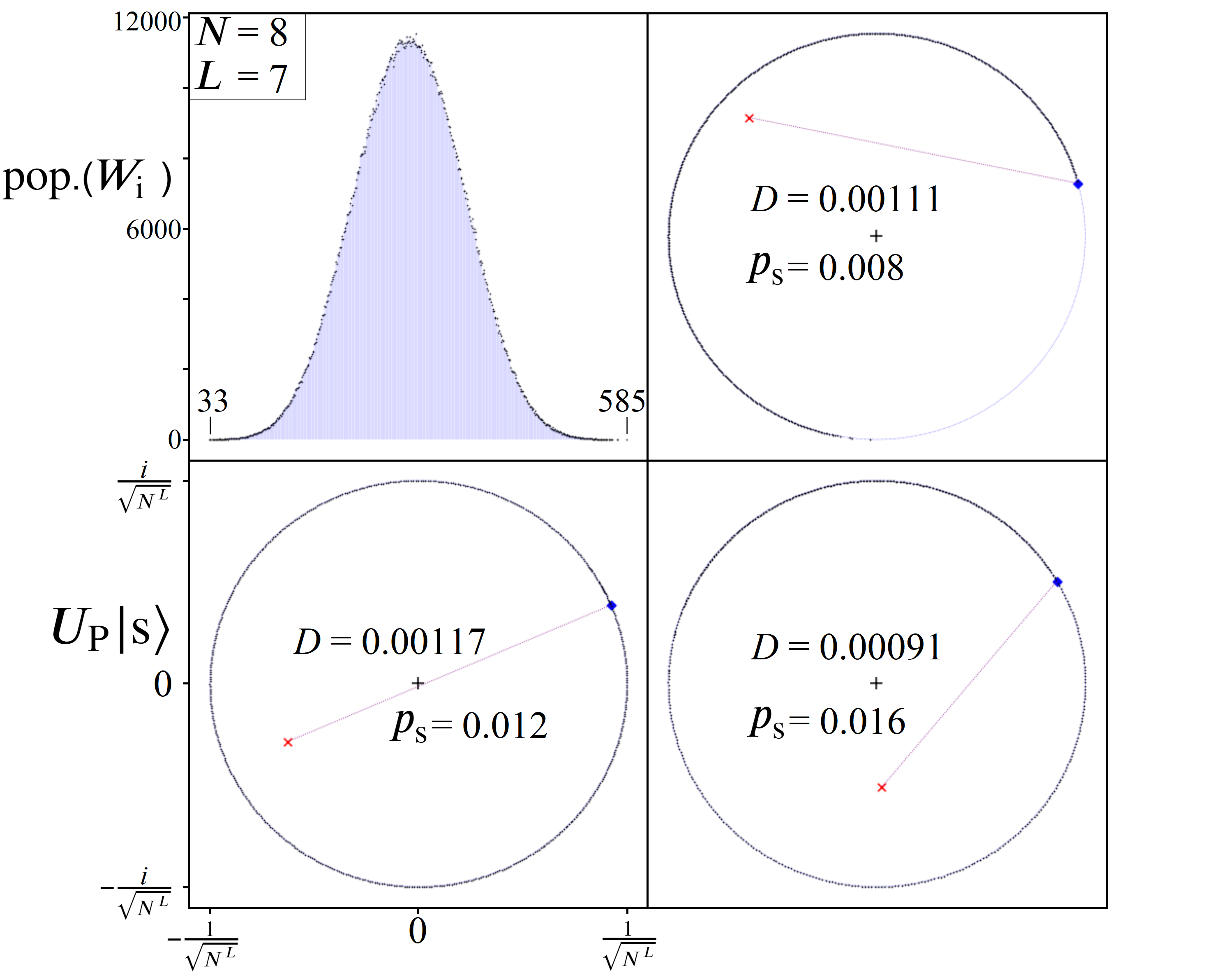}
		\caption{Illustration of the classical simulation technique used to determine the optimal $p_{\textrm{s}}$ value at each step by maximizing the distance between $| \textrm{P}_{\textrm{min}} \rangle$ and the mean point.}
		\label{F31}
	\end{figure}
	
	Figure \ref{F31} illustrates an example $\mathbb{W}$ distribution, along with three $p_{\textrm{s}}$ values and their effect on $|\Psi \rangle$ after the first application of $U_{\textrm{P}}$.  In each $U_{\textrm{P}} | \textrm{s} \rangle $ amplitude plot, the value of $D$ and $p_{\textrm{s}}$ are shown, along with a line connecting the locations of $\alpha_{\textrm{min}}$ and $\alpha_{\textrm{mean}}$.


\begin{thebibliography}{99}
		\bibitem{grover} L. K. Grover, A fast quantum mechanical algorithm for database search. arXiv: 9605043 (1996)
		\bibitem{boyer} M. Boyer, G. Brassard, P. Hoyer, A. Tapp, Tight bounds on quantum searching. Fortschritte der Physik \textbf{46}: 493-506 (1998)
		\bibitem{bennett} C. H. Bennett, E. Bernstein, G. Brassard, U. Vazirani, Strengths and Weaknesses of Quantum Computing. SIAM Journal on Computing \textbf{26}(5):1510-1523 (1997)
		\bibitem{farhi} E. Farhi and S. Gutmann, Analog analogue of a digital quantum computation. Phys. Rev. A \textrm{57}, 2403 (1998)
		\bibitem{brassard1} G. Brassard, P. Hoyer, A. Tapp, Quantum Counting, 25th Intl. Colloquium on Automata, Languages, and Programming (ICALP), LNCS 1443, pp. 820-831, (1998)
		\bibitem{brassard2}G. Brassard, P. Hoyer, M. Mosca, A. Tapp, Quantum Amplitude Amplification and Estimation. AMS Contemporary Mathematics \textbf{305}:53-74 (2002)
		\bibitem{childs} A. M. Childs and J. Goldstone, Spatial search by quantum walk. Phys. Rev. A \textbf{70}, 022314 (2004)
		\bibitem{ambainis} A. Ambainis, Variable time amplitude amplification and a faster quantum algorithm for solving systems of linear equations. arXiv: 1010.4458 (2010)
		\bibitem{singleton} R. L. Singleton Jr., M. L. Rogers, D. L. Ostby, Grover's Algorithm with Diffusion and Amplitude Steering. arXiv: 2110.11163 (2021)
		\bibitem{lloyd} S. Lloyd, Quantum search without entanglement. Phys. Rev. A \textbf{61}, 010301(R) (1999)
		\bibitem{viamontes} G. F. Viamontes, I. L. Markov, J. P. Hayes, Is Quantum Search Practical? arXiv: 0405001 (2004)
		\bibitem{reg} O. Regev and L. Schiff, Impossibility of a Quantum Speed-up with a Faulty Oracle. arXiv: 1202.1027 (2012)
		\bibitem{seidel} R. Seidel, C. K-U. Becker, S. Bock, N. Tcholtchev, I-D. Gheorge-Pop, M. Hauswirth, Automatic Generation of Grover Quantum Oracles for Arbitrary Data Structures. arXiv: 2110.07545  (2021)
		\bibitem{nielsen} M. A. Nielsen, I. L. Chuang, \textit{Quantum Computation and Quantum Information}, Cambridge University Press, pg. 249 (2000)
		\bibitem{long1} G. L. Long, W. L. Zhang, Y. S. Li, L. Niu, Arbitrary Phase Rotation of the Marked State Cannot Be Used for Grover's Quantum Search Algorithm. Commun. Theor. Phys. \textbf{32}(3), 335 (1999).
		\bibitem{long2} G. L. Long, Y. S. Li, W. L. Zhang, L. Niu, Phase matching in quantum searching. Phys. Lett. A \textbf{262}, 27-34 (1999).
		\bibitem{hoyer} P. Hoyer, Arbitrary phases in quantum amplitude amplification. Phys. Rev. A \textbf{62}, 052304 (2000)
		\bibitem{younes} A. Younes, Towards More Reliable Fixed Phase Quantum Search Algorithm. Applied Mathematics \& Information Sciences \textbf{1}(7), 10 (2013)
		\bibitem{li1} T. Li, W-S. Bao, W-Q. Lin, H. Zhang, X-Q. Fu, Quantum Search Algorithm Based on Multi-Phase. Chinese Phys. Lett. \textbf{31}(5), 050301 (2014)
		\bibitem{guo} Y. Guo, W. Shi, Y. Wang, J. Hu, Q-Learning-Based Adjustable Fixed-Phase Quantum Grover Search Algorithm. Journal of the Physical Society of Japan \textbf{86}, 024006 (2017)
		\bibitem{song} P. H. Song and I. Kim, Computational leakage: Grover's algorithm with imperfections. Eur. Phys. Jour. D \textbf{23}, 299-303 (2003)
		\bibitem{pomeransky} A. A. Pomeransky, O. V. Zhirov, D. L. Shepelyansky, Phase diagram for the Grover algorithm with static imperfections. Eur. Phys. Jour. D \textbf{31}, 131-135 (2004)
		\bibitem{janmark} J. Janmark, D. A. Meyer, T. G. Wong, Global Symmetry is Unnecessary for Fast Quantum Search. Phys. Rev. Lett. \textbf{112}, 210502 (2014)
		\bibitem{gutin} 	G. Gutin and A. P. Punnen, \textit{The Traveling Salesman Problem and Its Variations}, Springer New York  (2007)
		\bibitem{srinivasan} K. Srinivasan, S. Satyajit, B. K. Behera, P. K. Panigrahi, Efficient quantum algorithm for solving travelling salesman problem: An IBM quantum experience. arXiv:1805.10928 (2018)
		\bibitem{moylett} D. J. Moylett, N. Linden, A. Montanaro, Quantum speedup of the traveling-salesman problem for bounded-degree graphs. Phys. Rev. A \textbf{95}, 032323 (2017)
		\bibitem{martonak} R. Martoňák, G. E. Santoro, E. Tosatti, Quantum annealing of the traveling-salesman problem. Phys. Rev. E \textbf{70}, 057701 (2004)
		\bibitem{warren} R. H. Warren, Adapting the traveling salesman problem to an adiabatic quantum computer. Quantum Information Processing \textbf{12}, pgs 1781–1785 (2013)
		\bibitem{warren2} R. H. Warren, Solving the traveling salesman problem on a quantum annealer. SN Applied Sciences \textbf{2}, 75 (2020)
		\bibitem{chen} H. Chen, X. Kong, B. Chong, G. Qin, X. Zhou, X. Peng, J. Du, Experimental demonstration of a quantum annealing algorithm for the traveling salesman problem in a nuclear-magnetic-resonance quantum simulator. Phys. Rev. A \textbf{83}, 032314 (2011)
		\bibitem{bang} J. Bang, S. Yoo, J. Lim, J. Ryu, C. Lee, J. Lee, Quantum heuristic algorithm for traveling salesman problem. J. Korean Phys. Soc. \textbf{61}, 1944 (2012)
		\bibitem{kues}M. Kues et. al, On-chip generation of high-dimensional entangled quantum states and their coherent control. Nature \textbf{546} 622–626 (2017)
		\bibitem{low} P. J. Low, B. M. White, A. A. Cox, M. L. Day, C. Senko, Practical trapped-ion protocols for universal qudit-based quantum computing. Phys. Rev. Research \textbf{2}, 033128 (2020)
		\bibitem{yurtalan} M. A. Yurtalan, J. Shi, M. Kononenko, A. Lupascu, S. Ashhab, Implementation of a Walsh-Hadamard gate in a superconducting qutrit. Phys. Rev. Lett. \textbf{125}, 180504 (2020)
		\bibitem{lu} 	H-H. Lu, Z. Hu, M. S. Alshaykh, A. J. Moore, Y. Wang, P. Imany, A. M. Weiner, S. Kais, Quantum Phase Estimation with Time-Frequency Qudits in a Single Photon. Adv. Quantum Technol. 1900074 (2019)
		\bibitem{niu} M. Y. Niu, I. L. Chuang, J. H. Shapiro, Qudit-Basis Universal Quantum Computation Using $\chi^2$ Interactions. Phys. Rev. Lett. \textbf{120}, 160502 (2018)
		\bibitem{luo} M-X. Luo and X-J. Wang, Universal quantum computation with qudits. Sci. China Phys. Mech. Astron. \textbf{57}, 1712–1717 (2014)
		\bibitem{li2} B. Li, Z-H. Yu, S-M. Fei, Geometry of Quantum Computation with Qutrits. Scientific Reports \textbf{3}, 2594 (2013)
		\bibitem{lanyon} B. P. Lanyon et. al, Quantum computing using shortcuts through higher dimensions. Nature Physics \textbf{5}, 134–140 (2009)
		\bibitem{gokhale} P. Gokhale, J. M. Baker, C. Duckering, N. C. Brown, K. R. Brown, F. T. Chong, ISCA '19: Proceedings of the 46th International Symposium on Computer Architecture, 554–566 (2019)
		\bibitem{khan} 	F. S. Khan, M. Perkowski, Synthesis of multi-qudit Hybrid and d-valued Quantum Logic Circuits by Decomposition. Theoretical Computer Science \textbf{367}(3), 1 pgs. 336-346 (2006)
		\bibitem{muth} A. Muthukrishnan and C. R. Stroud Jr., Multi-valued Logic Gates for Quantum Computation. Phys. Rev. A \textbf{62}, 052309 (2000)
		\bibitem{daboul} J. Daboul, X. Wang, B. C. Sanders, Quantum gates on hybrid qudits. J. Phys. A: Math. Gen. \textbf{36} (14), 2525-2536 (2003)
		\bibitem{blok} M. S. Blok, V. V. Ramasesh, T. Schuster, K. O'Brien, J. M.~Kreikebaum, D. Dahlen, A. Morvan, B. Yoshida, N. Y. Yao, I. Siddiqi, Quantum Information Scrambling on a Superconducting Qutrit Processor. Phys. Rev. X \textbf{11}, 021010 (2021)
		\bibitem{hu} X-M. Hu, C. Zhang, B-H. Liu, Y. Cai, X-J. Ye, Y. Guo, W-B. Xing, C-X. Huang, Y-F. Huang, C-F. Li, G-C. Guo, Experimental High-Dimensional Quantum Teleportation. Phys. Rev. Lett. \textbf{125}, 230501 (2020)
		\bibitem{laplace} P. S. Laplace, Mémoire sur les approximations des formules qui sont fonctions de très grands nombres et sur leur application aux probabilités. Mémoires de l'Académie Royale des Sciences de Paris, \textbf{10} (1810)
		\bibitem{bernoulli} J. Bernoulli, \textit{Ars Conjectandi}, Basileae: Thurnisiorum. (1713)
		\bibitem{gauss} C. F. Gauss, \textit{Theoria Motus Corporum Coelestium in Sectionibus Conicis Solem Ambientium}, Hamburg: Friedrich Perthes and I.H. Besser (1809)
		\bibitem{shyamsundar} P. Shyamsundar, Non-Boolean Quantum Amplitude Amplification and Quantum Mean Estimation. arXiv: 2102.04975 (2021)
		\bibitem{satoh} T. Satoh, Y. Ohkura, R. V. Meter, Subdivided Phase Oracle for NISQ Search Algorithms. IEEE Transactions on Quantum Engineering (2020)
		\bibitem{bench} N. Benchasattabuse, T. Satoh, M. Hajdušek, R. V. Meter, Amplitude Amplification for Optimization via Subdivided Phase Oracle. arXiv:2205.00602 (2022)
		\bibitem{koch} D. Koch, L. Wessing, P. M. Alsing, Introduction to Coding Quantum Algorithms: A Tutorial Series Using Qiskit. arXiv:1903.04359 (2019)
		\bibitem{wang} Y. Wang, Z. Hu, B. C. Sanders, S. Kais, Qudits and High-Dimensional Quantum Computing.  Front. Phys. \textbf{10} (2020)
		\bibitem{qaoa} E. Farhi, J. Goldstone, S. Gutmann, A Quantum Approximate Optimization Algorithm. arXiv:1411.4028 (2014)
		\bibitem{qaoa2} S. Hadfield, Z. Wang, B. O'Gorman, E. G. Rieffel, D. Venturelli, R. Biswas, From the Quantum Approximate Optimization Algorithm to a Quantum Alternating Operator Ansatz. Algorithms \textbf{12}(2), 34 (2019)
		\bibitem{vqe} A. Peruzzo, J. McClean, P. Shadbolt, M-H. Yung, Z-Q. Zhou, P. J. Love, A. Aspuru-Guzik, J. L O'Brien, A variational eigenvalue solver on a quantum processor. Nature Communications \textbf{5}, 4213 (2014)
		\bibitem{ibmq}  IBM 7-Qubit Casablanca and Lagos Architectures, https://quantum-computing.ibm.com. Accessed July - Aug. 2021
		
		
		
		
		
		
	\end{thebibliography}
\end{document}